%% file: main.tex
\documentclass{article}

\usepackage{arxiv}

\usepackage{natbib}
\usepackage[utf8]{inputenc} % allow utf-8 input
\usepackage[T1]{fontenc}    % use 8-bit T1 fonts
\usepackage[hidelinks]{hyperref}
\usepackage{url}            % simple URL typesetting
\usepackage{booktabs}       % professional-quality tables
\usepackage{amsfonts}       % blackboard math symbols
\usepackage{nicefrac}       % compact symbols for 1/2, etc.
\usepackage{microtype}      % microtypography
\usepackage{lipsum}
\usepackage{graphicx}
\graphicspath{ {./images/} }

\input{packages}

\input{definitions}

\title{Rao--Blackwellized Markov Chain Monte Carlo\\ Light Transport}

\author{
  Sascha Holl\thanks{Corresponding author.}\\
  Max Planck Institute for Informatics\\
  Saarland University\\
  Saarbrücken, Germany\\
  \texttt{sascha.holl@gmail.com}\\
  \And
  Gurprit Singh\\
  Advanced Micro Devices, Inc. (AMD)\\
  Munich, Germany\\
  \texttt{gurprit.singh@amd.com}\\
  \And
  Hans-Peter Seidel\\
  Max Planck Institute for Informatics\\
  Saarbrücken, Germany\\
  \texttt{hpseidel@mpi-inf.mpg.de}\\
}

\begin{document}
\maketitle

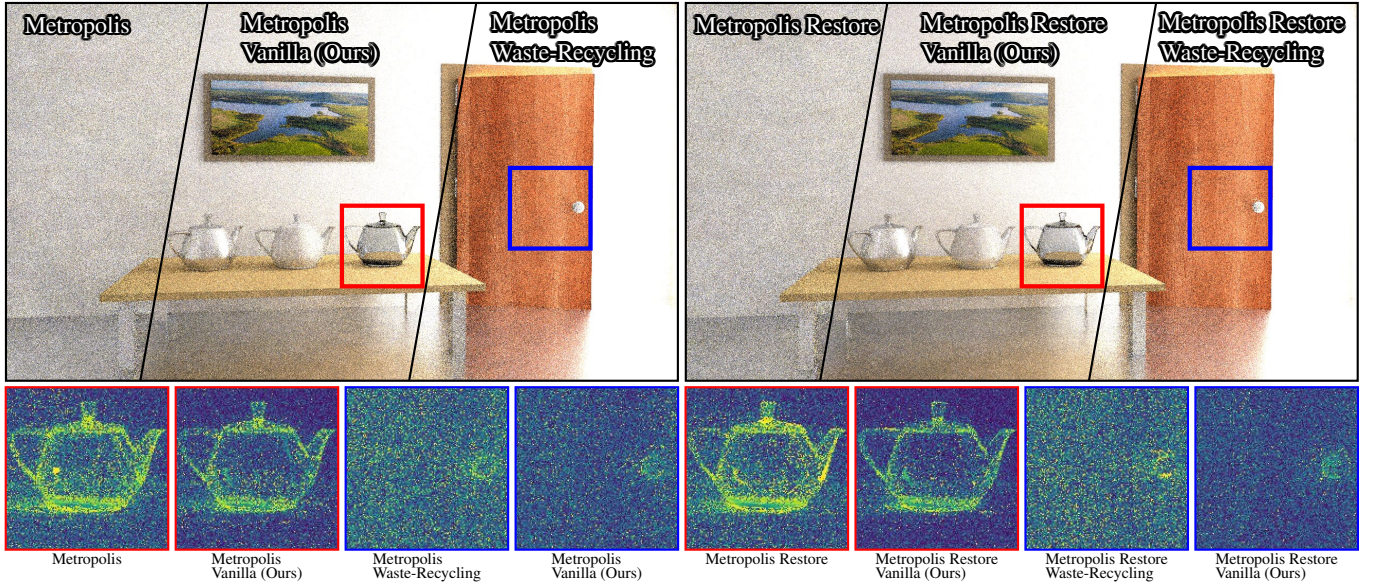
\begin{figure}[t]
    \centering
    \input{figures/teaser}
    
    \captionsetup{skip = 0pt}
    \caption{
        We introduce a computationally efficient and \textemdash\ compared to the traditional approach \textemdash\ superior \gls*{rb} technique \citep{robert2021rao} for both the \gls*{mh} algorithm \citep{hastings1970monte} and the recently proposed \gls*{jrlt} algorithm \citep{holl2025jrlt}. Applied to light transport simulation \citep{pharr2023pbrt}, the left rendering compares, in sequence, standard \gls*{mmlt} \citep{hachisuka2014multiplexed} (for simplicity called \emph{Metropolis}, following the terminology of \citet{holl2025jrlt}), our proposed \gls{rb} technique \emph{vanilla \acrlong{rb}}, and the \gls{rb} method commonly used in implementations of \gls{mh}-based light transport algorithms, \emph{waste-recycling}. The right rendering shows the corresponding comparison for \gls{jrlt}: Standard \gls{jrlt}, our vanilla \gls{rb} variant, and waste-recycling within the \gls{mh}-based local dynamics of \gls{jrlt}. The inset error maps show the corresponding \acrfull{mape} for each rendering.
    }
    \label{fig:teaser}
\end{figure}

\begin{abstract}
    In light transport simulation, \gls{mcmc} methods are particularly effective at exploring regions with complex lighting characteristics. However, estimator variance is a central concern across \gls{mc} methods in general. In light transport, high variance directly manifests as increased noise or, equivalently, longer rendering times at fixed image quality. Variance reduction techniques based on \gls{rb} have proven particularly effective. In practice, however, the \gls{rb} approach traditionally used in light transport, \emph{waste-recycling}, can yield little to no measurable variance reduction, a fact we empirically confirm in this work. Motivated by this lack of effective variance reduction, we introduce a novel \gls{rb} technique for the general-purpose \gls{mh} algorithm that is computationally efficient and achieves substantial variance reduction. We show that this method consistently outperforms waste-recycling in terms of both variance reduction and convergence speed. Building on this result, we adapt the proposed \gls{rb} approach to the recently introduced general-purpose Jump Restore algorithm, where it similarly achieves substantial variance reduction and accelerated convergence. Through extensive experiments in light transport simulation, we demonstrate that our \gls{rb} technique significantly outperforms the traditional approaches for both \gls{mh}-based light transport algorithms and \acrfull{jrlt}, under both equal-time and equal-sample-count comparisons.
\end{abstract}

\maketitle

% ----------------------------------------------------------------------------------------------------
% introduction
% ----------------------------------------------------------------------------------------------------
\section{Introduction}

\paragraph*{\gls{mcmc} in light transport}

In light transport, the computation of high-dimensional integrals is essential and is typically performed using \gls{mc} integration. Traditionally, this method involves generating independent samples within each pixel of the image space. However, a major drawback of this approach is that samples are generated regardless of their contribution to the final estimate. Even if a sample has no significant impact on the result, it is still drawn from a predefined importance distribution, without considering the value of the target density at the sampled location.
\gls{mcmc} methods offer a way to address this inefficiency. By constructing a Markov process, sample generation can be guided to better align with the target distribution, allowing for a more structured exploration of the underlying space.
\citet{veach1997thesis} introduced the \acrfull{mh} algorithm, a widely used \gls{mcmc} method, to the graphics community. Building on this pioneering work, numerous \gls{mh}-based \gls{mcmc} light transport algorithms have since been proposed. 
Beyond this, \citet{holl2025jrlt} recently introduced an alternative \emph{continuous-time} \gls{mcmc} approach, known as \emph{\acrfull{jrlt}}.
% Recently, \citet{holl2025jrlt} introduced an alternative \emph{continuous-time} \gls{mcmc} approach, known as \emph{\acrfull{jrlt}}. % This method constructs a pure-jump type Markov process \citep[Chapter~IV]{kallenberg2021probability}, where between jumps, the local dynamics are given by the continuous-time embedding of a \gls{mh} chain, and transitions occur after exponentially distributed holding times.

\paragraph*{Variance reduction}

In \gls{mcmc}-based light transport, estimator variance is a central concern, as high variance directly translates into increased noise or longer rendering times at equal image quality. For general applications, a widely used variance reduction technique is \acrlong{rb}, which replaces an estimator by its conditional expectation given a sufficient statistic, thereby producing an estimator that is at least as efficient and typically strictly superior.
In light transport, a \gls{rb} technique commonly referred to as \emph{waste-recycling} \citep{ceperley1977waste} is traditionally employed \citep[Section~4.6]{rioux2020delayed}. While this approach is computationally inexpensive, it does not guarantee variance reduction
.
% \citep{delmas2009waste}.
% In this paper, we propose an alternative \gls{rb} technique, \emph{vanilla \acrlong{rb}}, that remains computationally efficient while achieving a significant variance reduction and consistent improvements across all metrics in equal-time and equal-sample-count renderings.

\paragraph*{Contributions}

We empirically confirm that \emph{waste-recycling} does not necessarily lead to a significant reduction in estimator variance in light transport (\Cref{sec:numerical-study}). Motivated by this lack of effective variance reduction, we consider a well-known \gls{rb} technique for the general-purpose \gls{mh} algorithm, \emph{vanilla \acrlong{rb}}, which originally was computationally prohibitive for practical use. We therefore derive a computationally efficient variant (\Cref{sec:metropolis-hastings-vanilla}) and empirically demonstrate (\Cref{sec:numerical-study}) that it achieves substantial variance reduction and accelerated convergence across multiple standard metrics, under both equal-sample-count and equal-rendering-time comparisons. Finally, we adapt the vanilla \gls{rb} principle to the recently proposed general-purpose Jump Restore algorithm (\Cref{sec:jump-restore-vanilla-estimation}) and similarly demonstrate its substantial superiority
.
\section{Related work}\label{sec:related-work}

\paragraph*{Seminal work}

The rendering equation~\citep{kajiya1986rendering} is traditionally solved using \gls{mc} estimators such as ordinary path tracing~\citep{pharr2023pbrt} or its bidirectional extension~\citep{lafortune1996rendering,veach1994bidirectional}. While effective, these estimators draw samples without regard to their eventual contribution, limiting their ability to capture challenging transport effects. This motivated \citet{veach1997thesis} to introduce \gls{mcmc}-based rendering to light transport by adapting the \gls{mh} algorithm~\citep{metropolis1953equation,hastings1970monte}. Later, \citet{kelemen2002simple} replaced the path space representation by a Euclidean \emph{primary sample space}, significantly simplifying proposal design and making \gls{mh} practical for rendering. Since then, numerous \gls{mh} variants have emerged.

\paragraph*{Diffusion-based \gls{mh}}

Some of these variants correspond to the ordinary \gls{mh} algorithm, but equipped with specific proposal kernels. In the statistics community, many such proposal kernels are derived from time-discretized stochastic diffusion processes. Among the most popular are Langevin-based~\citep{roberts1996mala} and Hamiltonian-based~\citep{duane1987hmc} proposals, which make use of gradient information to improve local exploration of the target distribution. In rendering, these ideas were adapted by \citet{li2015anisotropic}, who employed Hamiltonian dynamics with first- and second-order derivatives, and by \citet{luan2020langevin}, who demonstrated that first-order gradients are already sufficient for producing guided proposals at lower computational cost.% Despite their benefits, these approaches remain demanding due to their reliance on derivative evaluations.

% \vspace{-.1cm}
% \paragraph*{General-purpose \gls{mh} variants}

% Other methods are genuine algorithmic extensions of \gls{mh} tailored to light transport. These include delayed rejection~\citep{mira2001delayed,rioux2020delayed}, which postpones rejecting proposals to reduce asymptotic variance; multiple-try \gls{mh}~\citep{liu2000multiple,segovia2007multiple}, which samples several proposal candidates to improve acceptance rates; charted \gls{mh}~\citep{marinari1992charted,pantaleoni2017charted}, which switches between space parameterizations; and reversible jump \gls{mh}~\citep{green1995reversible,bitterli2017reversible}, enabling transitions between state spaces of different dimensionality.% These methods increase the flexibility of \gls{mcmc} rendering, but still inherit the reversibility and backtracking tendency typical for \gls{mh}, which can slow convergence.

% % \paragraph*{Light-transport-specific \gls{mh} variants}

% % Beyond general-purpose methods, several techniques have been invented specifically for light transport. Examples include approaches fusing multiple proposal strategies~\citep{otsu2017fusing}, exploiting geometric structure~\citep{otsu2018geometry}, improving path space exploration using halfway-vector representations~\citep{hanika2015improved}, or advancing specular path sampling by constraining proposals to low-dimensional manifolds~\citep{jakob2012manifold}.

\paragraph*{\Gls{bdpt}}

\citet{hachisuka2014multiplexed} integrated \gls{bdpt} into \gls{mh}, enabling the algorithm to choose among multiple sampling techniques via strategy-dependent proposal kernels, while producing estimates in a manner reminiscent of multiple importance sampling. Most later \gls{mcmc}-based rendering algorithms adopt this multiplexed \gls{mh} variant as their basis.

\paragraph*{Stratification and global exploration}

A key challenge in \gls{mcmc}-based rendering is balancing efficient local exploration with global discovery. \citet{gruson2020stratified} addressed this issue by spawning multiple chains across predefined strata, thereby improving uniform convergence throughout the rendering.

\paragraph*{The Restore framework}

While \gls{mh} has long been the standard for \gls{mcmc}-based light transport simulation, \citet{wang2021regeneration} introduced with \emph{Restore} a novel general-purpose continuous-time \gls{mcmc} approach, in which fixed local dynamics terminate after a finite runtime and are regenerated at new locations afterwards. \citet{mckimm2024adaptive} later added adaptive regeneration mechanisms to minimize the number of eventual regenerations. Recently, \citet{holl2025jrlt} generalized the Restore framework substantially, allowing broader classes of local dynamics, state-dependent global dynamics (i.e., initial states from which local explorations originate), and ensuring applicability of Restore in the specific context of light transport.

\paragraph*{Variance reduction}

While \gls{bdpt} and stratification attempts also aim at variance reduction, there exist dedicated variance reduction techniques for \gls{mh} estimators in general-purpose applications. Among these, the class of \gls{rb} methods \citep{robert2021rao} is the most relevant for our work. In light transport, a \gls{rb} technique commonly referred to as \emph{waste-recycling} \citep{ceperley1977waste}, is traditionally employed \citep[Section~4.6]{rioux2020delayed}.
A fundamentally different \gls{rb} approach was introduced as \emph{vanilla \acrlong{rb}} in \citet{douc2011vanilla}. At the cost of substantial additional computational overhead in the light transport setting, this \gls{rb} technique can be shown to yield a strict variance reduction. The method proposed in this work builds on this approach, but departs from it in a way that avoids these costs, while still achieving significant variance reduction.

% ----------------------------------------------------------------------------------------------------
% nomenclature
% ----------------------------------------------------------------------------------------------------
% \input{nomenclature}

% ----------------------------------------------------------------------------------------------------
% Markov chain Monte Carlo
% ----------------------------------------------------------------------------------------------------
\section{Markov chain Monte Carlo}\label{sec:mcmc-preliminiaries}

\paragraph*{Basic principle}

Given a finite measure \setword{$\targetdistribution$}{inline:target-distribution}, \gls{mcmc} is a recipe for constructing an ergodic \citep{meyn1993markov} Markov process with invariant distribution~$\targetdistribution$.
This Markov process $(\process_\timepoint)_{\timepoint \in \timedomain}$, defined on either a discrete time domain $\timedomain=\mathbb N_0$ or a continuous time domain $\timedomain=[0,\infty)$, can subsequently be used to estimate the integral \begin{equation}\label{eq:integral}
    \integral:=\displaystyle\int\integrand\dd{\targetdistribution}
    %\targetdistribution\integrand:=\int\integrand\dd{\targetdistribution}
\end{equation} of a $\targetdistribution$-integrable function \setword{$\integrand$}{inline:integrand}. In fact, \begin{align}\label{eq:ergodic-theorem}
    \ergodicaverage_\timepoint\integrand:=\frac1\timepoint\left.\begin{cases}\displaystyle\sum_{\prevtimepoint=0}^{\timepoint-1}\integrand(\process_\prevtimepoint)&\text{, if }\timedomain=\mathbb N_0\\\displaystyle\int_0^\timepoint\integrand(\process_\prevtimepoint)\dd{\prevtimepoint}&\text{, if }\timedomain=[0,\infty)\end{cases}\right\}\xrightarrow{\timepoint\to\infty}
    %\targetdistribution\integrand
    \integral
\end{align} almost surely for all $\integrand\in\mathcal L^1(\targetdistribution)$ by \emph{Birkhoff's ergodic theorem} \cite[Theorem~25.6]{kallenberg2021probability}.
For a more detailed yet concise description of \gls{mcmc}, we refer the reader to \Cref{sec:appendix-mcmc-preliminiaries}.

\paragraph*{Assumption on the target distribution}

Throughout this work, we assume that $\targetdistribution$ admits a density with respect to a {reference measure \setword{$\referencemeasure$}{inline:reference-measure}; that is, \begin{equation}\label{eq:targetdensity-assumption-1}
    \targetdensity_\referencemeasure:=\displaystyle\int\targetdensity\dd{\referencemeasure}\in(0,\infty),
\end{equation} and \begin{equation}\label{eq:target-distribution-has-density}
    \targetdistribution(\pointset)=\frac1{\targetdensity_\referencemeasure}\displaystyle\int_{\pointset}\targetdensity\dd{\referencemeasure}
\end{equation} for some density $\targetdensity$.

% ----------------------------------------------------------------------------------------------------
% Rao-Blackwellization
% ----------------------------------------------------------------------------------------------------
\section{Rao--Blackwellization}\label{sec:rao-blackwellization}

A key challenge in \gls{mc} integration is reducing the variance of estimators. \gls{rb} is a classical and effective technique for this purpose. 
\gls{rb} is the practice of replacing an estimator by its conditional expectation given additional information. The neologism originates from the celebrated \emph{Rao--Blackwell theorem} \citep{rao1945information,blackwell1947conditional}, which
%, intuitively,
states that conditioning an estimator on a sufficient statistic improves estimation under any convex loss.

From a theoretical perspective, conditional expectation acts as an orthogonal projection in $L^2$ and therefore cannot increase variance \citep[Corollary~8.17]{klenke2020probability}. 
In particular, \begin{equation}\label{eq:rao-blackwell-theorem}
    \variance\left[S\mid\filtration\right]\ge\variance\left[S\mid\secondfiltration\right],
\end{equation} for any square-integrable random variable $S$ and any pair of information sets $\filtration\subseteq\secondfiltration$.
The intuition is straightforward: conditioning on richer observations incorporates more information and therefore cannot increase uncertainty. 
Instead, variability due to extraneous randomness is averaged out, resulting in reduced or, in the worst case, unchanged variance.

In particular, replacing an estimator $S$ by its conditional expectation with respect to a larger information set never increases variance. This follows immediately by taking $\filtration$ to be the information provided by the estimator $S$ itself. In \Cref{sec:rao-blackwellized-metropolis-hastings-algorithm,sec:rao-blackwellized-jump-restore-algorithm}, we exploit this
principle
to construct Rao--Blackwellized alternatives to the ergodic average estimator~\eqref{eq:ergodic-theorem}.

\section{Rao--Blackwellized Metropolis--Hastings}\label{sec:rao-blackwellized-metropolis-hastings-algorithm}

\subsection{Metropolis--Hastings algorithm}\label{sec:metropolis-hastings-algorithm}

The \acrlong{mh} algorithm is arguably the most popular and widely applicable MCMC method. It is an algorithmic construction of a Markov chain $(\process_\timepoint)_{\timepoint\in\mathbb N_0}$ with invariant distribution $\targetdistribution$. The procedure of simulating this chain up to a given time $\timepoint\in\mathbb N_0$ is summarized in \autoref{alg:metropolis-hastings}.

% \begin{algorithm}
%     \caption{Metropolis--Hastings algorithm\newline with proposal kernel $\proposalkernel$ and target distribution $\targetdistribution$.}\label{alg:metropolis-hastings}
%     \begin{algorithmic}[1]
%         \Require{Initial state $\point_0$ and sample count $\timepoint\in\mathbb N$.}
%         \Procedure{MetropolisHastingsUpdate}{$x$}
%             \State Sample $\secondpoint$ from $\proposalkernel(\State,\;\cdot\;)$;\AlgCommentLeft{generate the proposal}
%             \State Sample $\uniform$ from $\mathcal U_{[0,\:1)}$;\AlgCommentLeft{uniform distribution on $[0,1)$}
%             \If{$\left(\uniform<\acceptancefunction(\State,\secondpoint)\right)$}
%                 \State\label{line:metropolis-hastings-accept}\Return$\secondpoint$;\AlgCommentLeft{with probability $\acceptancefunction(\State,\secondpoint)$ return the proposal}
%             \EndIf
%             \State\label{line:metropolis-hastings-reject}\Return$\State$;\AlgCommentLeft{with probability $1-\acceptancefunction(\State,\secondpoint)$ reject the proposal}
%         \EndProcedure
%         \State
%         \State\textbf{for }$\left(\prevtimepoint=1;\prevtimepoint<\timepoint;\text{++}\prevtimepoint\right)$
%         \Indent
%             \State\label{line:metropolis-hastings-update}$\point_\prevtimepoint\text{ = }\textproc{MetropolisHastingsUpdate}(\point_{\prevtimepoint-1})$;
%         \EndIndent
%         \State\Return$\left(\point_0,\ldots,\point_{\timepoint-1}\right)$;
%     \end{algorithmic}
% \end{algorithm}
\begin{algorithm}
    \caption{Metropolis--Hastings algorithm\newline with proposal kernel $\proposalkernel$ and target distribution $\targetdistribution$.}\label{alg:metropolis-hastings}
    \begin{algorithmic}[1]
        \Require{Initial state $\point_0$ and sample count $\timepoint\in\mathbb N$.}
        \Ensure{Realization $\left(\point_0,\ldots,\point_{\timepoint-1}\right)$ of the \gls{mh} chain $(\process_\timepoint)_{\timepoint\in\mathbb N_0}$}
        \Procedure{MetropolisHastingsUpdate}{$x$}
            \State Sample $\secondpoint$ from $\proposalkernel(\point,\;\cdot\;)$;\AlgCommentLeft{generate the proposal}
            \State Sample $\uniform$ from $\mathcal U_{[0,\:1)}$;\AlgCommentLeft{uniform distribution on $[0,1)$}
            \If{$\left(\uniform<\acceptancefunction(\point,\secondpoint)\right)$:}\label{line:metropolis-hastings-accept-or-reject}
                \State
                \label{line:metropolis-hastings-accept}\Return$\secondpoint$;\AlgCommentLeft{with prob. $\acceptancefunction(\point,\secondpoint)$ accept proposal}
            \EndIf
            \State\label{line:metropolis-hastings-reject}\Return$\point$;\AlgCommentLeft{with prob. $1-\acceptancefunction(\point,\secondpoint)$ reject proposal}
        \EndProcedure
        % \State
        \State\textbf{for} $\left(\prevtimepoint=1;\prevtimepoint<\timepoint;\text{++}\prevtimepoint\right):$
        \Indent
            \State
            \label{line:metropolis-hastings-update}$\point_\prevtimepoint\text{ = }\textproc{MetropolisHastingsUpdate}(\point_{\prevtimepoint-1})$;
        \EndIndent
    \end{algorithmic}
\end{algorithm}

\paragraph*{Algorithmic description}

The user has to specify a \setword{\emph{proposal kernel}}{inline:proposal-kernel} $\proposalkernel$. For every state $\point$, $\proposalkernel(\point,\;\cdot\;)$ is a probability measure. Now, at each discrete time step, the algorithm is \emph{proposing} a state transition candidate $\secondpoint$ drawn from $\proposalkernel(\point,\;\cdot\;)$, where $\point$ is the current state of the chain generated so far. With probability $\acceptancefunction(\point,\secondpoint)$, where $\acceptancefunction$ is an \setword{\emph{acceptance function}}{inline:acceptance-function}, the \emph{proposal} $\secondpoint$ is \emph{accepted} (line~\ref{line:metropolis-hastings-accept}) and the current state is set to $\secondpoint$. With the opposite probability
%, $1-\acceptancefunction(\point,\secondpoint)$,
the proposal is \emph{rejected} (line~\ref{line:metropolis-hastings-reject}) and the current state will not be changed (cf. line~\ref{line:metropolis-hastings-update}).

% \paragraph*{Requirements}

% The initial state $\point_0$ in \autoref{alg:metropolis-hastings} may be chosen arbitrarily. The only theoretical requirement imposed on the proposal kernel $\proposalkernel$ for establishing the correctness of \autoref{alg:metropolis-hastings} is that the target distribution $\targetdistribution$ is absolutely continuous with respect to $\proposalkernel(\point,\,\cdot\,)$ for every state $\point$. This condition is intuitively reasonable, as it ensures that a proposal from $\proposalkernel(\point,\,\cdot\,)$ is able to reach any region where $\targetdistribution$ has positive measure.

\paragraph*{Acceptance function}

The mechanism ensuring that the Markov chain
% $(\process_\timepoint)_{\timepoint\in\mathbb N_0}$ constructed
generated by \autoref{alg:metropolis-hastings} is $\targetdistribution$-invariant is the
% acceptance/rejection step
acceptance step
in \autoref{line:metropolis-hastings-accept}. The acceptance function $\acceptancefunction$ cannot be arbitrary, but there is more than one valid choice. The one usually given is optimal with respect to the \emph{Peskun-Tierney ordering}~\citep{tierney1998note}.

To define it, we assume that the proposal kernel $\proposalkernel$ admits a density with respect to the same reference measure $\referencemeasure$ as $\targetdistribution$; that is, \begin{equation}\label{eq:proposal-kernel-has-density}
    \proposalkernel(\point,\pointset)=\displaystyle\int_{\pointset}\proposaldensity(\point,\;\cdot\;)\dd{\referencemeasure}
\end{equation} for some density $\proposaldensity$.
The acceptance function is then defined as \begin{equation}\label{eq:optimal-acceptance-function}
    \alpha(\point,\secondpoint):=\begin{cases}\displaystyle\min\left(1,\frac{\targetdensity(\secondpoint)\proposaldensity(\secondpoint,\point)}{\targetdensity(\point)\proposaldensity(\point,\secondpoint)}\right)&\text{, if }\targetdensity(\point)\proposaldensity(\point,\secondpoint)>0;\\1&\text{, otherwise.}\end{cases}
    % \alpha(\point,\secondpoint):=\min\left(1,\frac{\targetdensity(\secondpoint)\proposaldensity(\secondpoint,\point)}{\targetdensity(\point)\proposaldensity(\point,\secondpoint)}\right)
\end{equation}

\begin{definition}\label{def:metropolis-hastings}
    \autoref{alg:metropolis-hastings} with acceptance function \eqref{eq:optimal-acceptance-function} and the generated chain
    % $(\process_\timepoint)_{\timepoint\in\mathbb N_0}$
    are called \textbf{Metropolis--Hastings algorithm} and \textbf{Metropolis--Hastings chain with proposal kernel} $\bm\proposalkernel$ \textbf{and target distribution} $\bm\targetdistribution$, respectively.
\end{definition}

\subsection{Standard estimation}\label{sec:metropolis-hastings-standard-estimation}

% Disregarding potential optimization techniques \textemdash\ whether for variance reduction or any other metric \textemdash\ the standard ergodic average estimator \eqref{eq:ergodic-theorem} is used in practice.
Integral estimation using the \gls{mh} algorithm is usually performed using the
% standard
ergodic average estimator \eqref{eq:ergodic-theorem}.
That is, since the \gls{mh} chain is a time-\emph{discrete} Markov chain, the standard estimator of $\integral$ is
% given by
\begin{equation}\label{eq:standard-estimation}
	\ergodicaverage_\timepoint\integrand=\frac1\timepoint\displaystyle\sum_{\prevtimepoint=0}^{\timepoint-1}\integrand(\process_\prevtimepoint)\eqforall\timepoint\in\mathbb N.
\end{equation}

\subsection{Waste-Recycling}\label{sec:metropolis-hastings-waste-recycling}

% In implementations of \gls{mcmc}-based light transport algorithms, such as \citet{pbrt}, the standard estimator described in \autoref{sec:metropolis-hastings-standard-estimation}, which is given by the ergodic averages \eqref{eq:standard-estimation}, is typically replaced by a \gls{rb} technique known as \emph{waste-recycling} \citep{ceperley1977waste,delmas2009waste}.
In implementations of \gls{mcmc}-based light transport algorithms, such as \citet{pbrt}, the standard estimator \eqref{eq:standard-estimation}, is typically replaced by a \gls{rb} technique known as \emph{waste-recycling}~\citep{ceperley1977waste}.

\paragraph*{The augmented chain}

Waste-recycling is based on the following observation: The \gls{mh} \autoref{alg:metropolis-hastings} ultimately generates two chains: the \emph{proposal sequence} $(\secondprocess_\timepoint)_{\timepoint\in\mathbb N}$, which consists of proposals drawn from the proposal kernel $\proposalkernel$, and the actual \gls{mh} chain, which only contains the accepted states \textemdash\ including repetitions and the user-defined initial state $\process_0$. The key insight is that the \emph{augmented chain} \begin{equation}
    \thirdprocess_\timepoint:=\left(\process_{\timepoint-1},\secondprocess_\timepoint\right)\eqfor\timepoint\in\mathbb N,
\end{equation} formed by sequentially pairing the current state with the next proposal, is
% also
a time-homogeneous Markov chain. Its transition kernel is \begin{equation}
    \transitionkernel_{\textnormal{aug}}\left((\point,\secondpoint),\;\cdot\;\right):=\left(1-\acceptancefunction(\point,\secondpoint)\right)\delta_\point\otimes\proposalkernel+\acceptancefunction(\point,\secondpoint)\delta_\secondpoint\otimes\proposalkernel
    % \begin{split}
    %     \transitionkernel_{\textnormal{aug}}\left((\point,\secondpoint),\pointset_1\times\pointset_2\right)&:=\left(1-\acceptancefunction(\point,\secondpoint)\right)1_{\pointset_1}(\point)\proposalkernel(\point,\pointset_2)\\&\;\;\;\;\;\;\;\;\;+\acceptancefunction(\point,\secondpoint)1_{\pointset_1}(\secondpoint)\hphantom)\proposalkernel(\secondpoint,\pointset_2)
    % \end{split}
\end{equation} and its invariant distribution is \citep{rudolf2017importance}
% $\targetdistribution\otimes\proposalkernel$
\begin{equation}
    % \targetdistribution_{\textnormal{aug}}:=\targetdistribution\otimes\proposalkernel
    \targetdistribution_{\textnormal{aug}}(\pointset_1\times\pointset_2):=\displaystyle\int_{\pointset_1}\targetdistribution(\dd{\point})\proposalkernel(\point,\pointset_2)
\end{equation}
% as detailed in \citet{rudolf2017importance}.
% Here, \setword{$\delta_\point$}{inline:dirac-measure} denotes the Dirac measure at $\point$ and $\otimes$ the
% \setword{\emph{product}}{inline:product}~\citep{kallenberg2021probability,cinlar2011probability,douc2018markov,klenke2020probability}
% \setword{\emph{product}}{inline:product}~\citep{kallenberg2021probability,klenke2020probability}
% of transition kernels.

Thus, $(\thirdprocess_\timepoint)_{\timepoint\in\mathbb N}$ is a Markov chain whose invariant distribution has the target distribution $\targetdistribution$ as its first marginal. Consequentially, \begin{equation}
    \expectation\left[\thirdintegrand\left(\thirdprocess_{\timepoint+1}\right)\mid\thirdprocess_\timepoint\right]=
    % \left(\transitionkernel_{\textnormal{aug}}\thirdintegrand\right)\left(\thirdprocess_\timepoint\right)=
    \displaystyle\int\transitionkernel_{\textnormal{aug}}\left(\thirdprocess_\timepoint,\dd{(\point',\secondpoint')}\right)\thirdintegrand(\point',\secondpoint'),
\end{equation} for every bounded measurable function $\thirdintegrand$ on the product space, by the Markov property and hence \begin{equation}
    \expectation\left[\integrand\left(\process_\timepoint\right)\mid\thirdprocess_\timepoint\right]=\left(1-\acceptancefunction\left(\thirdprocess_\timepoint\right)\right)\integrand\left(\secondprocess_\timepoint\right)+\acceptancefunction\left(\thirdprocess_\timepoint\right)\integrand\left(\process_\timepoint\right),
\end{equation} for every bounded measurable function (depending only on the first coordinate corresponding to the current state).

\paragraph*{The estimator}

Based on these observations, we obtain a Rao--Blackwellized estimator via conditioning on the augmented chain:

\begin{definition}\label{def:waste-recycling}
    If $\integrand\in L^1(\targetdistribution)$, then
    \begin{equation}\label{eq:waste-recycling}
        \wasterecyclingestimator_\timepoint\integrand:=\frac1\timepoint\displaystyle\sum_{\prevtimepoint=1}^\timepoint\expectation\left[\integrand\left(\process_\prevtimepoint\right)\mid\thirdprocess_\prevtimepoint\right]\eqfor\timepoint\in\mathbb N
    \end{equation}
    is called \textbf{waste-recycling estimator of} $\bm\integral$.
\end{definition}

The terminology is intuitive: a generated proposal is never \emph{wasted}, even if it is rejected, it still contributes to the estimate \eqref{eq:waste-recycling}. In light transport simulation, where sample generation is computationally expensive, \eqref{eq:waste-recycling} is almost universally used in place of \eqref{eq:standard-estimation}; for example in \citet{pbrt}. % However, it is well-known \citep{delmas2009waste} that the waste-recycling estimator \eqref{eq:waste-recycling} is not guaranteed to reduce the variance compared to the standard estimator \eqref{eq:standard-estimation}.

\subsection{Vanilla \acrlong{rb} (Ours)}\label{sec:metropolis-hastings-vanilla}

By \eqref{eq:rao-blackwell-theorem}, replacing the ergodic average estimator \eqref{eq:ergodic-theorem} with the waste-recycling estimator \eqref{eq:waste-recycling} cannot \emph{increase} the variance.
% However, it also cannot be proven that it \emph{decreases} the variance.
However, it is well-known \citep{delmas2009waste} that it also cannot be proven that it \emph{decreases} the variance compared to the standard estimator \eqref{eq:standard-estimation}.
For that reason, an alternative \gls{rb} technique for the \gls{mh} algorithm was introduced in \citet{douc2011vanilla}. We subsequently explain the idea behind this method.

\paragraph*{Decomposition of the \gls{mh} chain into tours}

By construction of \autoref{alg:metropolis-hastings}, there exists an $[0,1)$-valued independent and identically uniformly-distributed process $(\uniformprocess _\timepoint)_{\timepoint\in\mathbb N}$ such that \begin{equation}
    \process_\timepoint=\displaystyle\begin{cases}
        \secondprocess_\timepoint&\text{if }\uniformprocess_\timepoint\le\acceptancefunction(\thirdprocess_\timepoint);\\
        \process_{\timepoint-1}&\text{otherwise}.
    \end{cases}
\end{equation} and $(\uniformprocess_\prevtimepoint)_{\prevtimepoint\ge\timepoint}$ is independent of $(\thirdprocess_1,\ldots,\thirdprocess_\timepoint)$ for all $\timepoint\in\mathbb N$.

The sequence $(\uniformprocess_\timepoint)_{\timepoint\in\mathbb N}$ corresponds to the uniform random numbers used to decide whether a proposal is accepted or rejected, as implemented in \autoref{line:metropolis-hastings-accept-or-reject} of \autoref{alg:metropolis-hastings}.
By means of $(\uniformprocess_\timepoint)_{\timepoint\in\mathbb N}$, we can define the time at which the $\sequenceindex$th proposal is accepted as
% \begin{equation}
%     \acceptancetime_\sequenceindex:=\begin{cases}
%         \inf\left\{\timepoint>\acceptancetime_{\sequenceindex-1}:\uniformprocess_\timepoint\le\acceptancefunction(\thirdprocess_\timepoint)\right\}&\text{if }\sequenceindex\in\mathbb N;\\
%         0&\text{if }\sequenceindex=0.
%     \end{cases}
% \end{equation}
\begin{equation}
    \acceptancetime_\sequenceindex:=\inf\left\{\timepoint>\acceptancetime_{\sequenceindex-1}:\uniformprocess_\timepoint\le\acceptancefunction(\thirdprocess_\timepoint)\right\},
\end{equation} where $\acceptancetime_0:=0$.
Here we interpret the user-defined initial state $\process_0$ as a proposal that is immediately accepted at time $0$. Consequently, the waiting time until the $\sequenceindex$th acceptance of a proposal is given by \begin{equation}
    \lifetime_\sequenceindex:=\begin{cases}
        \acceptancetime_{\sequenceindex+1}-\acceptancetime_\sequenceindex&\text{if }\acceptancetime_\sequenceindex<\infty;\\
        0&\text{otherwise}.
    \end{cases}
\end{equation} This definition accounts for the theoretical (but practically irrelevant, since degenerate) case that no further proposal is eventually accepted, in which case $\acceptancetime_\sequenceindex=\infty$ for all large enough $\sequenceindex\in\mathbb N$.

In the perspective introduced in \citet[Section~6.1]{holl2025jrlt}, each $\lifetime_\sequenceindex$ can be interpreted as the \emph{lifetime} of the $\sequenceindex$th \emph{tour}. In this context, by a tour, we mean a local exploration performed according to the Markov chain implicitly defined by the proposal kernel $\proposalkernel$, which continues until it is
interrupted
% (i.e., \emph{killed} or \emph{terminated} depending on your favorite perspective)
by a rejection.

Finally, these definitions yield the following description of the sequence of accepted states (without duplicates): \begin{equation}
    \acceptedchain_\sequenceindex:=\begin{cases}
        \process_{\acceptancetime_\sequenceindex}&\text{if }\acceptancetime_\sequenceindex<\infty;\\
        \process_{\acceptancetime_{\sequenceindex-1}}&\text{otherwise}.
    \end{cases}
\end{equation} That is, $\acceptancetime_\sequenceindex$ is the (discrete) time at which the $\sequenceindex$th proposal is accepted, and $\lifetime_\sequenceindex$ is the number of steps one has to wait thereafter until the next proposal is accepted. Moreover, $\acceptedchain_\sequenceindex$ is precisely the $\sequenceindex$th accepted proposal. Set-theoretically, the sequences $(\acceptedchain_\sequenceindex)_{\sequenceindex\in\mathbb N_0}$ and $(\process_\timepoint)_{\timepoint\in\mathbb N_0}$ contain the same states, except that the latter may exhibit consecutive repetitions due to the rejection of proposals.

\paragraph*{Alternative representation of the standard estimator}

Using these definitions, we can rewrite the standard estimator \eqref{eq:standard-estimation} at the time $\timepoint=\acceptancetime_\lastsequenceindex$ at which the $\lastsequenceindex$th proposal is accepted as
\begin{equation}\label{eq:ergodic-average-estimator-rewritten}
	\tilde\ergodicaverage_\lastsequenceindex\integrand:=\frac{\sum_{\sequenceindex=0}^{\lastsequenceindex-1}\lifetime_\sequenceindex\integrand(\acceptedchain_\sequenceindex)}{\sum_{\sequenceindex=0}^{\lastsequenceindex-1}\lifetime_\sequenceindex}=\ergodicaverage_{\acceptancetime_\lastsequenceindex}\integrand\;\;\;\text{on }\left\{\acceptancetime_\lastsequenceindex<\infty\right\}.
\end{equation}

\paragraph*{The vanilla \gls{rb} principle}

As in the waste-recycling case described in \autoref{sec:metropolis-hastings-waste-recycling}, the key idea is again to replace a stochastic quantity in this expression
%of the standard estimator
with a conditional expectation. More precisely, the aim is to replace the waiting time $\lifetime_\sequenceindex$ until acceptance of the $\sequenceindex$th proposal \textemdash\ which also can be interpreted as counting how many times the current state $\acceptedchain_\sequenceindex$ appears in the \gls{mh} chain \textemdash\ by its conditional expectation given that current state $\acceptedchain_\sequenceindex$. According to \citep[Lemma~1]{douc2011vanilla}, under the assumption that $\lastsequenceindex\in\mathbb{N}$ and $\acceptancetime_\lastsequenceindex<\infty$ (i.e., that at least $\lastsequenceindex$ proposals are eventually accepted), the waiting times $\lifetime_\sequenceindex$ are, for all $\sequenceindex\in\{0,\ldots,\lastsequenceindex-1\}$, conditionally geometrically distributed given $\acceptedchain_\sequenceindex$, with parameter $\expectedacceptance(\acceptedchain_\sequenceindex)$.
Here, \begin{equation}
	\expectedacceptance(\point):=\displaystyle\int\proposalkernel(\point,\dd{\secondpoint})\acceptancefunction(\point,\secondpoint)
\end{equation} is the expected acceptance probability when the current state of the \gls{mh} algorithm is $\point$.
% Here, \begin{equation}
% 	\expectedacceptance:=\proposalkernel\acceptancefunction=\int\proposalkernel(\;\cdot\;,\dd{\secondpoint})\acceptancefunction(\;\cdot\;,\secondpoint)
% \end{equation} and hence $\expectedacceptance(\point)$ is the expected acceptance probability when the current state of the \gls{mh} algorithm is $\point$: \begin{equation}
%     \expectedacceptance(\point)=\expectation\left[\acceptancefunction(\point,\secondprocess)\right],
% \end{equation} where $\secondprocess\sim\proposalkernel(\point,\;\cdot\;)$.
Specifically, we have
\begin{equation}\label{eq:proposal-acceptance-waiting-time-is-geometrically-distributed}
	\expectation\left[\lifetime_\sequenceindex\mid\acceptedchain_\sequenceindex\right]=\frac1{\expectedacceptance(\acceptedchain_\sequenceindex)}.
	% \expectation\left[\lifetime_\sequenceindex\mid\acceptedchain_\sequenceindex\right]={\expectedacceptance(\acceptedchain_\sequenceindex)}^{-1}.
\end{equation}

The idea is now to replace $\lifetime_\sequenceindex$ in \eqref{eq:ergodic-average-estimator-rewritten} by \eqref{eq:proposal-acceptance-waiting-time-is-geometrically-distributed}. As noted previously, such a replacement does not increase the variance by virtue of \eqref{eq:rao-blackwell-theorem}. The crucial addition shown in \citep{douc2011vanilla} is that this substitution yields a \emph{strict} variance reduction.
%
%However, two problems remain with this approach. First, it is not directly applicable in practice,
However, this approach is not directly applicable in practice, as $\expectedacceptance$ \textemdash\ except for pathological
%special
cases \textemdash\ is not available in closed form. This issue was already addressed in \citep{douc2011vanilla}, where the following unbiased estimator for ${\expectedacceptance(\acceptedchain_\sequenceindex)}^{-1}$ was proposed:

\begin{lemma}
    % If $\lastsequenceindex\in\mathbb N_0$ with $\lifetimesum_{\lastsequenceindex+1}<\infty$ and $\bigl(\tilde\secondprocess_\timepoint\bigr)_{\timepoint\in\mathbb N}$ is an independent conditionally given $\acceptedchain_\lastsequenceindex$ identically $\proposalkernel(\acceptedchain_\lastsequenceindex,\;\cdot\;)$-distributed process,
    If $\lastsequenceindex\in\mathbb N_0$ with $\lifetimesum_{\lastsequenceindex+1}<\infty$ and $\bigl(\tilde\secondprocess_\timepoint\bigr)_{\timepoint\in\mathbb N}$ is independent and conditionally, given $\acceptedchain_\lastsequenceindex$, identically $\proposalkernel(\acceptedchain_\lastsequenceindex,\;\cdot\;)$-distributed,
    then
    \begin{equation}\label{eq:unbiased-estimator-of-expected-acceptance}
        \vanillalifetime_\lastsequenceindex:=1+\displaystyle\sum_{\timepoint\in\mathbb N}\prod_{\prevtimepoint=1}^\timepoint\bigl(1-\acceptancefunction\bigl(\acceptedchain_\lastsequenceindex,\tilde\secondprocess_\prevtimepoint\bigr)\bigr)
    \end{equation}
    %is a well-defined unbiased estimator (conditionally given $\acceptedchain_\lastsequenceindex$) of ${\expectedacceptance(\acceptedchain_\lastsequenceindex)}^{-1}$; i.e.,
    is well-defined and
    \begin{equation}
        \expectation\left[\vanillalifetime_\lastsequenceindex\mid\acceptedchain_\lastsequenceindex\right]
        =
        \frac1{\expectedacceptance(\acceptedchain_\lastsequenceindex)}.
        % {\expectedacceptance(\acceptedchain_\lastsequenceindex)}^{-1}.
    \end{equation}
    \begin{proof}[Proof\textup:\nopunct]
        \citep[Lemma~1]{douc2011vanilla}
    \end{proof}
\end{lemma}

The sum in \eqref{eq:unbiased-estimator-of-expected-acceptance} is almost surely finite \citep{douc2011vanilla}. Moreover, note that all terms in the sum vanish as soon as a proposal is accepted with probability~1.
% , i.e., when $\acceptancefunction$ evaluates to~1.
However, in practice, evaluating this sum may still require generating an impractically large number of proposals from $\proposalkernel(\acceptedchain_\lastsequenceindex,\;\cdot\;)$ at the current state $\acceptedchain_\lastsequenceindex$.

For better understanding, recall that in \gls{mh}, proposals from\newline $\proposalkernel(\acceptedchain_\lastsequenceindex,\;\cdot\;)$ must be generated anyway at the current state $\acceptedchain_\lastsequenceindex$, albeit only until one of them is accepted. After that, the algorithm transitions to the accepted proposal, which becomes the new current state, and hence no further sample from $\proposalkernel(\acceptedchain_\lastsequenceindex,\;\cdot\;)$ needs to be computed.

In \citep[Proposition~1]{douc2011vanilla}, the authors proposed a truncation scheme that terminates the infinite sum in \eqref{eq:unbiased-estimator-of-expected-acceptance} after a fixed number of terms, allowing for a trade-off between computational cost and the accuracy of the estimate of $1/\expectedacceptance(\acceptedchain_\lastsequenceindex)$. Nevertheless, the resulting method in \citep{douc2011vanilla} still requires generating proposals beyond those already necessary for \gls{mh} execution (i.e., until a proposal is accepted). While this is acceptable in many applications due to the significant variance reduction it provides, in the context of light transport we face the major issue that proposal generation is a highly costly task, rendering even a simple truncation attempt impractical.

% We therefore propose a modification of the approach introduced in \citep{douc2011vanilla}.
We therefore modify the approach of \citet{douc2011vanilla}.
Instead of estimating $1/\expectedacceptance(\acceptedchain_\lastsequenceindex)$ using additional samples from $\proposalkernel(\acceptedchain_\lastsequenceindex,\cdot)$, we base the estimate solely on the $\lifetime_\lastsequenceindex$ proposals $\secondprocess_{\acceptancetime_{\lastsequenceindex+1}},\ldots,\secondprocess_{\acceptancetime_{\lastsequenceindex+\lifetime_\lastsequenceindex}}$ that are
necessarily
generated anyway until the next state $\acceptedchain_{\lastsequenceindex+1}$ is accepted. This yields the following estimator of $1/\expectedacceptance(\acceptedchain_\lastsequenceindex)$:

% We therefore propose the following modification of the approach introduced in \citep{douc2011vanilla}. Instead of estimating $1/\expectedacceptance(\acceptedchain_\lastsequenceindex)$ by generating additional samples from $\proposalkernel(\acceptedchain_\lastsequenceindex,\;\cdot\;)$, we estimate it solely based on the $\lifetime_\lastsequenceindex$ proposals $\secondprocess_{\acceptancetime_{\lastsequenceindex+1}},\ldots,\secondprocess_{\acceptancetime_{{\lastsequenceindex+1}}}$ that are necessarily generated anyway until the next state $\acceptedchain_{\lastsequenceindex+1}$ is accepted. This leads to the following estimator of $1/\expectedacceptance(\acceptedchain_\lastsequenceindex)$:

\begin{definition}
    If $\lastsequenceindex\in\mathbb N_0$ with $\lifetimesum_{\lastsequenceindex+1}<\infty$, then \begin{equation}\label{eq:vanilla-weight}
        \ourvanillalifetime_\lastsequenceindex:=1+\displaystyle\sum_{\timepoint=1}
        % ^{\lifetime_\lastsequenceindex-1}
        ^{\lifetime_\lastsequenceindex}
        \prod_{\prevtimepoint=1}^\timepoint\left(1-\acceptancefunction\left(\acceptedchain_\lastsequenceindex,\secondprocess_{\acceptancetime_\lastsequenceindex+\prevtimepoint}\right)\right).
    \end{equation}
\end{definition}

In \eqref{eq:vanilla-weight}, the sequence $(\secondprocess_\timepoint)_{\timepoint\in\mathbb N}$
% still
denotes the proposal sequence generated during \gls{mh} execution (as in \autoref{sec:metropolis-hastings-waste-recycling}) and hence no further samples from $\proposalkernel(\acceptedchain_\lastsequenceindex,\;\cdot\;)$ need to be drawn.
The heuristic behind estimating $1/\expectedacceptance(\acceptedchain_\lastsequenceindex)$ by $\ourvanillalifetime_\lastsequenceindex$ is straightforward: since $\timepoint=\lifetime_\lastsequenceindex$ corresponds to the time at which the proposal $\secondprocess_{\acceptancetime_\lastsequenceindex+\timepoint}$ is accepted, the product $\prod_{\prevtimepoint=1}^{\timepoint}\left(1-\acceptancefunction\bigl(\acceptedchain_\lastsequenceindex,\secondprocess_{\acceptancetime_\lastsequenceindex+\prevtimepoint}\bigr)\right)$ is on average close to $0$, rendering any additional contributions to the running sum from subsequent proposals negligible.
We formalize our approach in the following definition:

\begin{definition}\label{def:vanilla-estimator}
    If $\integrand\in\mathcal L^1(\targetdistribution)$, then \begin{equation}\label{eq:vanilla-estimation}
        \vanillaestimator_\lastsequenceindex\integrand:=\frac{\sum_{\sequenceindex=0}^{\lastsequenceindex-1}\ourvanillalifetime_\sequenceindex\integrand(\acceptedchain_\sequenceindex)}{\sum_{\sequenceindex=0}^{\lastsequenceindex-1}\ourvanillalifetime_\sequenceindex}\eqfor\lastsequenceindex\in\mathbb N
    \end{equation} is called the \textbf{vanilla Rao--Blackwellized estimator of} $\bm\integral$.
\end{definition}

We emphasize that, despite the name, our estimator in \autoref{def:vanilla-estimator} is not identical to the one proposed in \citet{douc2011vanilla}. We adopt this terminology to acknowledge the origin and inspiration of our approach in their work. Crucially, our variant is specifically adapted to the light transport setting and can be applied without incurring any noticeable additional computational cost.

We note that this estimator has a ratio-of-sums structure, similar to \emph{self-normalized importance sampling} estimators \citep{rudolf2017importance}.

\paragraph*{Bias characterization and control}

Strictly speaking, the vanilla Rao--Blackwellized estimator introduced in \autoref{def:vanilla-estimator} constitutes an \emph{inexact} \gls{mcmc} method \citep{eberle2023inexact}.
In Section~B of the supplemental, we derive an explicit expression for the induced bias of the underlying holding time estimator \eqref{eq:vanilla-weight}.
While a theoretical characterization of how this bias propagates to the vanilla Rao--Blackwellized estimator \autoref{def:vanilla-estimator} is left for future work, our empirical results in \autoref{sec:numerical-study} consistently show no measurable bias in practice.

% Strictly speaking, the vanilla Rao--Blackwellized estimator introduced in \autoref{def:vanilla-estimator} constitutes an \emph{inexact} \gls{mcmc} method \citep{eberle2023inexact}. In Section~B of the supplemental, we derive an explicit expression for the induced bias and derive how it can be controlled in practice. Finally, we explain why this bias is purely theoretical, does not manifest in the empirical results presented in \autoref{sec:numerical-study}, and is numerically negligible in the light transport setting.

The Rao--Blackwellized \gls{mh} algorithm for estimating the integral $\integral$ in \eqref{eq:integral} is summarized in \eqref{alg:rao-blackwellized-metropolis-hastings-algorithm}. The $\textbf{if}$-branches querying the \gls{rb} technique to be used can clearly be evaluated at compile time.

\begin{algorithm}
    \caption{Rao--Blackwellized \acrlong{mh} estimator with proposal kernel $\proposalkernel$ and target distribution $\targetdistribution$.}\label{alg:rao-blackwellized-metropolis-hastings-algorithm}
    \begin{algorithmic}[1]
        % \Require{Initial state $\point_0$, sample count $\timepoint\in\mathbb N$ and \gls{rb} technique $\textnormal{RaoBlackwellization}\in\left\{\textnormal{None},\textnormal{Vanilla},\textnormal{WasteRecycling}\right\}$.}
        \Require{Initial state $\point_0$ and sample count $\timepoint\in\mathbb N$.}
        \Ensure{Estimate $\tilde\integral$ of $\integral$.}
        \State$\acceptancetime=\tilde\integral=0$;
        \If{$\left(\textnormal{RaoBlackwellization == Vanilla}\right)$:}
            \State
            $\ourvanillalifetime=\vanillaproduct=1$;
        \EndIf
        \For{$\left(\prevtimepoint=1;;\textnormal{++}\prevtimepoint\right)$:}
            \If{$\left(\textnormal{RaoBlackwellization == None}\right)$:}
                \State
                $\acceptancetime\mathrel+=1$;
                \State
                $\tilde\integral\mathrel+=\integrand(\point_{\prevtimepoint-1})$;
            \EndIf
            \If{$\left(\textnormal{RaoBlackwellization != Vanilla}\textbf{ and }\prevtimepoint\ge\timepoint\right)$:}
                \State
                \textbf{break};
            \EndIf
            \State Sample $\secondpoint$ from $\proposalkernel(\point_{\prevtimepoint-1},\;\cdot\;)$;
            \If{$\left(\textnormal{RaoBlackwellization == Vanilla}\right)$:}
                \State
                $\ourvanillalifetime\mathrel+=\vanillaproduct\mathrel*=1-\alpha$;
            \EndIf
            \If{$\left(\textnormal{RaoBlackwellization == WasteRecycling}\right)$:}
                \State$\acceptancetime\mathrel+=1$;
                \State$\tilde\integral\mathrel+=(1-\acceptancefunction(\point_{\prevtimepoint-1},\secondpoint))\integrand(\point_{\prevtimepoint-1})+\acceptancefunction(\point_{\prevtimepoint-1},\secondpoint)\integrand(\secondpoint)$;
            \EndIf
            \State Sample $\uniform$ from $\mathcal U_{[0,\:1)}$;\AlgCommentLeft{uniform distribution on $[0,1)$}
            \If{$\left(\uniform<\acceptancefunction(\point_{\prevtimepoint-1},\secondpoint)\right)$:}
                \If{$\left(\textnormal{RaoBlackwellization == Vanilla}\right)$:}
                    \State$\acceptancetime\mathrel+=\ourvanillalifetime$;
                    \State
                    $\tilde\integral\mathrel+=\ourvanillalifetime\integrand(\point_{\prevtimepoint-1})$;
                    \If{$\left(\prevtimepoint+1\ge\timepoint\right)$:}
                        \State
                        \textbf{break};
                    \EndIf
                    \State$\ourvanillalifetime=\vanillaproduct=1$;
                \EndIf
                \State$\point_\prevtimepoint=\secondpoint$;\AlgCommentLeft{with probability $\acceptancefunction(\point_{\prevtimepoint-1},\secondpoint)$ accept the proposal}
            \Else:
                \State
                $\point_\prevtimepoint=\point_{\prevtimepoint-1}$;\AlgCommentLeft{with probability $1-\acceptancefunction(\point_{\prevtimepoint-1},\secondpoint)$ reject the proposal}
                % \If{$\left(\textnormal{RaoBlackwellization == Vanilla}\right)$:}
                %     %\State
                %     $\ourvanillalifetime\mathrel+=\vanillaproduct\mathrel*=1-\alpha$;
                % \EndIf
            \EndIf
        \EndFor
        \State\Return$\tilde\integral/\lifetimesum$;
    \end{algorithmic}
\end{algorithm}

\section{Rao--Blackwellized Jump Restore}\label{sec:rao-blackwellized-jump-restore-algorithm}

\subsection{Jump Restore algorithm}\label{sec:jump-restore-algorithm}

The Jump Restore algorithm was introduced to the graphics community in \citet{holl2025jrlt} as a novel continuous-time \gls{mcmc} method. It is designed to combine rapid \emph{local exploration} with efficient \emph{global discovery} of the target distribution $\targetdistribution$.

\paragraph*{Local dynamics}

The \emph{local dynamics} of the Jump Restore algorithm are given by a user-defined Markov kernel $\transitionkernel$. Conceptually, $\transitionkernel$ represents the dynamics of a Markov chain $(\markovchain_\timepoint)_{\timepoint\in\mathbb N_0}$ that is well suited for rapid local exploration of the state space. Such a Markov chain may, for instance, be given by an existing \gls{mcmc} method.

\paragraph*{Embedding into continuous time}

The Restore framework \citep{holl2025jrlt} operates in continuous time. For this reason, the Markov chain $(\markovchain_\timepoint)_{\timepoint\in\mathbb N_0}$ is embedded into continuous time by \emph{holding} each state for a random duration. These holding times must be chosen to be exponentially distributed in order for the resulting continuous-time process $(\localprocess_\timepoint)_{\timepoint\ge0}$ to remain Markovian. Accordingly, the process remains in state $\point$ for a time distributed as $\operatorname{Exp}(\holdingrate(\point))$, where $\holdingrate$ is a strictly positive \emph{holding rate}. To preserve the infinitesimal behavior of $(\markovchain_\timepoint)_{\timepoint\in\mathbb N_0}$, one must choose $\holdingrate=1$ \citep[Section~7.1]{holl2025jrlt}. For clarity, however, we retain a general holding rate $\holdingrate$.

\paragraph*{Killing}

Each local exploration evolves according to the dynamics of $(\localprocess_\timepoint)_{\timepoint\ge0}$ and is \emph{killed} (or \emph{terminated}) after a finite time. Termination is governed by a continuous-time clock that decays with a time-dependent exponential rate \begin{equation}
    [0,\infty)\ni\timepoint\mapsto\killingrate\bigl(\localprocess_\timepoint\bigr).
\end{equation} The killing rate $\killingrate$ is chosen inversely proportional to the target density $\targetdensity$, and is therefore higher in low-density regions.
The execution of a single local exploration up to its termination is referred to as a \emph{tour}, and its termination time is called the \emph{lifetime} of the tour.

\paragraph*{Global dynamics}

The global dynamics of the Jump Restore algorithm are also given by a user-defined Markov kernel $\regenerationdistribution$. Upon termination of a local exploration, a new local exploration \textemdash\ again evolving according to the dynamics $(\localprocess_\timepoint)_{\timepoint\ge0}$ \textemdash\ is started. The spawn location of this new local exploration is distributed according to $\regenerationdistribution(\point,\;\cdot\;)$, conditional on the event that the exit point (i.e., the last state visited immediately before termination) of the previous local exploration was $\point$. Since $\regenerationdistribution$ describes the spatial \emph{transfer} of local explorations, it is referred to as a \emph{transfer kernel}.

Formally, following \citet[Section~7.2]{holl2025jrlt}, we
% obtain the following definition:
have:

\begin{definition}\label{def:jump-restore-process}
    If $\combinedrate:=\holdingrate+\killingrate$, then the pure-jump type Markov process with transition kernel \begin{equation}\label{eq:pure-jump-type-transition-rule}
    	\displaystyle\frac{\holdingrate(\point)}{\combinedrate(\point)}\transitionkernel(\point,\;\cdot\;)+\frac{\killingrate(\point)}{\combinedrate(\point)}\regenerationdistribution(\point,\;\cdot\;),
    \end{equation} is called the \textbf{Jump Restore process with local dynamics }$\bm{(\markovchain_\timepoint)_{\timepoint\in\mathbb N_0}}$\textbf{, global dynamics }$\bm\regenerationdistribution$\textbf{, and target distribution }$\bm\targetdistribution$.
\end{definition}

\paragraph*{Ensuring invariance}

In order to ensure invariance of the Jump Restore process, the killing rate $\killingrate$ cannot be chosen arbitrarily. To state the definition that guarantees invariance, we assume that the transfer kernel $\regenerationdistribution$ admits a density with respect to the same reference measure $\referencemeasure$ as $\targetdistribution$; that is, \begin{equation}
    \regenerationdistribution(\point,\pointset)=\displaystyle\int_{\pointset}\regenerationdensity(\point,\;\cdot\;)\dd{\referencemeasure}
\end{equation} for some density $\regenerationdensity$.
If $\regenerationdistribution$ is state-independent, then the choice of killing rate that renders the Jump Restore process $\targetdistribution$-invariant is
% given by
\begin{equation}\label{eq:killing-rate}
    \killingrate:=\expectedlifetime\targetdensity_\referencemeasure\regenerationdensity/\targetdensity
    %\frac{\regenerationdensity}{\targetdensity},
\end{equation} where $\expectedlifetime>0$ is an arbitrary constant. It represents the reciprocal of the expected lifetime of a tour \citep[Section~6.2]{holl2025jrlt}.
For the general definition of the killing rate $\killingrate$ in the case of state-dependent transfer kernels $\regenerationdistribution$, we refer to \citet[Section~6.2]{holl2025jrlt}.

\paragraph*{On-the-fly computation of the normalization constant}

Section~C of the supplemental shows how the normalization constant $\targetdensity_\referencemeasure$ can be absorbed into $\expectedlifetime$ and subsequently estimated from a run of the Jump Restore algorithm, eliminating the need for a separate bootstrapping phase as required by \gls{mh}-based algorithms.

% \paragraph*{Parallelizability}

% If the transfer kernel $\regenerationdistribution$ is state-independent, then all local explorations may be executed simultaneously, which leads to the high degree of parallelizability of the Jump Restore algorithm.

\paragraph*{Numerical simulation}

%Given \autoref{def:jump-restore-process},
The Jump Restore process can be simulated like any other pure-jump Markov process. The algorithm proposed in \citet[Algorithm~7.1]{holl2025jrlt} is reproduced in \autoref{alg:jump-restore-algorithm}.% Since it is conceptually more natural to think in terms of \emph{tours} rather than \emph{samples}, we have formulated \autoref{alg:jump-restore-algorithm} so that the number $\thirdsequenceindex\in\mathbb N$ of tours can be specified directly.

\begin{algorithm}
    \caption{Jump Restore algorithm with local dynamics $\transitionkernel$, global dynamics $\regenerationdistribution$ and target distribution $\targetdistribution$.}\label{alg:jump-restore-algorithm}
    \begin{algorithmic}[1]
        \Require{Initial state $\point_0$ and tour count $\thirdsequenceindex\in\mathbb N$.}
        \Ensure{Realization of the Jump Restore process up to tour $\thirdsequenceindex$.}
        \For{$\left(\sequenceindex=1,\secondsequenceindex=1;;\text{++}\sequenceindex\right)$}:
            \State\label{line:sample-holding-time}Sample $\timepoint_1$ from $\operatorname{Exp}\left(\holdingrate(\point_{\sequenceindex-1})\right)$;\AlgCommentLeft{holding time of current state $\point_{\sequenceindex-1}$}
            \State\label{line:sample-killing-time}Sample $\timepoint_2$ from $\operatorname{Exp}\left(\killingrate\left(\point_{\sequenceindex-1}\right)\right)$;\AlgCommentLeft{time till next termination attempt}%\AlgCommentLeft{$\operatorname{Exp}(0):=\delta_\infty$}
            \If{$\left(\timepoint_1<\timepoint_2\right)$:}\label{line:killing}\AlgCommentLeft{next local state transition before termination attempt}
            % \State\{\AlgCommentLeft{next local state transition before termination attempt}
                \State$\Delta\lifetime_\sequenceindex=\timepoint_1$;
                % \If{$\left(\sequenceindex==\thirdsequenceindex\right)$:}\label{line:return-if-sample-count-reached}
                %     \State\Return$\left(\left(\Delta\lifetime_1,\point_0\right),\ldots,\left(\Delta\lifetime_\sequenceindex,\point_{\sequenceindex-1}\right)\right)$;
                % \EndIf
                \State\label{line:simulate-mixture-transition-embeddeding-if}Sample $\point_\sequenceindex$ from $\transitionkernel\left(\point_{\sequenceindex-1},\;\cdot\;\right)$;\AlgCommentLeft{local state transition}
            \Else:\AlgCommentLeft{termination before next local state transition}
                \State$\Delta\lifetime_\sequenceindex=\timepoint_2$;
                % \If{$\left(\sequenceindex==\thirdsequenceindex\right)$:}\label{line:return-if-sample-count-reached2}
                %     \State\Return$\left(\left(\Delta\lifetime_1,\point_0\right),\ldots,\left(\Delta\lifetime_\sequenceindex,\point_{\sequenceindex-1}\right)\right)$;
                % \EndIf
                \If{$\left(\secondsequenceindex==\thirdsequenceindex\right)$:}\label{line:return-if-tour-count-reached}
                    \State
                    \Return$\left(\left(\Delta\lifetime_1,\point_0\right),\ldots,\left(\Delta\lifetime_\sequenceindex,\point_{\sequenceindex-1}\right)\right)$;
                \EndIf
                \State$\text{++}\secondsequenceindex$;
                \State\label{line:simulate-mixture-transition-embeddeding-else}Sample $\point_\sequenceindex$ from $\regenerationdistribution\left(\point_{\sequenceindex-1},\;\cdot\;\right)$;\AlgCommentLeft{global state transition}
            \EndIf
        \EndFor
    \end{algorithmic}
\end{algorithm}

In \autoref{sec:jump-restore-standard-estimation}, we describe how the output of \autoref{alg:jump-restore-algorithm} is traditionally used to estimate the integral $\integral$. In \autoref{sec:jump-restore-vanilla-estimation}, we then apply the vanilla \gls{rb} principle to derive a vanilla Rao--Blackwellized estimator variant. These constructions apply to general choices of local dynamics. In \autoref{sec:jump-restore-with-metropolis-hastings-local-dynamics}, we specialize to the case where the local dynamics are given by a \gls{mh} chain. In this setting, the waste-recycling idea can also be adopted, leading to a waste-recycling estimator, which we briefly describe in \autoref{sec:jump-restore-waste-recycling}.

\subsection{Standard estimation}\label{sec:jump-restore-standard-estimation}

Integral estimation using the general Restore algorithm \citep{holl2025jrlt} is typically performed via the standard ergodic average estimator \eqref{eq:ergodic-theorem}; that is, since the Restore process is a time-\emph{continuous} Markov process $(\concatenatedprocess_\timepoint)_{\timepoint\ge0}$, the standard estimator of $\integral$ is given by
\begin{equation}\label{eq:restore-standard-estimation}
    \ergodicaverage_\timepoint\displaystyle\integrand=\frac1\timepoint\int_0^\timepoint\integrand(\concatenatedprocess_\prevtimepoint)\dd{\prevtimepoint}.
\end{equation}
% By construction in \autoref{def:restore-process}, \begin{equation}\label{eq:restore-standard-estimation}
%     \frac1{\lifetimesum_\thirdsequenceindex}\int_0^{\lifetimesum_\thirdsequenceindex}\integrand(\concatenatedprocess_\timepoint)\dd{\timepoint}\approx\int\integrand\dd{\targetdistribution}.
% \end{equation}

% \paragraph*{Integral estimation}

% The integrals on the right-hand side of \eqref{eq:tour-integral} can be approximated using the left-hand rectangle rule. For this purpose, it is crucial to observe that, since it is defined as the continuous-time embedding of a discrete-time Markov chain, the local process $(\localprocess_\timepoint)_{\timepoint\ge0}$ is piecewise constant.

In the special case of the \emph{Jump} Restore process from \autoref{def:jump-restore-process}, the process $(\concatenatedprocess_\timepoint)_{\timepoint\ge0}$ is piecewise constant and fully characterized by the sequence $(\left(\Delta\lifetime_1,\point_0\right),\ldots,\left(\Delta\lifetime_\lastsequenceindex,\point_{\lastsequenceindex-1}\right))$ produced by \autoref{alg:jump-restore-algorithm}. In fact,
\begin{equation}
    \concatenatedprocess_\prevtimepoint(\outcome)=\point_{\secondsequenceindex-1}\eqforall\prevtimepoint\in\left[\displaystyle\sum_{\sequenceindex=1}^{\secondsequenceindex-1}\Delta\lifetime_\sequenceindex,\displaystyle\sum_{\sequenceindex=1}^{\secondsequenceindex}\Delta\lifetime_\sequenceindex\right),
\end{equation}
% \begin{equation}
%     \concatenatedprocess_\prevtimepoint(\outcome)=\begin{cases}
%         \point_0&\text{, if }\prevtimepoint\in[0,\Delta\lifetime_1);\\
%         \point_1&\text{, if }\prevtimepoint\in[\Delta\lifetime_1,\Delta\lifetime_1+\Delta\lifetime_2);\\
%         \;\vdots\\
%         \point_{\lastsequenceindex-1}&\text{, if }\prevtimepoint\in\left[\sum_{\sequenceindex=1}^{\lastsequenceindex-1}\Delta\lifetime_\sequenceindex,\sum_{\sequenceindex=1}^\lastsequenceindex\Delta\lifetime_\sequenceindex\right)
%     \end{cases}
% \end{equation}
for an outcome $\outcome$, and the
% ergodic average
standard
estimator \eqref{eq:restore-standard-estimation} reduces to
\begin{equation}\label{eq:jump-restore-standard-estimation}
	\left(\ergodicaverage_{\lifetimesum_\thirdsequenceindex}\right)(\outcome)=\left(\displaystyle\sum_{\sequenceindex=1}^\lastsequenceindex\Delta\lifetime_\sequenceindex\right)^{-1}\displaystyle\sum_{\sequenceindex=1}^\lastsequenceindex\Delta\lifetime_\sequenceindex\integrand\left(\point_{\sequenceindex-1}\right),
    %\approx\int\integrand\dd{\targetdistribution},
\end{equation}
where $\lifetimesum_\thirdsequenceindex(\outcome)=\sum_{\sequenceindex=1}^\lastsequenceindex\Delta\timepoint_\sequenceindex$ \citep[Section~7.3]{holl2025jrlt}.

\subsection{Vanilla \acrlong{rb} (Ours)}\label{sec:jump-restore-vanilla-estimation}

For the same output $(\left(\Delta\lifetime_1,\point_0\right),\ldots,\left(\Delta\lifetime_\lastsequenceindex,\point_{\lastsequenceindex-1}\right))$ of \autoref{alg:jump-restore-algorithm}, the holding times satisfy $\probabilitymeasure\left[\Delta\lifetime_\sequenceindex\in\;\cdot\;\mid\point_\sequenceindex\right]=\operatorname{Exp}\left(\combinedrate(\point_\sequenceindex)\right)$ 
% \begin{equation}
%     \probabilitymeasure\left[\Delta\lifetime_\sequenceindex\in\;\cdot\;\mid\point_\sequenceindex\right]=\operatorname{Exp}\left(\combinedrate(\point_\sequenceindex)\right),
% \end{equation}
and therefore \begin{equation}\label{eq:jump-restore-rao-blackwellization}
    \expectation\left[\Delta\lifetime_\sequenceindex\mid\point_\sequenceindex\right]=
    % {\combinedrate(\point_\sequenceindex)}^{-1}
    \frac1{\combinedrate(\point_\sequenceindex)}.
\end{equation}

Consequently, continuing the vanilla \acrlong{rb} principle by replacing the time increments in \eqref{eq:jump-restore-standard-estimation} by their conditional expectations, $\Delta\lifetime_\sequenceindex$ can be substituted with $1/\combinedrate(\point_\sequenceindex)$, yielding
% what we refer to as the 
a
\emph{vanilla Rao--Blackwellized estimator} of the Jump Restore algorithm.

\subsection{Jump Restore for improving mixture proposal \gls{mh}}\label{sec:jump-restore-with-metropolis-hastings-local-dynamics}

The Jump Restore algorithm is particularly useful in applications where the classical \gls{mh} algorithm is used with a \emph{mixture} proposal% kernel
\begin{equation}\label{eq:mixture-proposal-kernel}
    \proposalkernel_\largestepprobability(\point,\;\cdot\;):=(1-\largestepprobability)\smallstepkernel(\point,\;\cdot\;)+\largestepprobability\regenerationdistribution(\point,\;\cdot\;).
\end{equation}
% \begin{equation}\label{eq:mixture-proposal-kernel}
%     \proposalkernel_\largestepprobability(\point,\;\cdot\;):=(1-\largestepprobability)\underexplanation{\textnormal{local}}\smallstepkernel(\point,\;\cdot\;)+\largestepprobability\underexplanation{\textnormal{global}}\regenerationdistribution(\point,\;\cdot\;).
%     \vspace{-.175cm}
% \end{equation}
In such a traditional \gls{mh} algorithm, \emph{local} (or \emph{small-scale}) moves according to a transition kernel $\smallstepkernel$ are proposed with probability $1-\largestepprobability$, while \emph{global} (or \emph{large-scale}) moves are proposed with probability $\largestepprobability\in[0,1]$ according to another transition kernel $\regenerationdistribution$. Accordingly, $\largestepprobability$ is referred to as the \emph{large step probability}.

% \paragraph*{Practical relevance}

% This is precisely the setting encountered in \gls{mh}-based light transport algorithms such as \citet{hachisuka2014multiplexed}, \citet{luan2020langevin}, or \citet{li2015anisotropic}. There, one typically works in the so-called \emph{primary sample space} \citep{kelemen2002simple} \textemdash\ a unit hypercube \textemdash\ where the small-scale kernel $\smallstepkernel$ is given by a toroidally wrapped Gaussian perturbation of the current state, and the large-scale distribution $\regenerationdistribution$ is taken to be the uniform distribution over the state space.

% Due to its practical relevance — including the scenario considered in \autoref{sec:numerical-study}, which also satisfies this assumption — and subtle yet important implementation differences that enable parallelization, we assume in the following description of the \gls{rb} Jump Restore estimator that the global dynamics $\regenerationdistribution$ are state-independent. The necessary adaptation to the general case is evident from \autoref{alg:jump-restore-algorithm}.

% We also assume that $\regenerationdistribution$ admits a density $\regenerationdensity$ with respect to the same reference measure $\referencemeasure$ with respect to which $\targetdistribution$ possesses the (typically unnormalized) density $\targetdensity$.

\paragraph*{Jump Restore with \gls{mh} local dynamics}

We now consider the Jump Restore algorithm with local dynamics given by the \gls{mh} chain with proposal kernel $\proposalkernel_0$, global dynamics $\regenerationdistribution$ \textemdash\ $\regenerationdistribution$ being the same Markov kernel as in \eqref{eq:mixture-proposal-kernel} \textemdash\ and target distribution $\targetdistribution$.

%, in order to combine the \gls{rb} technique from \eqref{eq:jump-restore-rao-blackwellization} with the \gls{rb} techniques for \gls{mh} local dynamics introduced in \autoref{sec:rao-blackwellized-metropolis-hastings-algorithm}.

% Under these assumptions, the definition of the killing rate \eqref{eq:killing-rate} simplifies to
% \begin{equation}\label{eq:killing-rate-simplified}
% 	\killingrate:=\expectedlifetime\targetdensity_\referencemeasure\frac\regenerationdensity\targetdensity,
% \end{equation}
% where
% \begin{equation}\label{eq:target-density-normalization-constant}
% 	\targetdensity_\referencemeasure:=\int\targetdensity\dd{\referencemeasure}\in(0,\infty)
% \end{equation}
% denotes the normalization constant of $\targetdensity$; see \citet[Section~6.2]{holl2025jrlt}. In this case, the constant $\expectedlifetime>0$ can be chosen arbitrarily. It represents the reciprocal of the expected lifetime of an individual tour; see \citet[Section~6.2]{holl2025jrlt}. Formally, \begin{equation}\label{eq:expected-lifetime}
%     \expectation_\regenerationdistribution\left[\lifetime_\sequenceindex\right]=\frac1\expectedlifetime,
% \end{equation} where $\expectation_\regenerationdistribution$ denotes expectation with respect to the probability measure according to which the $\sequenceindex$th tour is respawned at a location sampled from $\regenerationdistribution$.

% \authorcomment[sholl]{TODO: Resolve conflicting usage of the symbol $c$.}

\subsection{Waste-Recycling}\label{sec:jump-restore-waste-recycling}

Given that the local dynamics are now assumed to be given by a \gls{mh} chain, we can adopt the waste-recycling idea from \autoref{sec:metropolis-hastings-waste-recycling}. Specifically, we construct a waste-recycling estimator by replacing $\integrand(\point_{\sequenceindex-1})$ with $(1-\acceptancefunction(\point_{\sequenceindex-1},\secondpoint))\integrand(\point_{\sequenceindex-1})+\acceptancefunction(\point_{\sequenceindex-1},\secondpoint)\integrand(\secondpoint)$, where $\secondpoint$ denotes the current \gls{mh} proposal.
Overall, we obtain a \emph{Rao--Blackwellized} Jump Restore algorithm with \gls{mh} local dynamics by combining the \gls{rb} techniques for \gls{mh} introduced in \autoref{sec:rao-blackwellized-metropolis-hastings-algorithm} with the Jump Restore algorithm in \autoref{alg:jump-restore-algorithm}. The resulting generic algorithm is given in \autoref{alg:rao-blackwellized-jump-restore-algorithm}.

% Overall, we construct a \emph{Rao--Blackwellized} Jump Restore algorithm with $\gls{mh}$ local dyncamics by combining the \gls{rb} techniques for the \gls{mh} algorithm introduced in \autoref{sec:rao-blackwellized-metropolis-hastings-algorithm} with the Jump Restore algorithm in \autoref{alg:jump-restore-algorithm}. For the vanilla \gls{rb} technique, we additionally apply \eqref{eq:jump-restore-rao-blackwellization} to average the time intervals. The resulting generic algorithm is summarized in \autoref{alg:rao-blackwellized-jump-restore-algorithm}.

% Due to our assumption of state-independent global dynamics $\regenerationdistribution$, all tour simulations in \autoref{line:simulate-tour} can be executed in parallel. Moreover, as in \autoref{alg:rao-blackwellized-metropolis-hastings-algorithm}, the \texttt{if}-branches querying the \gls{rb} technique to be used can clearly be evaluated at compile time.

\begin{algorithm}
    \caption{Rao--Blackwellized Jump Restore estimator with \gls{mh} local dynamics $\transitionkernel$, global dynamics $\regenerationdistribution$ and target distribution $\targetdistribution$.}\label{alg:rao-blackwellized-jump-restore-algorithm}
    \begin{algorithmic}[1]
        % \Require{Tour count $\lastsequenceindex\in\mathbb N$ and \gls{rb} technique\newline $\textnormal{RaoBlackwellization}\in\left\{\textnormal{None},\textnormal{Vanilla},\textnormal{WasteRecycling}\right\}$.}
        %\Require{Tour count $\lastsequenceindex\in\mathbb N$. \textbf{Output:} Estimate $\tilde\integral$ of $\integral$.}
        % \Ensure{Estimate $\tilde\integral$ of $\integral$.}
        \Procedure{SimulateTour}{\vphantom x}
            \State$\lifetime=0$;
            \State$\tilde\integral=0$;
            % \State$\lifetime=\tilde\integral=0$;
            \If{$\left(\textnormal{RaoBlackwellization == Vanilla}\right)$:}
                \State
                $\ourvanillalifetime=\vanillaproduct=1$;
            \EndIf
            \State Sample $\point_0$ from $\regenerationdistribution$;
            \For{$\left(\sequenceindex=1;;\textnormal{++}\sequenceindex\right)$:}
                \State Sample $\Delta\lifetime_1$ from $\operatorname{Exp}\left(\holdingrate(\point_{\sequenceindex-1})\right)$ and $\Delta\lifetime_2$ from $\operatorname{Exp}\left(\killingrate(\point_{\sequenceindex-1})\right)$;
                % \State Sample $\Delta\lifetime_1\sim\operatorname{Exp}\left(\holdingrate(\point_{\sequenceindex-1})\right)$ and $\Delta\lifetime_2\sim\operatorname{Exp}\left(\killingrate(\point_{\sequenceindex-1})\right)$;
                \If{$\Delta\lifetime_1<\Delta\lifetime_2$:}
                    \If{$\left(\textnormal{RaoBlackwellization == None}\right)$:}
                        \State$\lifetime\mathrel+=\Delta\lifetime_1$;
                        \State$\tilde\integral\mathrel+=\Delta\lifetime_1\integrand(\point_{\sequenceindex-1})$;
                    \EndIf                
                    \State Sample $\secondpoint$ from $\proposalkernel(\point_{\sequenceindex-1},\;\cdot\;)$;
                    \If{$\left(\textnormal{RaoBlackwellization == Vanilla}\right)$:}
                        %\State$\ourvanillalifetime\mathrel+=\vanillaproduct\mathrel*=1-\acceptancefunction(\point_{\sequenceindex-1},\secondpoint)$;
                        \State$\vanillaproduct\mathrel*=1-\acceptancefunction(\point_{\sequenceindex-1},\secondpoint)$;
                        \State
                        $\ourvanillalifetime\mathrel+=\vanillaproduct$;
                    \EndIf
                    \If{$\left(\textnormal{RaoBlackwellization == WasteRecycling}\right)$:}
                        \State$\lifetime\mathrel+=\Delta\lifetime_1$;
                        \State$\tilde\integral\mathrel+=\Delta\lifetime_1\left((1-\acceptancefunction(\point_{\sequenceindex-1},\secondpoint))\integrand(\point_{\sequenceindex-1})+\acceptancefunction(\point_{\sequenceindex-1},\secondpoint)\integrand(\secondpoint)\right)$;
                    \EndIf
                    \State Sample $\uniform$ from $\mathcal U_{[0,\:1)}$;\AlgCommentLeft{uniform distribution on $[0,1)$}
                    \If{$\left(\uniform<\acceptancefunction(\point_{\sequenceindex-1},\secondpoint)\right)$:}
                        \If{$\left(\textnormal{RaoBlackwellization == Vanilla}\right)$:}
                            % \State$\vanillaweight=\ourvanillalifetime/\combinedrate(\point_{\sequenceindex-1})$;
                            % \State$\lifetime\mathrel+=\vanillaweight$;
                            % \State$\tilde\integral\mathrel+=\vanillaweight\integrand(\point_{\sequenceindex-1})$;
                            \State$\lifetime\mathrel+=\ourvanillalifetime/\combinedrate(\point_{\sequenceindex-1})$;
                            \State$\tilde\integral\mathrel+=\ourvanillalifetime/\combinedrate(\point_{\sequenceindex-1})\integrand(\point_{\sequenceindex-1})$;
                            \State$\ourvanillalifetime=\vanillaproduct=1$;
                        \EndIf
                        \State$\point_\sequenceindex=\secondpoint$;\AlgCommentLeft{with probability $\acceptancefunction(\point_{\sequenceindex-1},\secondpoint)$ accept the proposal}
                    \Else:
                        \State
                        $\point_\sequenceindex=\point_{\sequenceindex-1}$;\AlgCommentLeft{with probability $1-\acceptancefunction(\point_{\sequenceindex-1},\secondpoint)$ reject}
                        % \If{$\left(\textnormal{RaoBlackwellization == Vanilla}\right)$:}
                        %     \State$\ourvanillalifetime\mathrel+=\vanillaproduct\mathrel*=1-\acceptancefunction(\point_{\sequenceindex-1},\secondpoint)$;
                        % \EndIf
                    \EndIf
                \Else:
                    \If{$\left(\textnormal{RaoBlackwellization == Vanilla}\right)$:}
                        % \State$\vanillaweight=\ourvanillalifetime/\combinedrate(\point_{\sequenceindex-1})$
                        % \State$\lifetime\mathrel+=\vanillaweight$;
                        % \State$\tilde\integral\mathrel+=\vanillaweight\integrand(\point_{\sequenceindex-1})$;
                        \State$\lifetime\mathrel+=\ourvanillalifetime/\combinedrate(\point_{\sequenceindex-1})$;
                        \State$\tilde\integral\mathrel+=\ourvanillalifetime/\combinedrate(\point_{\sequenceindex-1})\integrand(\point_{\sequenceindex-1})$;
                    \Else:
                        \State
                        $\lifetime\mathrel+=\Delta\lifetime_1$;
                        $\tilde\integral\mathrel+=\Delta\lifetime_1\integrand(\point_{\sequenceindex-1})$;
                    \EndIf
                    \State\Return$(\lifetime,\tilde\integral)$;
                \EndIf
            \EndFor
        \EndProcedure
        \State$\lifetimesum=0$;
        \State$\tilde\integral=0$;
        % \State$\lifetimesum=\tilde\integral=0$;
        \For{$\left(\sequenceindex=0;\sequenceindex<\lastsequenceindex;\textnormal{++}\sequenceindex\right)$:}
            \State
            $(\lifetimesum,\tilde\integral)\mathrel+=\textproc{SimulateTour}()$;\label{line:simulate-tour}
        \EndFor
        \State\Return$\tilde\integral/\lifetimesum$;
    \end{algorithmic}
\end{algorithm}

% ----------------------------------------------------------------------------------------------------
% 
% ----------------------------------------------------------------------------------------------------
\section{Numerical study}\label{sec:numerical-study}

\paragraph*{Setup}

In our evaluation, we considered three \gls{mh}-based light transport algorithms and their corresponding \gls{jrlt} variants. Following the terminology introduced in \citet{holl2025jrlt}, we refer to these methods as \emph{Metropolis}, \emph{\gls{mala}}, and \emph{\gls{hmc}}, and to their \gls{jrlt} counterparts as \emph{Metropolis Restore}, \emph{\gls{mala} Restore}, and \emph{\gls{hmc} Restore}. All methods were used with the parameter settings reported therein.

% Metropolis, \gls{mala}, and \gls{hmc} correspond to the algorithms introduced in \citet{hachisuka2014multiplexed}, \citet{luan2020langevin}, and \citet{li2015anisotropic}, respectively. These \gls{mh}-based light transport algorithms are derived from their classical counterparts for general-purpose applications in Euclidean state spaces \citep{metropolis1953equation,roberts1996mala,duane1987hmc}. They differ only in the choice of proposal kernel, which is always derived from time-discretized diffusion processes, namely Brownian motion \citep{karatzas1998brownian}, Langevin dynamics \citep{roberts1996mala}, and Hamiltonian dynamics \citep{duane1987hmc}, in this order.
%
% For each of these methods,
We compared the three estimators studied in this work. For the \gls{mh}-based methods Metropolis, \gls{mala}, and \gls{hmc}, these are standard estimation (\autoref{sec:metropolis-hastings-standard-estimation}), our
% \gls{rb} technique
vanilla \acrlong{rb} (\autoref{sec:metropolis-hastings-vanilla}), and the traditionally used \gls{rb} technique waste-recycling (\autoref{sec:metropolis-hastings-waste-recycling}). We refer to the corresponding variants as \emph{Metropolis}, \emph{Metropolis Vanilla}, and \emph{Metropolis Waste-Recycling}, with the naming applied mutatis mutandis to \gls{mala} and \gls{hmc}. % Algorithmically, a generic implementation supporting all these techniques is given in \autoref{alg:rao-blackwellized-metropolis-hastings-algorithm}.

For the \gls{jrlt} methods Metropolis Restore, \gls{mala} Restore, and \gls{hmc} Restore,
% we compared the corresponding \gls{jrlt} variants of the same three estimators. Standard estimation
standard estimation
is described in \autoref{sec:jump-restore-standard-estimation}, while our vanilla \gls{rb} and waste-recycling are described in \autoref{sec:jump-restore-vanilla-estimation} and \autoref{sec:jump-restore-waste-recycling}, respectively. We refer to these variants as \emph{Metropolis Restore}, \emph{Metropolis Restore Vanilla}, and \emph{Metropolis Restore Waste-Recycling}, again with the naming applied mutatis mutandis to \gls{mala} Restore and \gls{hmc} Restore. % A generic algorithmic implementation supporting all these techniques is provided in \autoref{alg:rao-blackwellized-jump-restore-algorithm}.
Generic implementations supporting all evaluated algorithms were given in  \autoref{alg:rao-blackwellized-metropolis-hastings-algorithm} and \autoref{alg:rao-blackwellized-jump-restore-algorithm}.

% \paragraph{Evaluation}

% We selected a diverse set of test scenes with varying lighting characteristics that are known to be challenging to render. We implemented \autoref{alg:rao-blackwellized-metropolis-hastings-algorithm} and \autoref{alg:rao-blackwellized-jump-restore-algorithm} in the rendering systems \citet{pbrt} and \citet{dpt}.

% For all scenes, we report quantitative results in the form of plots showing the evolution of the MSE, MRSE, MAPE, and the empirical variance as functions of sample count and rendering time, as well as error tables at fixed sample counts and fixed rendering times. In addition, we provide qualitative visual comparisons, including MAPE images.

% As the results show, waste recycling almost never leads to a perceptible reduction in any of the metrics. In contrast, our vanilla \gls{rb} techniques achieve significant reductions across all metrics.

\paragraph*{Rendering environment}

% We implemented our methods in the \textsc{pbrt-v4} \cite{pharr2023pbrt} and \textsc{lmc} \cite{luan2020langevin} rendering system and applied the \gls{mcmc} methods to both direct and indirect lighting.
We implemented our methods in \textsc{pbrt-v4} \cite{pharr2023pbrt} and \textsc{lmc} \cite{luan2020langevin} and applied the \gls{mcmc} methods to both direct and indirect lighting.

\paragraph*{Error metrics}

We assessed several quantitative metrics:
% the $L^2$-error (i.e., MSE),
%MSE,
%MRSE, MAPE and empirical variance.
MSE, MRSE, MAPE and
% empirical
variance.
% Reference images were generated using \gls{bdpt} with $2^{20}$ samples per pixel.
References use \gls{bdpt} with $2^{20}$ samples per pixel.
%
% For each method, all metrics are averaged over 100 realizations with % identical
% equal
% random seeds, such that differences arise solely from the % estimator construction.
To assess potential bias, for each method we average all metrics over 100 realizations using identical random seeds, ensuring that observed differences arise solely from estimator construction.

\paragraph*{Test scenes}

We evaluated a diverse set of scenes exhibiting different light transport characteristics. The scenes were drawn from multiple sources: \textsc{Glass of Water}, \textsc{Salle de bain}, and \textsc{Veach, Ajar} from \citet{bitterli2016resources}; \textsc{Torus} from \citet{lmc}; \textsc{Swimming Pool} from \citet{rioux2020delayed}; and \textsc{Zero Day} from \citet{pharr2025scenes}.

% \vspace{-.1cm}
% \paragraph*{Evaluation}
% Qualitative comparisons are presented in \Cref{fig:teaser,fig:pool,fig:torus,fig:bathroom2,fig:glass-of-water,fig:zero-day} as well as in Section H of the supplemental, covering all test scenes introduced above.
% %
% These results show that our vanilla \gls{rb} estimators consistently achieve higher visual fidelity than standard estimation, whereas waste recycling rarely yields a perceptible improvement.
% %
% Quantitative results in \Cref{fig:error} corroborate this finding: both the
% % $L^2$-error and the empirical variance
% MSE and variance
% decrease more rapidly over time for the vanilla \gls{rb} variants.
% %
% Detailed tables reporting absolute errors at equal-sample-count and equal-rendering-time are provided in Sections~G and~H of the supplemental, together with plots showing the evolution of the MSE, MRSE, MAPE, and the empirical variance as functions of sample count and rendering time.

\section{Conclusion}

Qualitative comparisons presented in \Cref{fig:teaser,fig:pool,fig:torus,fig:bathroom2,fig:glass-of-water,fig:zero-day} and Section~H of the supplemental demonstrate that our vanilla \gls{rb} estimators consistently achieve higher visual fidelity than standard estimation, while waste-recycling rarely yields a perceptible improvement.
These observations are corroborated by the quantitative results in \Cref{fig:error} and Sections~G and~H of the supplemental, which provide detailed error tables at equal-sample-count and equal-rendering-time, as well as plots of all evaluation metrics as functions of sample count and rendering time.
Across all metrics, errors decrease substantially faster when using our proposed vanilla \gls{rb} estimators, establishing them as a generally applicable \gls{rb} technique that consistently outperforms the current state-of-the-art in rendering.

% \begin{figure}
%     \input{figures/bathroom}
%     \captionsetup{skip = 0pt}
%     \caption{Equal-rendering-time comparison (120\,s) of Metropolis (left), Metropolis Restore (middle), and \gls{erpt} (right) for the \textsc{Contemporary Bathroom} scene provided by \citep{bitterli2016resources}.}
%     \label{fig:bathroom}
% \end{figure}

% \input{plots/kitchen}
% \input{plots/c0}

% ----------------------------------------------------------------------------------------------------
% bibliography
% ----------------------------------------------------------------------------------------------------
% \bibliographystyle{plainnat}
\bibliographystyle{ACM-Reference-Format}
\bibliography{bibliography}

\newpage
% ----------------------------------------------------------------------------------------------------
% additional results
% ----------------------------------------------------------------------------------------------------

% \begin{figure}
%     \input{figures/pool}
%     \captionsetup{skip = 0pt}
%     \caption{Equal-rendering-time comparison (20\,s) of ordinary (left), vanilla Rao--Blackwellized (middle), and waste-recycled (right) \gls{mala} for the \textsc{Swimming Pool} scene provided by \citet{rioux2020delayed}.}
%     \label{fig:pool}
% \end{figure}

% \begin{figure}[t]
%     \input{figures/veach_torus_error}
%     \captionsetup{skip = 0pt}
%     \caption{$L^2$-error and empirical variance over rendering time in seconds for the \textsc{Veach, ajar} and \textsc{Torus} scene depicted in \autoref{fig:veach} and \autoref{fig:torus}, respectively.}
%     \label{fig:veach-torus-error}
% \end{figure}

\begin{figure}
    \centering
    \includegraphics[width=\linewidth]{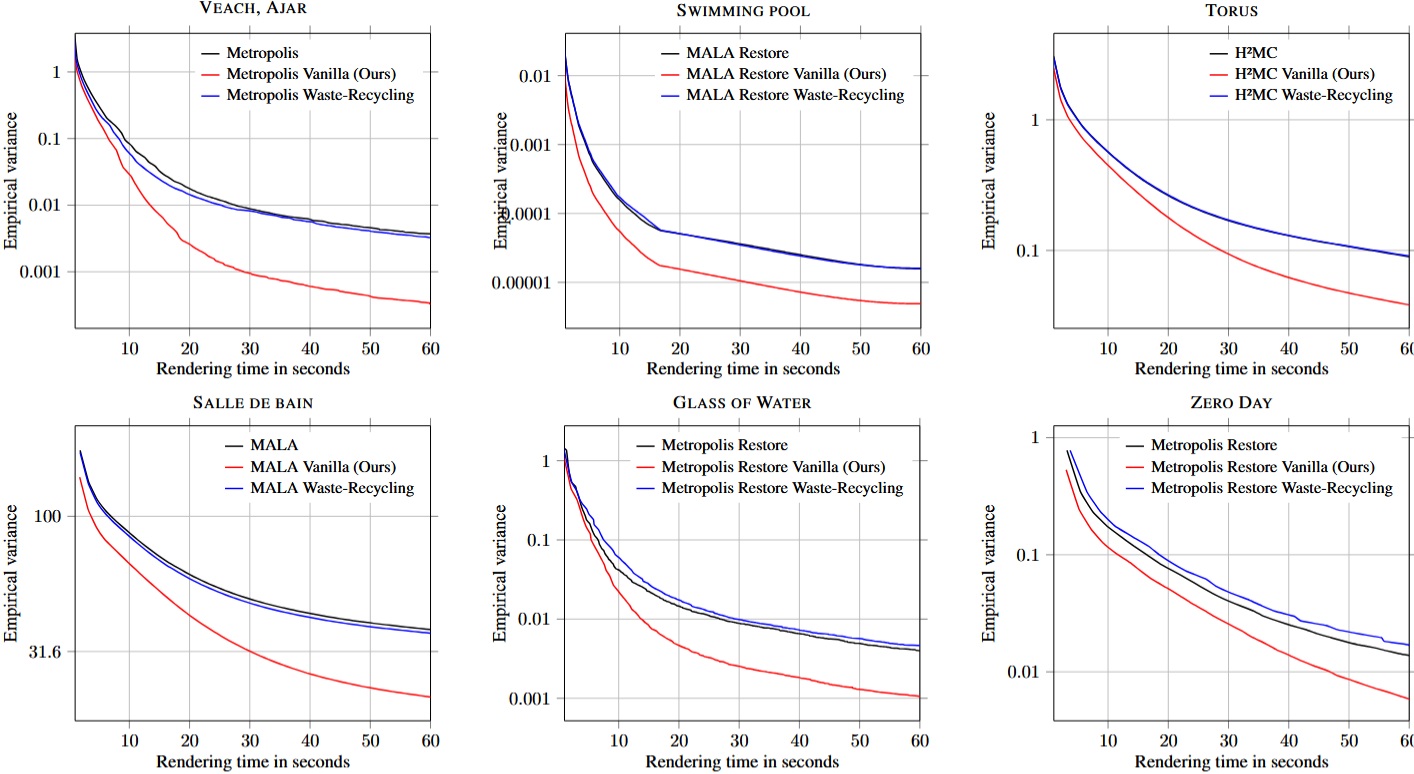}
    \caption{
        Empirical variance as a function of rendering time (up to 60\,s), averaged over 100 realizations, for the scenes shown in \Cref{fig:teaser,fig:pool,fig:torus,fig:bathroom2,fig:glass-of-water,fig:zero-day}. In some cases, the variance reduction achieved by waste-recycling compared to standard estimation is so small that the two plots almost overlap. In contrast, our vanilla \gls{rb} consistently leads to a clear variance reduction.
    }
    \label{fig:error}
\end{figure}

\begin{figure}
    \input{figures/pool/MALA\_Restore_new_layout}
    \captionsetup{skip = 0pt}
    \caption{Equal-rendering-time comparison (20\,s) of ordinary (left), vanilla Rao--Blackwellized (middle), and waste-recycled (right) MALA Restore for the \textsc{Swimming Pool} scene provided by \citet{rioux2020delayed}. The inset error maps show the corresponding MAPE for each rendering.}
    \label{fig:pool}
\end{figure}

% \begin{figure}[t]
%     \input{plots/error_metropolis}
%     \captionsetup{skip = 0pt}
%     \caption{Equal-rendering-time comparison (up to 60 s, avg.\ over 100 realizations) for the \textsc{Swimming Pool} and \textsc{Torus} scenes depicted in \Cref{fig:pool,fig:torus}.}
%     \label{fig:error-metropolis}
% \end{figure}

\clearpage

\begin{figure}
    \input{figures/torus/hmc_new_layout}
    \captionsetup{skip = 0pt}
    \caption{Equal-rendering-time comparison (20\,s) of ordinary (left), vanilla Rao--Blackwellized (middle), and waste-recycled (right) \gls{hmc} for the \textsc{Torus} scene provided by \citet{lmc}. The inset error maps show the corresponding MAPE for each rendering.}
    \label{fig:torus}
\end{figure}

\begin{figure}
    \input{figures/bathroom2/mala_new_layout}
    \captionsetup{skip = 0pt}
    \caption{Equal-rendering-time comparison (20\,s) of ordinary (left), vanilla Rao--Blackwellized (middle), and waste-recycled (right) Metropolis for the \textsc{Veach, ajar} scene provided by \citet{bitterli2016resources}. The inset error maps show the corresponding MAPE for each rendering.}
    \label{fig:bathroom2}
\end{figure}
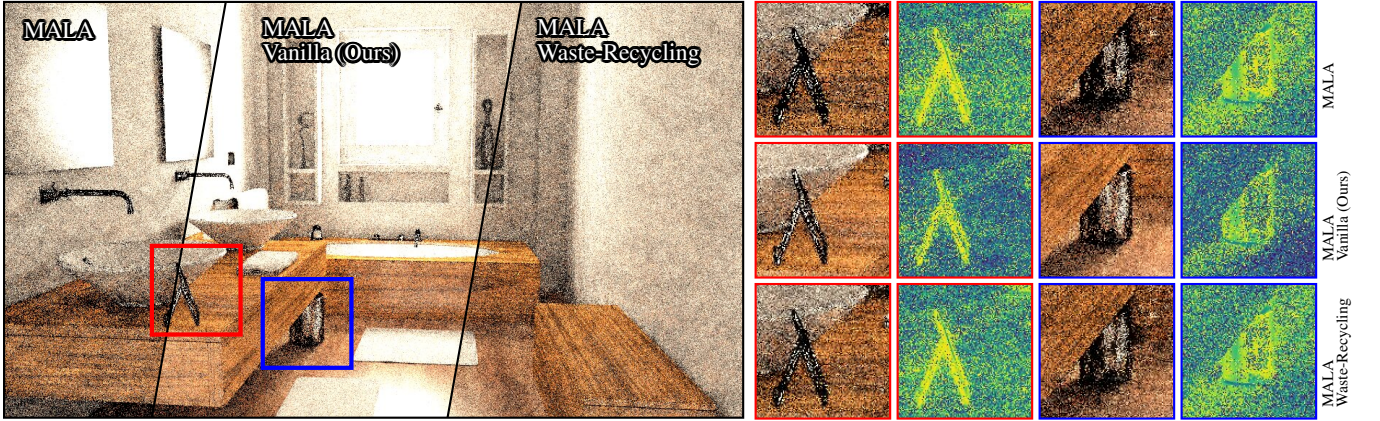

% \begin{figure}[t]
%     \input{plots/error_veach_ajar_glass_of_water}
%     \captionsetup{skip = 0pt}
%     \caption{Equal-rendering-time comparison (up to 60 s, avg.\ over 100 realizations) for the \textsc{Veach, Ajar} and \textsc{Glass of Water} scenes depicted in \Cref{fig:veach-ajar,fig:glass-of-water}.}
%     \label{fig:error-veach-ajar-glass-of-water}
% \end{figure}

\clearpage

\begin{figure}
    \input{figures/glass_of_water/Metropolis\_Restore_new_layout}
    \captionsetup{skip = 0pt}
    \caption{Equal-rendering-time comparison (20\,s) of ordinary (left), vanilla Rao--Blackwellized (middle), and waste-recycled (right) Metropolis Restore for the \textsc{Glass of Water} scene provided by \citet{bitterli2016resources}. The inset error maps show the corresponding MAPE for each rendering.}
    \label{fig:glass-of-water}
\end{figure}

\begin{figure}
    \input{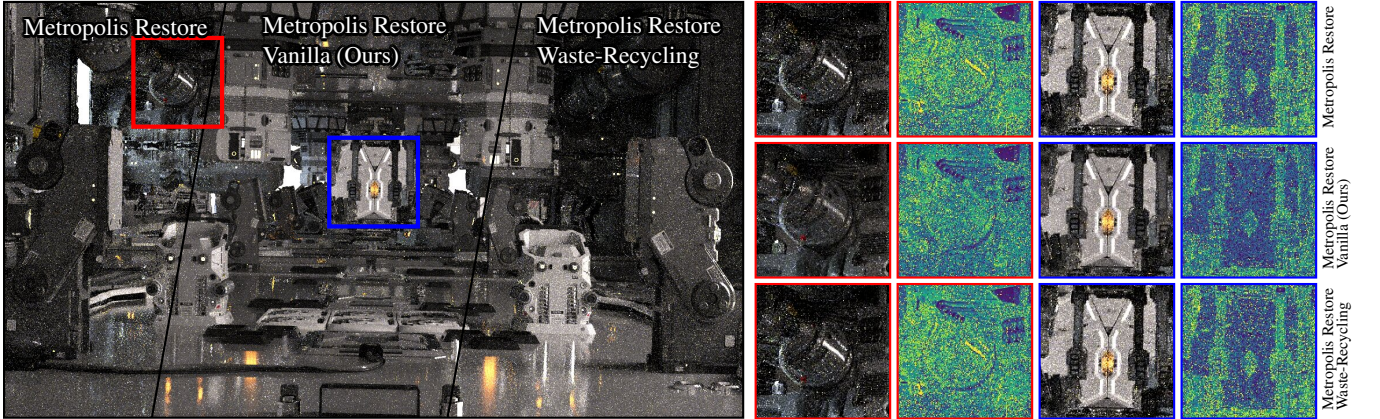}
    \captionsetup{skip = 0pt}
    \caption{Equal-rendering-time comparison (20\,s) of ordinary (left), vanilla Rao--Blackwellized (middle), and waste-recycled (right) Metropolis Restore for the \textsc{Zero Day} scene provided by \citet{pharr2025scenes}. The inset error maps show the corresponding MAPE for each rendering.}
    \label{fig:zero-day}
\end{figure}

\clearpage
\appendix
\counterwithin{figure}{section}

% ----------------------------------------------------------------------------------------------------
% Markov chain Monte Carlo
% ----------------------------------------------------------------------------------------------------
\section{Preliminaries on Markov chain Monte Carlo}\label{sec:appendix-mcmc-preliminiaries}

% \subsection{Basic principle}

Given a finite measure \setword{$\targetdistribution$}{inline:target-distribution}, \acrfull{mcmc} is a technique for estimating the integral \begin{equation}\label{eq:integral}
    \integral:=\int\integrand\dd{\targetdistribution}
    %\targetdistribution\integrand:=\int\integrand\dd{\targetdistribution}
\end{equation} of a $\targetdistribution$-integrable function \setword{$\integrand$}{inline:integrand}. More precisely, it is a recipe for constructing an ergodic Markov process with invariant distribution~$\targetdistribution$.

\paragraph*{Markov process}

A process is a state system evolving over time. Usually, the time domain $\timedomain$ is either discrete, $\timedomain=\mathbb N_0$, or continuous, $\timedomain=[0,\infty)$. Informally, the process is said to be Markov, if at any fixed point in time, the evolution of the process does only depend on the present state, but not on the past. %A formal definition is given in \autoref{def:markov-process}.
For a general introduction to Markov processes, see \citet{kallenberg2021probability,cinlar2011probability,douc2018markov,ethier2009markov,klenke2020probability}.

\paragraph*{Invariance}

A Markov process $(\process_\timepoint)_{\timepoint\in\timedomain}$ is $\targetdistribution$-invariant if and only if
once $\process_\prevtimepoint$ is distributed according to $\targetdistribution$ at a certain time point $\prevtimepoint\in\timedomain$, every state $\process_\timepoint$ at a future time point $\timepoint\in\timedomain\cap(\prevtimepoint,\infty]$ will be distributed according to $\targetdistribution$ as well. That is, the distribution of a state is stationary in time after it once coincided with $\targetdistribution$.

\paragraph*{Ergodicity}

Ergodicity ensures that the long time average of an observation is effectively equal to space averaging with respect to the invariant distribution. More precisely, given that the invariant distribution $\targetdistribution$ actually exists, ergodicity is equivalent to enforcing that if $\process_0$ is distributed according to $\targetdistribution$, then \begin{align}\label{eq:ergodic-theorem}
    \ergodicaverage_\timepoint\integrand:=\frac1\timepoint\left.\begin{cases}\displaystyle\sum_{\prevtimepoint=0}^{\timepoint-1}\integrand(\process_\prevtimepoint)&\text{, if }\timedomain=\mathbb N_0\\\displaystyle\int_0^\timepoint\integrand(\process_\prevtimepoint)\dd{\prevtimepoint}&\text{, if }\timedomain=[0,\infty)\end{cases}\right\}\xrightarrow{\timepoint\to\infty}
    %\targetdistribution\integrand
    \integral
\end{align} almost surely for all $\integrand\in\mathcal L^1(\targetdistribution)$.

% ----------------------------------------------------------------------------------------------------
% Bias characterization and control of the vanilla Rao--Blackwellization estimator
% ----------------------------------------------------------------------------------------------------
\section{Bias characterization and control of the vanilla Rao--Blackwellization estimator}

Strictly speaking, the vanilla \acrlong{rb} estimator introduced in Definition~5.4 constitutes an \emph{inexact} \gls{mcmc} method \citep{eberle2023inexact}.
% In this section, we derive an explicit expression for the induced bias and derive how this bias can be controlled in practice. Finally, we explain why this bias is purely theoretical, does not manifest in the empirical results presented in Section 7, and is numerically negligible in the light transport setting.
In this section, we derive an explicit expression for the induced bias of the underlying holding time estimator (Equation 24 of the main paper). While a theoretical characterization of how this bias propagates to the vanilla Rao--Blackwellized estimator (Equation 25 of the main paper) is left for future work, our empirical results in Sections~E, G and H consistently show no measurable bias in practice. Comparisons against reference images generated via BDPT can be examined for further inspection using the interactive viewer shipped together with this document.

For the analysis, we fix $\lastsequenceindex\in\mathbb N_0$ for the remainder of this section and consider the time $\lifetimesum_{\lastsequenceindex+1}$ at which a proposal is accepted for the $(\lastsequenceindex+1)$-th time. For simplicity, we assume $\lifetimesum_{\lastsequenceindex+1}<\infty$, which holds with probability one outside of degenerate scenarios. This assumption is made without loss of generality, as all statements can alternatively be understood as holding conditionally on $\{\lifetimesum_{\lastsequenceindex+1}<\infty\}$.

Immediately before this time, the current state is $\acceptedchain_{\lastsequenceindex}$. The proposal accepted at this time, as well as all proposals rejected in sequence beforehand, are conditionally distributed according to $\proposalkernel(\acceptedchain_{\lastsequenceindex},\;\cdot\;)$ given $\acceptedchain_{\lastsequenceindex}$.

To keep the notation as simple as possible for the subsequent analysis, let $(\tilde\secondprocess_\timepoint)_{\timepoint\in\mathbb N}$ be an independent process that is independent of $\process_0$ and $(\secondprocess_\timepoint)_{\timepoint\in\mathbb N}$ and is identically conditionally distributed according to $\proposalkernel(\acceptedchain_{\lastsequenceindex},\;\cdot\;)$ given $\acceptedchain_{\lastsequenceindex}$; i.e., \begin{equation}
    \probabilitymeasure\left[\tilde\secondprocess_\timepoint\in\;\cdot\;\mid\acceptedchain_{\lastsequenceindex}\right]
    =\proposalkernel(\acceptedchain_{\lastsequenceindex},\;\cdot\;)\eqforall\timepoint\in\mathbb N.
\end{equation} We continue to denote the proposal sequence generated by the Metropolis--Hastings algorithm by $(\secondprocess_\timepoint)_{\timepoint\in\mathbb N}$. Define \begin{equation}
    \vanillaproduct_\timepoint:=\prod_{\prevtimepoint=1}^\timepoint\left(1-\acceptancefunction\left(\acceptedchain_{\lastsequenceindex},\secondprocess_{\lifetimesum_{\lastsequenceindex}+\prevtimepoint}\right)\right)
\end{equation} and \begin{equation}
    \tilde\vanillaproduct_\timepoint:=\prod_{\prevtimepoint=1}^\timepoint\left(1-\acceptancefunction\left(\acceptedchain_{\lastsequenceindex},\tilde\secondprocess_\prevtimepoint\right)\right)
\end{equation} for $\timepoint\in\mathbb N$.
Recall that the only novel quantities in the vanilla \acrlong{rb} estimator from Definition~5.4 are the $\hat\lifetime_\sequenceindex$. By definition, \begin{equation}
    \hat\lifetime_\sequenceindex=1+\sum_{\timepoint=1}
    % ^{\lifetime_\sequenceindex-1}
    ^{\lifetime_\sequenceindex}
    \vanillaproduct_\timepoint
\end{equation} for all $\sequenceindex\in\{1,\ldots,\lastsequenceindex\}$. The quantitites $\hat\lifetime_\sequenceindex$ are \emph{not} unbiased estimators (conditionally given $\acceptedchain_\sequenceindex$) of ${\expectedacceptance(\acceptedchain_\sequenceindex)}^{-1}$. However, they can be completed to an unbiased estimator as follows:

\begin{lemma}\label{lem:completion}
    Let $\sequenceindex\in\{1,\ldots,\lastsequenceindex\}$. Then \begin{equation}
        \overline\lifetime_\sequenceindex:=\hat\lifetime_\sequenceindex+\vanillaproduct_{\lifetime_\sequenceindex-1}\sum_{\timepoint\in\mathbb N}\tilde\vanillaproduct_\timepoint
    \end{equation} is a well-defined unbiased estimator (conditionally given $\acceptedchain_\sequenceindex$) of ${\expectedacceptance(\acceptedchain_\sequenceindex)}^{-1}$; i.e., \begin{equation}
        \expectation\left[\overline\lifetime_\sequenceindex\mid\acceptedchain_\sequenceindex\right]
        =\frac1{\expectedacceptance(\acceptedchain_\sequenceindex)}.
    \end{equation}
    \begin{proof}[Proof\textup:\nopunct]
        By construction, this follows immediately from Lemma~5.1.
    \end{proof}
\end{lemma}

In view of \autoref{lem:completion}, the bias introduced by using the $\hat\lifetime_\sequenceindex$ in the vanilla \acrlong{rb} estimator can thus be characterized by the error terms \begin{equation}
    \vanillaerror_\sequenceindex:=\overline\lifetime_\sequenceindex-\hat\lifetime_\sequenceindex
\end{equation} for $\sequenceindex\in\{1,\ldots,\lastsequenceindex\}$. To this end, define \begin{equation}
    r_2:=\int\proposalkernel(\;\cdot\;,\dd{\secondpoint})\left|1-\acceptancefunction(\;\cdot\;,\secondpoint)\right|^2.
\end{equation}

By construction, $r_2(\point)$ is the second moment of the rejection probability under the proposal $\proposalkernel(\point,\;\cdot\;)$, i.e.,
$r_2(\point)=\expectation\left[\left|1-\acceptancefunction(\point,\secondprocess)\right|^2\right]$
where $\secondprocess\sim\proposalkernel(\point,\;\cdot\;)$, and it quadratically measures how far a typical proposal is from immediate acceptance. With this we obtain the following result:

\begin{theorem}\label{thm:error}
    Let $\sequenceindex\in\{1,\ldots,\lastsequenceindex\}$. Then
    \begin{equation}
        \expectation\left[\vanillaerror_\sequenceindex\mid\acceptedchain_\sequenceindex\right]
        =\frac{1-\expectedacceptance(\acceptedchain_\sequenceindex)}{1-r_2(\acceptedchain_\sequenceindex)}.
    \end{equation}
    \begin{proof}[Proof\textup:\nopunct]
        First,
        \begin{equation}\label{eq:error1}
            \begin{split}
                \expectation\left[\vanillaproduct_{\lifetime_\sequenceindex-1}\mid\acceptedchain_\sequenceindex\right]
                &=\sum_{\timepoint\in\mathbb N_0}\expectation\left[1_{\{\lifetime_\sequenceindex=\timepoint+1\}}\vanillaproduct_{\lifetime_\sequenceindex-1}\mid\acceptedchain_\sequenceindex\right]\\
                &=\frac{\expectedacceptance(\acceptedchain_\sequenceindex)}{1-r_2(\acceptedchain_\sequenceindex)},
            \end{split}
        \end{equation}
        where
        \begin{equation}
            \begin{split}
                \expectation\left[1_{\{\lifetime_\sequenceindex=\timepoint+1\}}\vanillaproduct_{\lifetime_\sequenceindex-1}\mid\acceptedchain_\sequenceindex\right]
                &=\expectation\left[\vanillaproduct_\timepoint^2\acceptancefunction\left(\acceptedchain_\sequenceindex,\secondprocess_{\lifetimesum_\sequenceindex+\timepoint+1}\right)\mid\acceptedchain_\sequenceindex\right]\\
                &=\expectation\left[\vanillaproduct_\timepoint^2\mid\acceptedchain_\sequenceindex\right]
                  \expectation\left[\acceptancefunction\left(\acceptedchain_\sequenceindex,\secondprocess_{\lifetimesum_\sequenceindex+\timepoint+1}\right)\mid\acceptedchain_\sequenceindex\right]\\
                &={r_2(\acceptedchain_\sequenceindex)}^{\timepoint}\expectedacceptance(\acceptedchain_\sequenceindex)
            \end{split}
        \end{equation}
        for all $\timepoint\in\mathbb N_0$. The degenerate case $r_2(\point)=1$ can be excluded without loss of generality, since in this case no proposal would be accepted with probability one. Moreover,
        \begin{equation}\label{eq:error2}
            \expectation\left[\sum_{\timepoint\in\mathbb N}\tilde\vanillaproduct_\timepoint\;\middle|\;\acceptedchain_\sequenceindex\right]
            =\frac{1-\expectedacceptance(\acceptedchain_\sequenceindex)}{\expectedacceptance(\acceptedchain_\sequenceindex)}
        \end{equation}
        by Lemma~1.9. Combining \eqref{eq:error1} and \eqref{eq:error2} yields the claimed identity.
    \end{proof}
\end{theorem}

As is evident from the proof of \autoref{thm:error}, $r_2(\point)$ controls the exponential decay rate of the product terms that constitute the tail of $\overline\lifetime_\sequenceindex$. From the identity in \autoref{thm:error}, we can directly conclude that the bias introduced by the truncation $\hat\lifetime_\sequenceindex$ is negligible if
\begin{enumerate}[(i)]
    \item $r_2(\acceptedchain_\sequenceindex)$ is small; or
    \item $\expectedacceptance(\acceptedchain_\sequenceindex)$ is close to $1$ (high acceptance probability at state $\acceptedchain_\sequenceindex$).
\end{enumerate}

Accordingly, if the proposal kernel induces local moves for which the second moment of the rejection probability under $\proposalkernel(\acceptedchain_\sequenceindex),\;\cdot\;)$ is small, the contribution of the omitted tail terms is numerically negligible.

This theoretical intuition based on small-scale perturbations is illustrative and does not aim to cover the full range of proposal mechanisms used in practice. However, this condition is typically satisfied even for algorithms such as MALA or H²MC. While a formal proof is beyond the scope of this document, empirical evidence is given in Sections E and G–H, which show convergence toward the reference and clear advantage of the vanilla Rao--Blackwellization estimator over the baseline estimators.

% This condition is naturally satisfied in the light transport setting, where proposals are constructed as small-scale Gaussian perturbations of the current state and therefore induce only mild and low-variance changes in the Metropolis--Hastings acceptance probability. This explains the empirical results reported in Section~7, which show a clear advantage of the vanilla \acrlong{rb} estimator over the baseline estimators.

\section{Importance sampling and\\ estimation of the normalization constant}

In practice, the integral of interest to be estimated is often not with respect to the given target (probability) distribution $\targetdistribution$, but rather with respect to the reference measure $\referencemeasure$ with respect to which $\targetdistribution$ is absolutely continuous. This assumption~\textemdash\; that $\targetdistribution$ is absolutely continuous with respect to a reference measure $\referencemeasure$~\textemdash\; has already been necessary for the formulation of the Metropolis--Hastings algorithm in \autoref{sec:metropolis-hastings-algorithm} and the Jump Restore algorithm in \autoref{sec:jump-restore-algorithm}.

More concretely, we are not actually interested in the integral $\integral$ in \eqref{eq:integral}, but rather in
\begin{equation}\label{eq:importance-sampling}
	\int\secondintegrand\dd{\referencemeasure}=\targetdensity_\referencemeasure\int\frac\secondintegrand\targetdensity\dd{\targetdistribution}.
\end{equation}

At this point, the normalization constant $\targetdensity_\referencemeasure$ of $\targetdensity$ becomes explicitly required. In Metropolis--Hastings applications — including Metropolis--Hastings-based light transport implementations such as in \citet{pbrt} — this constant is computed during a \emph{bootstrapping} phase using traditional Monte Carlo methods (or, in the context of light transport, \emph{path tracing}).

In fact, in the Jump Restore algorithm this additional bootstrapping stage is not necessary. By definition of the killing rate (\eqref{eq:killing-rate} of the main paper), the constant $\expectedlifetime>0$ occurring in it, represents the reciprocal of the expected lifetime of an individual tour; see \citet[Section~6.2]{holl2025jrlt}. Formally, \begin{equation}\label{eq:expected-lifetime}
    \expectation_\regenerationdistribution\left[\lifetime_\sequenceindex\right]=\frac1\expectedlifetime,
\end{equation} where $\expectation_\regenerationdistribution$ denotes expectation with respect to the probability measure according to which the $\sequenceindex$th tour is respawned at a location sampled from $\regenerationdistribution$.

Upon relinquishing the option to prescribe the expected lifetime of individual tours via the constant $\expectedlifetime$, the normalization constant $\targetdensity_\referencemeasure$ can instead be absorbed into the parameter $\expectedlifetime$ by defining \begin{equation}
	\tilde\expectedlifetime:=\expectedlifetime\targetdensity_\referencemeasure.
\end{equation} 
Accordingly, we define the killing rate as an alternative to \eqref{eq:killing-rate} by
\begin{equation}\label{eq:killing-rate-modified}
	\killingrate:=\tilde\expectedlifetime\frac\regenerationdensity\targetdensity,
\end{equation}
where again $\tilde\expectedlifetime>0$ can be chosen arbitrarily. By definition and equation~\eqref{eq:expected-lifetime}, the normalization constant $\targetdensity_\referencemeasure$ is now recovered via
\begin{equation}\label{eq:normalization-constant-by-expectedlifetime}
	\targetdensity_\referencemeasure=\tilde\expectedlifetime\expectation_\regenerationdistribution\left[\lifetime_\sequenceindex\right].
\end{equation}
Since
\begin{equation}
	\expectation_\regenerationdistribution\left[\lifetime_\sequenceindex\right]=\frac1\thirdsequenceindex\sum_{\secondsequenceindex=1}^\thirdsequenceindex\lifetime_\secondsequenceindex=\frac{\lifetimesum_\thirdsequenceindex}\thirdsequenceindex,
\end{equation}
we obtain, by combining \eqref{eq:importance-sampling}, \eqref{eq:jump-restore-ergodic-average}, and \eqref{eq:normalization-constant-by-expectedlifetime},
\begin{equation}\label{eq:jump-restore-auto-normalization}
	\int\secondintegrand\dd{\referencemeasure}=\frac{\tilde\expectedlifetime}\thirdsequenceindex\int_0^{\lifetimesum_\thirdsequenceindex}\frac\secondintegrand\targetdensity(\concatenatedprocess_\timepoint)\dd{\timepoint}.
\end{equation}
Evidently, the normalization constant $\targetdensity_\referencemeasure$ of the target distribution $\targetdistribution$ no longer appears in this expression and is thus not required for computing the estimator. Now, letting $\left(\left(\Delta\timepoint_1,\point_0\right),\ldots,\left(\Delta\timepoint_\lastsequenceindex,\point_{\lastsequenceindex-1}\right)\right)$ denote an output of \autoref{alg:jump-restore-algorithm} again, we conclude by combining \eqref{eq:jump-restore-auto-normalization} and \eqref{eq:jump-restore-standard-estimation} that
\begin{equation}\label{eq:jump-restore-standard-estimation}
	\frac{\tilde\expectedlifetime}\thirdsequenceindex\sum_{\sequenceindex=1}^\lastsequenceindex\Delta\timepoint_\sequenceindex\integrand\left(\point_{\sequenceindex-1}\right)\approx\int\integrand\dd{\targetdistribution}.
\end{equation}
Remarkably, when using the definition \eqref{eq:killing-rate-modified} of the killing rate, there is even no longer any need to track the lifetime estimates $\tilde\lifetime$ in \autoref{alg:rao-blackwellized-jump-restore-algorithm}.

In our numerical results, we employed this definition of the killing rate. Consequently, the reported computation times implicitly include, to some extent, the cost of estimating the normalization constant. In contrast, we refrained from including the bootstrapping times for the Metropolis--Hastings-based algorithms in the reported computation times. Accounting for those times fairly would be difficult, as the additional computational effort depends heavily on the number of samples chosen during the bootstrapping phase: a substantial increase in runtime may occur, despite the fact that a reasonably accurate estimate of the normalization constant might already be achievable with significantly fewer samples.

\section{Implementation Details and Hardware Setup}

\paragraph*{Parameter choices}

Across all considered Metropolis--Hastings variants \textemdash\ Metropolis, MALA, and H²MC \textemdash\ we used a large step probability of $\largestepprobability=0.3$, a value known to perform well in practice. To account for burn-in, we discarded the first 10{,}000 iterations before collecting estimates.

For the corresponding Jump Restore methods \textemdash\ Metropolis Restore, MALA Restore, and H²MC Restore \textemdash\ we set $\expectedlifetime=1$ in \eqref{eq:killing-rate}.

\paragraph*{Hardware setup}

All renderings were performed on a system equipped with two AMD EPYC~7702 processors, each providing 64~cores and 128~hardware threads. The CPUs operate at a clock speed of 2–3.3\,GHz and are paired with 2048\,GiB of DDR4 ECC memory running at 3200\,MHz. All computations were executed entirely on the CPU. Since our hardware limited us to at most 256 concurrent threads, the results reported in \autoref{sec:numerical-study} may further improve on systems that support a higher degree of parallelism.

\section{Statistical experiments}

\subsection{Metropolis--Hastings algorithm}

In this section, following \citet[Section~4]{douc2011vanilla}, we consider three statistical toy examples that illustrate and visualize the variance reduction achieved by our vanilla \acrlong{rb} estimator, compared to standard estimation and waste recycling, in the context of the Metropolis--Hastings algorithm.

We begin with an Metropolis--Hastings algorithm with proposal kernel $\proposalkernel(\point,\;\cdot\;)=\mathcal N(\point,\varsigma^2)$ \textemdash\ that is, a classical Metropolis algorithm \textemdash\ and target distribution $\targetdistribution=\mathcal N(0,1)$. In \autoref{tab:gaussian-gaussian}, we report estimates of $\integral$ obtained with our vanilla \acrlong{rb} estimator $\vanillaestimator_\timepoint\integrand$, the standard ergodic average estimator $\ergodicaverage_\timepoint\integrand$, and the waste-recycling estimator $\wasterecyclingestimator_\timepoint\integrand$ for four different integrands $\integrand$ and four choices of the standard deviation $\varsigma$.
In each column, separated by “/”, we report the ratio of the variances of $\vanillaestimator_\timepoint\integrand$ and $\ergodicaverage_\timepoint\integrand$ \textemdash\ that is, $\operatorname{Var}\left[\vanillaestimator_\timepoint\integrand\right]/\operatorname{Var}\left[\ergodicaverage_\timepoint\integrand\right]$ \textemdash\ as well as the ratio of the variances of $\wasterecyclingestimator_\timepoint\integrand$ and $\ergodicaverage_\timepoint\integrand$ \textemdash\ that is, $\operatorname{Var}\left[\wasterecyclingestimator_\timepoint\integrand\right]/\operatorname{Var}\left[\ergodicaverage_\timepoint\integrand\right]$. The smaller a value below $1$, the larger the variance reduction compared to standard estimation via $\ergodicaverage_\timepoint\integrand$. All values are computed from 1{,}000 independent realizations and correspond to $\timepoint=100$ iterations.

\setlength{\tabcolsep}{3pt}
\begin{table}[H]
    \centering
    \captionsetup{skip = 0pt}
    \caption{Ratio of the empirical variances of our vanilla Rao--Blackwellization estimator $\vanillaestimator_\timepoint\integrand$ and the standard estimator $\ergodicaverage_\timepoint\integrand$ of $\integral$ at $\timepoint=100$ for the Metropolis--Hastings algorithm with proposal kernel $\proposalkernel(\point,\;\cdot\;)=\mathcal N(\point,\varsigma^2)$ and initial state drawn from the target distribution $\targetdistribution=\mathcal N(0,1)$. For comparison, the values shown after “/” report the corresponding ratio between the waste-recycling estimator $\wasterecyclingestimator_\timepoint\integrand$ and the standard estimator. The computation has been performed using $1{,}000$ independent realizations.}
    \label{tab:gaussian-gaussian}
    \begin{tabular}{lcccc}
        \toprule
        $\integrand(\point)$ 
        & $\point$ 
        & $\point^2$ 
        & $1_{\{\;\point\;>\;0\;\}}$ 
        & $\expectedacceptance(\point)$ \\
        \midrule
        $\varsigma=0.1$
        & \textbf{0.982} / 0.998
        & \textbf{0.975} / 0.998
        & \textbf{0.983} / 0.999
        & \textbf{0.971} / 0.983 \\
        
        $\varsigma=2$   
        & \textbf{0.792} / 0.893 
        & \textbf{0.716} / 0.817
        & \textbf{0.803} / 0.945
        & \textbf{0.787} / 0.924 \\
        
        $\varsigma=5$   
        & \textbf{0.789} / 0.941 
        & \textbf{0.803} / 0.905
        & \textbf{0.757} / 0.912
        & \textbf{0.784} / 0.915 \\
        
        $\varsigma=7$   
        & \textbf{0.763} / 0.992 
        & \textbf{0.795} / 0.960
        & \textbf{0.743} / 0.942
        & \textbf{0.719} / 0.932 \\
        \bottomrule
    \end{tabular}
\end{table}

In \autoref{fig:gaussian-gaussian}, we further illustrate the gain provided by our vanilla \acrlong{rb} estimator $\vanillaestimator_\timepoint\integrand$ over the standard estimator $\ergodicaverage_\timepoint\integrand$. To this end, we simulated the Metropolis--Hastings algorithm for $250$ independent realizations up to $\timepoint=1{,}000$ iterations. For both estimators, we report the 90\% interquantile range as well as the full range.

\begin{figure}
    \centering
    \includegraphics[width = .95\linewidth]{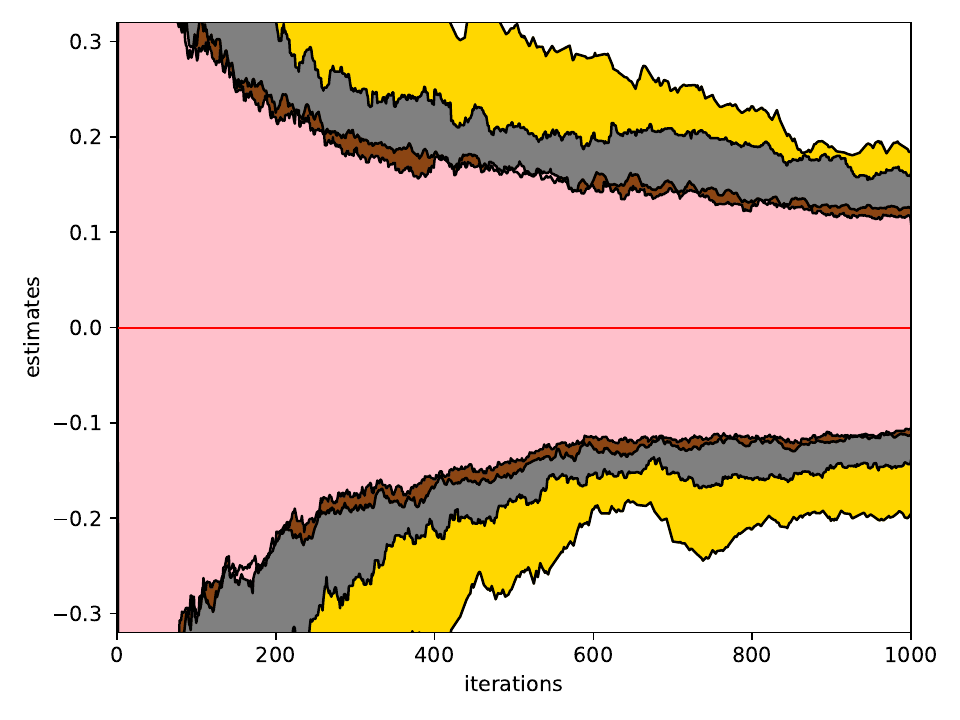}
    \captionsetup{skip = 0pt}
    \caption{
        Overlay of 250 i.i.d. realizations of the standard estimator $\ergodicaverage_\timepoint\integrand$ (gold) and the vanilla \acrlong{rb} estimator $\vanillaestimator_\timepoint\integrand$ (gray) for $\integrand(\point)=\point$, corresponding to estimation of the target mean. Also shown are the 90\% interquantile ranges of the standard estimator (brown) and the vanilla \acrlong{rb} estimator (pink). All results are reported up to $\timepoint=1{,}000$ iterations for the Metropolis--Hastings algorithm with proposal kernel $\proposalkernel(\point,\cdot)=\mathcal N(\point,\sqrt{10})$ and target distribution $\targetdistribution=\mathcal N(0,1)$.
    }
    \label{fig:gaussian-gaussian}
\end{figure}

In \autoref{tab:cauchy-gaussian} and \autoref{fig:cauchy-gaussian}, we repeat this experiment for the proposal kernel $\proposalkernel(\point,\;\cdot\;)=\mathcal C(0,\gamma)$ (a Cauchy distribution with location $0$ and scale $\gamma>0$). In \autoref{tab:exp-exp}, we additionally consider an experiment with proposal kernel $\proposalkernel(\point,\;\cdot\;)=\operatorname{Exp}(\gamma)$ and target distribution $\targetdistribution=\operatorname{Exp}(1)$.

\begin{table}[H]
    \centering
    \captionsetup{skip = 0pt}
    \caption{Ratio of the empirical variances of our vanilla Rao--Blackwellization estimator $\vanillaestimator_\timepoint\integrand$ and the standard estimator $\ergodicaverage_\timepoint\integrand$ of $\integral$ at $\timepoint=100$ for the Metropolis--Hastings algorithm with proposal kernel $\proposalkernel(\point,\;\cdot\;)=\mathcal C(0,\gamma)$ (Cauchy distribution with mean $0$ and scale $\gamma>0$) and initial state drawn from the target distribution $\targetdistribution=\mathcal N(0,1)$. For comparison, the values shown after “/” report the corresponding ratio between the waste-recycling estimator $\wasterecyclingestimator_\timepoint\integrand$ and the standard estimator. The computation was performed using $1{,}000$ independent realizations.}
    \label{tab:cauchy-gaussian}
    \begin{tabular}{lcccc}
        \toprule
        $\integrand(\point)$ 
        & $\point$ 
        & $\point^2$ 
        & $1_{\{\;\point\;>\;0\;\}}$ 
        & $\expectedacceptance(\point)$ \\
        \midrule
        $\gamma=0.25$ 
        & \textbf{0.622} / 0.953
        & \textbf{0.750} / 0.895
        & \textbf{0.688} / 0.920
        & \textbf{0.699} / 0.875 \\
        
        $\gamma=0.5$   
        & \textbf{0.749} / 0.898
        & \textbf{0.613} / 0.715
        & \textbf{0.782} / 0.904
        & \textbf{0.788} / 0.899 \\
        
        $\gamma=1$   
        & \textbf{0.795} / 0.896
        & \textbf{0.673} / 0.774
        & \textbf{0.897} / 0.915
        & \textbf{0.884} / 0.986 \\
        
        $\gamma=2$   
        & \textbf{0.792} / 0.893
        & \textbf{0.657} / 0.758
        & \textbf{0.850} / 0.992
        & \textbf{0.837} / 0.975 \\
        \bottomrule
    \end{tabular}
\end{table}

\begin{figure}
    \centering
    \includegraphics[width = .95\linewidth]{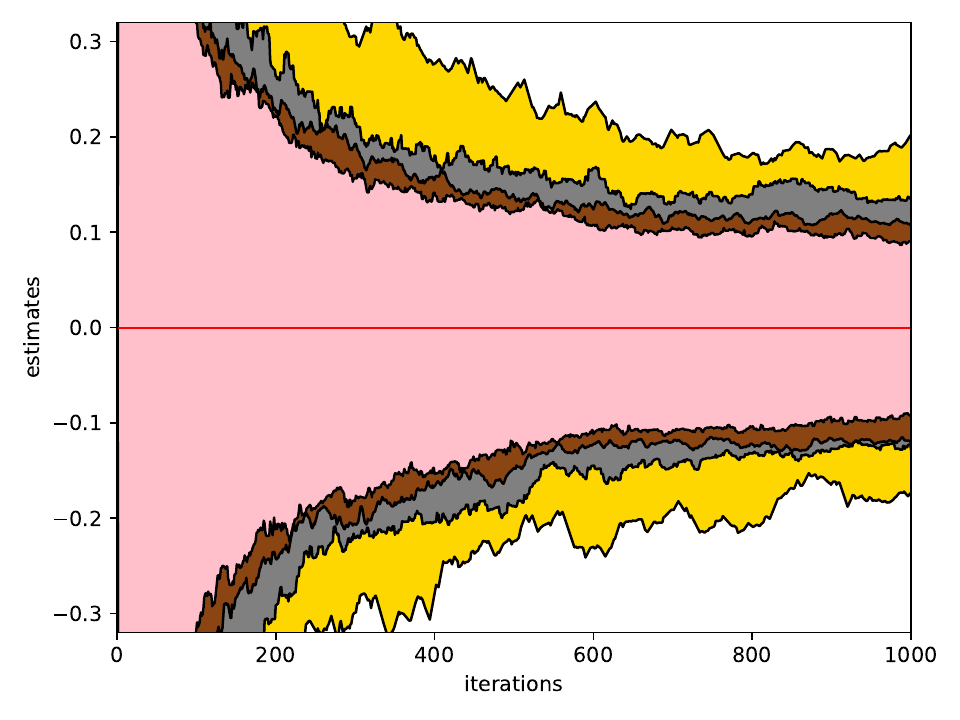}
    \captionsetup{skip = 0pt}
    \caption{
        Overlay of 250 i.i.d. realizations of the standard estimator $\ergodicaverage_\timepoint\integrand$ (gold) and the vanilla \acrlong{rb} estimator $\vanillaestimator_\timepoint\integrand$ (gray) for $\integrand(\point)=\point$, corresponding to estimation of the target mean. Also shown are the 90\% interquantile ranges of the standard estimator (brown) and the vanilla \acrlong{rb} estimator (pink). All results are reported up to $\timepoint=1{,}000$ iterations for the Metropolis--Hastings algorithm with proposal kernel $\proposalkernel(\point,\cdot)=\mathcal C(0,0.25)$ (Cauchy distribution with mean $0$ and scale $0.25$) and target distribution $\targetdistribution=\mathcal N(0,1)$.
    }
    \label{fig:cauchy-gaussian}
\end{figure}

\begin{table}[H]
    \centering
    \captionsetup{skip = 0pt}
    \caption{Ratio of the empirical variances of our vanilla Rao--Blackwellization estimator $\vanillaestimator_\timepoint\integrand$ and the standard estimator $\ergodicaverage_\timepoint\integrand$ of $\integral$ at $\timepoint=100$ for the Metropolis--Hastings algorithm with proposal kernel $\proposalkernel(\point,\;\cdot\;)=\operatorname{Exp}(\gamma)$ and initial state drawn from the target distribution $\targetdistribution=\operatorname{Exp}(1)$. For comparison, the values shown after “/” report the corresponding ratio between the waste-recycling estimator $\wasterecyclingestimator_\timepoint\integrand$ and the standard estimator. The computation was performed using $1{,}000$ independent realizations.}
    \label{tab:exp-exp}
    \begin{tabular}{lcccc}
        \toprule
        $\integrand(\point)$ 
        & $\point$ 
        & $\point^2$ 
        & $1_{\{\;\point\;>\;1\;\}}$ 
        & $\expectedacceptance(\point)$ \\
        \midrule
        $\gamma=0.9$ 
        & \textbf{0.811} / 0.912
        & \textbf{0.731} / 0.833
        & \textbf{0.881} / 0.983
        & \textbf{0.882} / 0.983 \\
        
        $\gamma=.5$   
        & \textbf{0.578} / 0.680
        & \textbf{0.423} / 0.625
        & \textbf{0.694} / 0.795
        & \textbf{0.665} / 0.766 \\
        
        $\gamma=.3$   
        & \textbf{0.674} / 0.775
        & \textbf{0.517} / 0.619
        & \textbf{0.749} / 0.855
        & \textbf{0.727} / 0.829 \\
        
        $\gamma=.1$   
        & \textbf{0.703} / 0.971
        & \textbf{0.748} / 0.895
        & \textbf{0.680} / 0.998
        & \textbf{0.696} / 0.980 \\
        \bottomrule
    \end{tabular}
\end{table}

Across all experiments, the results consistently demonstrate a significant variance reduction when using our vanilla \acrlong{rb} estimator compared to standard estimation, as well as a substantially stronger variance reduction than that achieved by waste recycling relative to standard estimation.

\subsection{Jump Restore algorithm}

We performed analogous experiments, but with the Jump Restore algorithm and the corresponding variants of the standard, waste-recycling and our vanilla \acrlong{rb} estimator as generically described in Algorithm 6.2. For all experiments, we used a state-independent transfer kernel $\regenerationdistribution=\mathcal N(0,2)$.

The results are summarized in \autoref{tab:restore-gaussian-gaussian}, \autoref{fig:restore-gaussian-gaussian}, \autoref{tab:restore-cauchy-gaussian} and \autoref{fig:restore-cauchy-gaussian}. As they reveal, the gain from going over to our vanilla \acrlong{rb} estimator is even more pronounced in the Jump Restore algorithm than it already was in the results of the Metropolis--Hastings algorithm in the preceding section.

\begin{table}[t]
    \centering
    \caption{Ratio of the empirical variances of our vanilla Rao--Blackwellization estimator $\vanillaestimator_\timepoint\integrand$ and the standard estimator $\ergodicaverage_\timepoint\integrand$ of $\integral$ at $\timepoint=\lifetimesum_{20}$ (corresponding to the simulation of 20 tours) for the Jump Restore algorithm with global dynamics $\regenerationdistribution=\mathcal N(0,2)$ and target distribution $\targetdistribution=\mathcal N(0,1)$. The local dynamics were given by the Metropolis--Hastings algorithm with proposal kernel $\proposalkernel(\point,\;\cdot\;)=\mathcal N(\point,\varsigma)$ and target distribution $\targetdistribution$. For comparison, the values shown after “/” report the corresponding ratio between the waste-recycling estimator $\wasterecyclingestimator_\timepoint\integrand$ and the standard estimator. The computation was performed using $1{,}000$ independent realizations.}
    \label{tab:restore-gaussian-gaussian}
    \begin{tabular}{lcccc}
        \toprule
        $\integrand(\point)$ 
        & $\point$ 
        & $\point^2$ 
        & $1_{\{\;\point>0\;\}}$ 
        & $\expectedacceptance(\point)$ \\
        \midrule
        $\varsigma=0.1$ 
        & \textbf{0.898} / 0.999
        & \textbf{0.795} / 0.997
        & \textbf{0.858} / 0.999
        & \textbf{0.935} / 0.992 \\
        
        $\varsigma=2$   
        & \textbf{0.626} / 0.831
        & \textbf{0.740} / 0.999
        & \textbf{0.657} / 0.888
        & \textbf{0.762} / 0.991 \\
        
        $\varsigma=5$   
        & \textbf{0.607} / 0.900
        & \textbf{0.730} / 0.994
        & \textbf{0.630} / 0.930
        & \textbf{0.620} / 0.939 \\
        
        $\varsigma=7$   
        & \textbf{0.645} / 0.897
        & \textbf{0.723} / 0.984
        & \textbf{0.677} / 0.899
        & \textbf{0.633} / 0.947 \\
        \bottomrule
    \end{tabular}
\end{table}

\begin{figure}
    \centering
    \includegraphics[width = .95\linewidth]{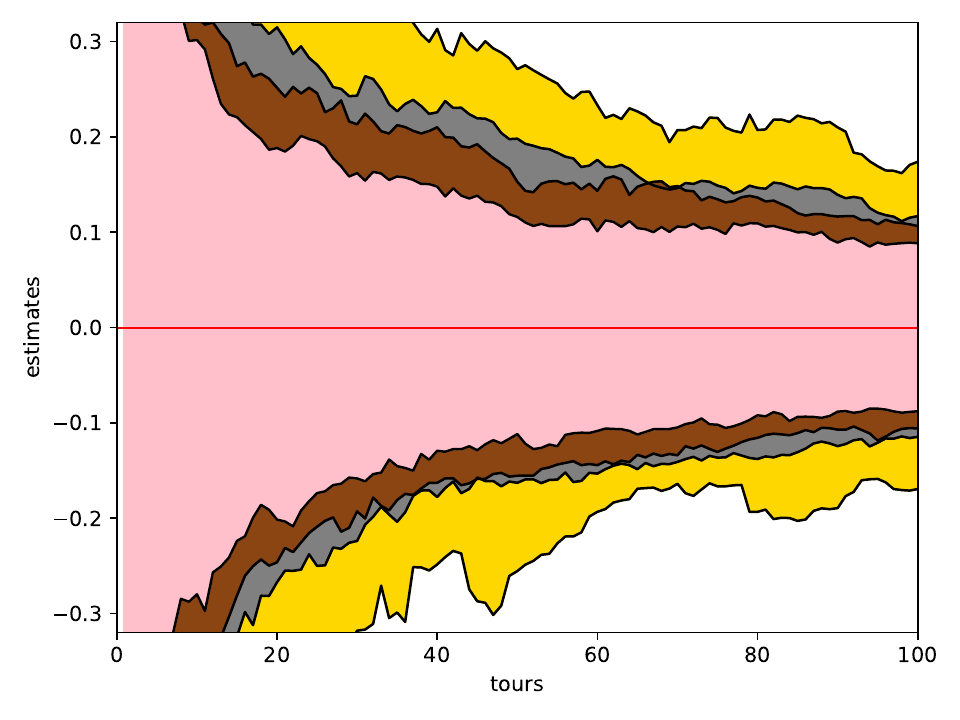}
    \captionsetup{skip = 0pt}
    \caption{
        Overlay of 250 i.i.d. realizations of the standard estimator $\ergodicaverage_\timepoint\integrand$ (gold) and the vanilla \acrlong{rb} estimator $\vanillaestimator_\timepoint\integrand$ (gray) for $\integrand(\point)=\point$, corresponding to estimation of the target mean. Also shown are the 90\% interquantile ranges of the standard estimator (brown) and the vanilla \acrlong{rb} estimator (pink). All results are reported up to $\timepoint=1{,}00$ tours for the Jump Restore algorithm with global dynamics $\regenerationdistribution=\mathcal N(0, 2)$ and target distribution $\targetdistribution=\mathcal N(0,1)$. The local dynamics are given by the Metropolis--Hastings algorithm with proposal kernel $\proposalkernel(\point,\cdot)=\mathcal N(\point,\sqrt{10})$ and target distribution $\targetdistribution$. The constant $\expectedlifetime$ in the definition in Equation (29) of the killing rate was chosen as $\expectedlifetime=0.1$.
    }
    \label{fig:restore-gaussian-gaussian}
\end{figure}

\begin{table}[t]
    \centering
    \caption{Ratio of the empirical variances of our vanilla Rao--Blackwellization estimator $\vanillaestimator_\timepoint\integrand$ and the standard estimator $\ergodicaverage_\timepoint\integrand$ of $\integral$ at $\timepoint=\lifetimesum_{20}$ (corresponding to the simulation of 20 tours) for the Jump Restore algorithm with global dynamics $\regenerationdistribution=\mathcal N(0,2)$ and target distribution $\targetdistribution=\mathcal N(0,1)$. The local dynamics were given by the Metropolis--Hastings algorithm with proposal kernel $\proposalkernel(\point,\;\cdot\;)=\mathcal C(0,\gamma)$ (Cauchy distribution with mean $0$ and scale $\gamma>0$) and target distribution $\targetdistribution$. For comparison, the values shown after “/” report the corresponding ratio between the waste-recycling estimator $\wasterecyclingestimator_\timepoint\integrand$ and the standard estimator. The computation was performed using $1{,}000$ independent realizations.}
    \label{tab:restore-cauchy-gaussian}
    \begin{tabular}{lcccc}
        \toprule
        $\integrand(\point)$ 
        & $\point$ 
        & $\point^2$ 
        & $1_{\{\;\point>0\;\}}$ 
        & $\expectedacceptance(\point)$ \\
        \midrule
        $\gamma=0.25$ 
        & \textbf{0.500} / 0.959
        & \textbf{0.523} / 0.992
        & \textbf{0.470} / 0.953
        & \textbf{0.802} / 0.969 \\
        
        $\gamma=0.5$   
        & \textbf{0.457} / 0.883
        & \textbf{0.618} / 0.991
        & \textbf{0.502} / 0.949
        & \textbf{0.864} / 0.989 \\
        
        $\gamma=1$   
        & \textbf{0.487} / 0.802
        & \textbf{0.702} / 0.999
        & \textbf{0.606} / 0.935
        & \textbf{0.819} / 0.997 \\
        
        $\gamma=2$   
        & \textbf{0.536} / 0.806
        & \textbf{0.664} / 0.993
        & \textbf{0.629} / 0.934
        & \textbf{0.706} / 0.991254 \\
        \bottomrule
    \end{tabular}
\end{table}

\begin{figure}
    \centering
    \includegraphics[width = .95\linewidth]{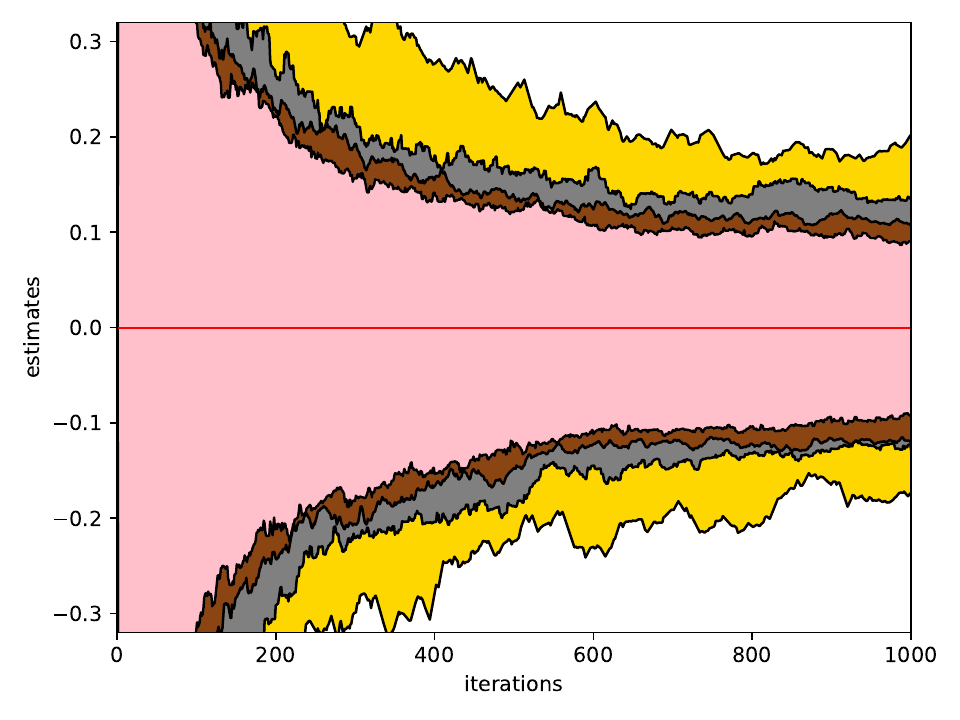}
    \captionsetup{skip = 0pt}
    \caption{
        Overlay of 250 i.i.d. realizations of the standard estimator $\ergodicaverage_\timepoint\integrand$ (gold) and the vanilla \acrlong{rb} estimator $\vanillaestimator_\timepoint\integrand$ (gray) for $\integrand(\point)=\point$, corresponding to estimation of the target mean. Also shown are the 90\% interquantile ranges of the standard estimator (brown) and the vanilla \acrlong{rb} estimator (pink). All results are reported up to $\timepoint=1{,}00$ tours for the Jump Restore algorithm with global dynamics $\regenerationdistribution=\mathcal N(0, 2)$ and target distribution $\targetdistribution=\mathcal C(0,0.25)$ (Cauchy distribution with mean $0$ and scale $0.25$). The local dynamics are given by the Metropolis--Hastings algorithm with proposal kernel $\proposalkernel(\point,\cdot)=\mathcal N(\point,\sqrt{10})$ and target distribution $\targetdistribution$. The constant $\expectedlifetime$ in the definition in Equation (29) of the killing rate was chosen as $\expectedlifetime=0.1$.
    }
    \label{fig:restore-cauchy-gaussian}
\end{figure}

\newpage
\section{Index of notation}

\input{nomenclature}

\end{document}

%% file: packages.tex
\usepackage[section]{algorithm}
\usepackage[noend]{algpseudocode}
\usepackage{amssymb, amsthm}
\usepackage{bm}
\usepackage{caption}
\usepackage{contour}
\usepackage[shortlabels]{enumitem}
\usepackage[acronym]{glossaries}
\usepackage{hyperref}
\usepackage{natbib}
\usepackage{pgfplots}
\usepackage{physics}
\usepackage{subcaption}
\usepackage{tabularx}
\usepackage{tikz}
\usepackage{tikzfill}
\usepackage{xspace}

 \usepackage{booktabs,multirow,graphicx}

\usepackage[nameinlink,noabbrev]{cleveref}
\usepackage{needspace}

\usepackage{soul}

%% file: definitions.tex
\newacronym{mc}{MC}{Monte Carlo}
\newacronym{mcmc}{MCMC}{Markov chain Monte Carlo}

\newacronym{mh}{MH}{Metropolis-Hastings}
\newacronym{mala}{MALA}{Metropolis-adjusted Langevin algorithm}
\newacronym{hmc}{H²MC}{Hamiltonian Monte Carlo}

\newacronym{mmlt}{MMLT}{Multiplexed Metropolis Light Transport}

\newacronym{jrlt}{JRLT}{Jump Restore Light Transport}
\newacronym{rb}{RB}{Rao--Blackwellization}

\newacronym{mse}{MSE}{mean squared error}
\newacronym{mrse}{MRSE}{mean relative squared error}
\newacronym{mape}{MAPE}{mean absolute percentage error}

\newacronym{bdpt}{BDPT}{bidirectional path tracing}
\newacronym{spp}{SPP}{samples per pixel}

\glsdisablehyper

\let\originalleft\left
\let\originalright\right
\renewcommand{\left}{\mathopen{}\mathclose\bgroup\originalleft}
\renewcommand{\right}{\aftergroup\egroup\originalright}

\newcommand\declaresymbol[2]{\newcommand{#1}{\TextOrMath{$#2$\xspace}{#2}}}

\declaresymbol\vanillaproduct P
\declaresymbol\vanillaerror E
\declaresymbol\covariance\Sigma

\declaresymbol\measurablespace E
\declaresymbol\measurablesystem{\mathcal E}

\declaresymbol\process X
\declaresymbol\secondprocess Y
\declaresymbol\thirdprocess Z

\declaresymbol\markovchain M
\declaresymbol\transitionkernel\kappa

\declaresymbol\uniformprocess U

\declaresymbol\pointset B

\declaresymbol\probabilityspace\Omega
\declaresymbol\probabilitysystem{\mathcal A}
\declaresymbol\probabilitymeasure{\operatorname P}
\declaresymbol\expectation{\operatorname E}
\declaresymbol\variance{\operatorname{Var}}
\declaresymbol\filtration{\mathcal F}
\declaresymbol\secondfiltration{\mathcal G}

\declaresymbol\targetdistribution\pi
\declaresymbol\referencemeasure\lambda
\declaresymbol\targetdensity p

\declaresymbol\regenerationdistribution\mu
\declaresymbol\regenerationdensity{\MakeLowercase\uniformprocess}

\declaresymbol\sequenceindex i
\declaresymbol\secondsequenceindex j
\declaresymbol\thirdsequenceindex m
\declaresymbol\lastsequenceindex n

\declaresymbol\lifetimesum\sigma
\declaresymbol\acceptancetime\lifetimesum
\declaresymbol\lifetime\tau

\declaresymbol\integral I
\declaresymbol\integrand f
\declaresymbol\secondintegrand g
\declaresymbol\thirdintegrand h

\declaresymbol\timedomain T
\declaresymbol\timepoint t
\declaresymbol\prevtimepoint s

\declaresymbol\point x
\declaresymbol\secondpoint y
\declaresymbol\uniform u

\declaresymbol\ergodicaverage A
\declaresymbol\vanillaestimator V
\declaresymbol\wasterecyclingestimator W

\declaresymbol\vanillalifetime{\tilde\lifetime}
\declaresymbol\ourvanillalifetime{\hat\lifetime}
\declaresymbol\vanillaweight\rho

\declaresymbol\acceptancefunction\alpha
\declaresymbol\proposalkernel Q
\declaresymbol\proposaldensity q
\declaresymbol\acceptedchain\zeta
\declaresymbol\expectedacceptance\varrho

\declaresymbol\concatenatedprocess X
\declaresymbol\localprocess Y
\declaresymbol\expectedlifetime c
\declaresymbol\localgenerator L
\declaresymbol\genericoperator T

\declaresymbol\largestepprobability\ell
\declaresymbol\smallstepkernel\zeta
\declaresymbol\holdingrate h
\declaresymbol\killingrate k
\declaresymbol\combinedrate c

\declaresymbol\generator L
\declaresymbol\expvariable\xi

\declaresymbol\outcome\omega

\newcommand\eqfor{\quad\text{for }}
\newcommand\eqforall{\quad\text{for all }}

\makeatletter
\newcommand{\setword}[2]{%
    \phantomsection
    #1\def\@currentlabel{\unexpanded{#1}}\label{#2}%
}
\makeatother

\makeatletter
\patchcmd{\ALG@step}{\addtocounter{ALG@line}{1}}{\refstepcounter{ALG@line}}{}{}
\newcommand{\ALG@lineautorefname}{Line}
\makeatother

\theoremstyle{definition}
\newtheorem{theorem}{Theorem}[section]
\newtheorem{lemma}{Lemma}[section]
\newtheorem{definition}{Definition}[section]

\definecolor{DarkGreen}{rgb}{0, .6, 0}
\newcommand{\AlgCommentTemplate}[2]{\hfill{\fontsize{7}{6}\selectfont\textcolor{DarkGreen}{\text{#1\;#2}}}}

\newcommand{\AlgCommentLeft}[1]{\AlgCommentTemplate{$\leftarrow$}{#1}}

\algdef{SE}[SUBALG]{Indent}{EndIndent}{}{\algorithmicend\ }%
\algtext*{Indent}
\algtext*{EndIndent}

\algrenewcommand\algorithmicrequire{\textbf{Input:}}
\algrenewcommand\algorithmicensure{\textbf{Output:}}
\algnewcommand\algorithmicforeach{\textbf{for each}}
\algdef{S}[FOR]{ForEach}[1]{\algorithmicforeach\ #1\ \algorithmicdo}

\renewcommand\algorithmicdo{}

%% file: figures/teaser.tex
%!TEX root = ../main.tex

% #1 = x position
% #2 = y position
% #3 = anchor
% #3 = first line
% #4 = second line
\newcommand{\MethodLabelAt}[5]{%
  \node[anchor = #3] at (#1, #2) {%
    \shortstack[l]{%
      \contour{black}{\small\textcolor{white}{#4}}\\[-0.2ex]%
      \contour{black}{\small\textcolor{white}{#5}}%
    }%
  };%
}

% Two-line tiny label for tables (left-aligned)
\newcommand{\TableMethodLabel}[2]{%
  \begin{tabular}{@{}l@{}}%
    \tiny #1\\[-0.1ex]%
    \tiny #2%
  \end{tabular}%
}

% \pgfmathsetmacro\width{1024}
% \pgfmathsetmacro\height{576}
\pgfmathsetmacro\width{1280}
\pgfmathsetmacro\height{720}
\pgfmathsetmacro\aspectratio{16 / 9}

\newcommand\scene{veach_ajar}
\newcommand\firstmethod{Metropolis}
\newcommand\secondmethod{Metropolis\_Vanilla\_Ours}
\newcommand\thirdmethod{Metropolis\_Waste-Recycling}
\newcommand\firstmethoddisplaynamea{Metropolis}
\newcommand\firstmethoddisplaynameb{\vphantom|}
\newcommand\secondmethoddisplaynamea{Metropolis}
\newcommand\secondmethoddisplaynameb{Vanilla (Ours)}
\newcommand\thirdmethoddisplaynamea{Metropolis}
\newcommand\thirdmethoddisplaynameb{Waste-Recycling}

\newcommand\fourthmethod{Metropolis\_Restore}
\newcommand\fifthmethod{Metropolis\_Restore\_Vanilla\_Ours}
\newcommand\sixthmethod{Metropolis\_Restore\_Waste-Recycling}
\newcommand\fourthmethoddisplaynamea{Metropolis Restore}
\newcommand\fourthmethoddisplaynameb{\vphantom|}
\newcommand\fifthmethoddisplaynamea{Metropolis Restore}
\newcommand\fifthmethoddisplaynameb{Vanilla (Ours)}
\newcommand\sixthmethoddisplaynamea{Metropolis Restore}
\newcommand\sixthmethoddisplaynameb{Waste-Recycling}

\pgfmathsetmacro\croplength{.12}
\pgfmathsetmacro\firstcropxmin{.5}
\pgfmathsetmacro\firstcropymin{.25}
\pgfmathsetmacro\secondcropxmin{.75}
\pgfmathsetmacro\secondcropymin{.35}
% \pgfmathsetmacro\firstcropxmin{.45}
% \pgfmathsetmacro\firstcropymin{.0175}
% \pgfmathsetmacro\secondcropxmin{.7}
% \pgfmathsetmacro\secondcropymin{.15}
% \pgfmathsetmacro\firstcropxmin{.41}
% \pgfmathsetmacro\firstcropymin{.03}
% \pgfmathsetmacro\secondcropxmin{.75}
% \pgfmathsetmacro\secondcropymin{.35}
\newcommand\firstcropcolor{red}
\newcommand\secondcropcolor{blue}
\pgfmathsetmacro\thirdcropxmin{.5}
\pgfmathsetmacro\thirdcropymin{.25}
\pgfmathsetmacro\fourthcropxmin{.75}
\pgfmathsetmacro\fourthcropymin{.35}
% \pgfmathsetmacro\thirdcropxmin{.45}
% \pgfmathsetmacro\thirdcropymin{.0175}
% \pgfmathsetmacro\fourthcropxmin{.7}
% \pgfmathsetmacro\fourthcropymin{.15}
% \pgfmathsetmacro\thirdcropxmin{.41}
% \pgfmathsetmacro\thirdcropymin{.03}
% \pgfmathsetmacro\fourthcropxmin{.75}
% \pgfmathsetmacro\fourthcropymin{.35}
\newcommand\thirdcropcolor{red}
\newcommand\fourthcropcolor{blue}

\newcommand{\SplitLeftTwoImages}[2]{
    \begin{scope}
       \clip (0, 0) -- (.7 * \aspectratio, 0) -- (.3 * \aspectratio, 1) -- (0, 1) -- cycle;
        \path[fill overzoom image = results/#1.jpg] (0, 0) rectangle ( \aspectratio, 1);
    \end{scope}
   \begin{scope}
       \clip (.7 * \aspectratio, 0) -- (\aspectratio, 0) -- (\aspectratio, 1) -- (0.3 * \aspectratio, 1) -- cycle;
       \path[fill overzoom image = results/#2.jpg] (0,0) rectangle (\aspectratio, 1);
   \end{scope}
   \begin{scope}
        \draw[black, thick] (0, 0) rectangle (\aspectratio, 1);
        \draw[draw=black,thick] (.3 * \aspectratio, 1) -- (.7 * \aspectratio, 0);
   \end{scope}
}

\newcommand{\SplitRightTwoImages}[2]{
    \begin{scope}
       \clip (0, 0) -- (.3 * \aspectratio, 0) -- (.7 * \aspectratio, 1) -- (0, 1) -- cycle;
        \path[fill overzoom image = results/#1.jpg] (0, 0) rectangle ( \aspectratio, 1);
    \end{scope}
   \begin{scope}
       \clip (.3 * \aspectratio, 0) -- (\aspectratio, 0) -- (\aspectratio, 1) -- (0.7 * \aspectratio, 1) -- cycle;
       \path[fill overzoom image = results/#2.jpg] (0,0) rectangle (\aspectratio, 1);
   \end{scope}
   \begin{scope}
        \draw[black, thick] (0, 0) rectangle (\aspectratio, 1);
        \draw[draw=black,thick] (.3 * \aspectratio, 0) -- (.7 * \aspectratio, 1);
   \end{scope}
}

\newcommand{\SplitThreeImages}[3]{
    \begin{scope}
       \clip (0, 0) -- (.2 * \aspectratio, 0) -- (.3 * \aspectratio, 1) -- (0, 1) -- cycle;
        \path[fill overzoom image = results/#1.jpg] (0, 0) rectangle (\aspectratio, 1);
    \end{scope}
    \begin{scope}
       \clip (.2 * \aspectratio, 0) -- (0.6 * \aspectratio, 0) -- (0.7 * \aspectratio, 1) -- (0.3 * \aspectratio, 1) -- cycle;
        \path[fill overzoom image = results/#2.jpg] (0, 0) rectangle (\aspectratio, 1);
    \end{scope}
   \begin{scope}
       \clip (.6 * \aspectratio, 0) -- (\aspectratio, 0) -- (\aspectratio, 1) -- (0.7 * \aspectratio, 1) -- cycle;
       \path[fill overzoom image = results/#3.jpg] (0,0) rectangle (\aspectratio, 1);
   \end{scope}
   \begin{scope}
        \draw[black, thick] (0, 0) rectangle (\aspectratio, 1);
        \draw[draw=black,thick] (.2 * \aspectratio, 0) -- (.3 * \aspectratio, 1);
        \draw[draw=black,thick] (.6 * \aspectratio, 0) -- (.7 * \aspectratio, 1);
   \end{scope}
}

%   - #1: scene name (without extension; image is results/#1.jpg)
%   - #2: crop x (normalized, 0–1; 0=left)
%   - #3: crop y (normalized, 0–1; 0=bottom; note: image origin is lower left)
%   - #4: crop length (normalized)
%   - #5: color (for drawing the outline)
\newcommand{\CropImage}[5]{%
    \begingroup
        % Convert normalized crop to pixel values
        \pgfmathsetmacro\length{#4 * \width}
        \pgfmathsetmacro\xminval{#2 * \width}
        \pgfmathsetmacro\xmaxval{#2 * \width + \length}
        \pgfmathsetmacro\yminval{(1 - #3) * \height - \length}
        \pgfmathsetmacro\ymaxval{(1 - #3) * \height}

        \begin{scope}
            % Use fill with a rectangle, not a node
            \path[fill overzoom image = results/#1.jpg] (0, 0) rectangle (1, 1);
            % Now clip to viewport
            \clip (0, 0) rectangle (1, 1);
            \begin{scope}[xscale = 1, yscale = 1]
                \pgfmathsetmacro{\trimleft}{\xminval}
                \pgfmathsetmacro{\trimbottom}{\height - \ymaxval}  % Flip y-axis because trim starts from top-left
                \pgfmathsetmacro{\trimright}{\width - \xmaxval}
                \pgfmathsetmacro{\trimtop}{\yminval}
                
                \node at (0.5,0.5) {%
                    \includegraphics[
                        width = 2.15cm, height = 2.15cm,
                        trim = {\trimleft bp} {\trimbottom bp} {\trimright bp} {\trimtop bp},
                        clip
                    ]{results/#1.jpg}%
                };
            \end{scope}
        \end{scope}

        \begin{scope}
            \draw[#5, thick] (0, 0) rectangle (1, 1);
        \end{scope}
    \endgroup
}

\def\arraystretch{.5}
\setlength\tabcolsep{1pt}
\begin{tabularx}{\linewidth}{cccccccc}
\multicolumn{4}{c}{
\begin{tikzpicture}[scale = 5]
    \SplitThreeImages{\scene/\firstmethod}{\scene/\secondmethod}{\scene/\thirdmethod}
    \begin{scope}
        \draw[\firstcropcolor, ultra thick]
        (\aspectratio * \firstcropxmin, \firstcropymin)
        rectangle
        (\aspectratio * \firstcropxmin + \aspectratio * \croplength, \firstcropymin + \aspectratio * \croplength);
        
        \draw[\secondcropcolor, ultra thick]
        (\aspectratio * \secondcropxmin, \secondcropymin)
        rectangle
        (\aspectratio * \secondcropxmin + \aspectratio * \croplength, \secondcropymin + \aspectratio * \croplength);
    \end{scope}
    \begin{scope}
        \contourlength{.1em}
        \MethodLabelAt{.025}{.9}{west}{\firstmethoddisplaynamea}{\firstmethoddisplaynameb}
        \MethodLabelAt{.6}{.9}{west}{\secondmethoddisplaynamea}{\secondmethoddisplaynameb}
        \MethodLabelAt{1.26}{.9}{west}{\thirdmethoddisplaynamea}{\thirdmethoddisplaynameb}
    \end{scope}
\end{tikzpicture}
}
&
\multicolumn{4}{c}{
\begin{tikzpicture}[scale = 5]
    \SplitThreeImages{\scene/\fourthmethod}{\scene/\fifthmethod}{\scene/\sixthmethod}
    \begin{scope}
        \draw[\firstcropcolor, ultra thick]
        (\aspectratio * \firstcropxmin, \firstcropymin)
        rectangle
        (\aspectratio * \firstcropxmin + \aspectratio * \croplength, \firstcropymin + \aspectratio * \croplength);
        
        \draw[\secondcropcolor, ultra thick]
        (\aspectratio * \secondcropxmin, \secondcropymin)
        rectangle
        (\aspectratio * \secondcropxmin + \aspectratio * \croplength, \secondcropymin + \aspectratio * \croplength);
    \end{scope}
    \begin{scope}
        \contourlength{.1em}
        \MethodLabelAt{.00005}{.9}{west}{\fourthmethoddisplaynamea}{\fourthmethoddisplaynameb}
        \MethodLabelAt{.6}{.9}{west}{\fifthmethoddisplaynamea}{\fifthmethoddisplaynameb}
        \MethodLabelAt{1.23}{.9}{west}{\sixthmethoddisplaynamea}{\sixthmethoddisplaynameb}
    \end{scope}
\end{tikzpicture}
}
\\
\begin{tikzpicture}[scale = 2.15]
    \CropImage{\scene/\firstmethod-MAPE}\firstcropxmin\firstcropymin\croplength{\firstcropcolor}
\end{tikzpicture}
&
\begin{tikzpicture}[scale = 2.15]
    \CropImage{\scene/\secondmethod-MAPE}\firstcropxmin\firstcropymin\croplength{\firstcropcolor}
\end{tikzpicture}
&
\begin{tikzpicture}[scale = 2.15]
    \CropImage{\scene/\thirdmethod-MAPE}\secondcropxmin\secondcropymin\croplength{\secondcropcolor}
\end{tikzpicture}
&
\begin{tikzpicture}[scale = 2.15]
    \CropImage{\scene/\secondmethod-MAPE}\secondcropxmin\secondcropymin\croplength{\secondcropcolor}
\end{tikzpicture}
&
\begin{tikzpicture}[scale = 2.15]
    \CropImage{\scene/\fourthmethod-MAPE}\thirdcropxmin\thirdcropymin\croplength{\firstcropcolor}
\end{tikzpicture}
&
\begin{tikzpicture}[scale = 2.15]
    \CropImage{\scene/\fifthmethod-MAPE}\thirdcropxmin\thirdcropymin\croplength{\thirdcropcolor}
\end{tikzpicture}
&
\begin{tikzpicture}[scale = 2.15]
    \CropImage{\scene/\sixthmethod-MAPE}\fourthcropxmin\fourthcropymin\croplength{\fourthcropcolor}
\end{tikzpicture}
&
\begin{tikzpicture}[scale = 2.15]
    \CropImage{\scene/\fifthmethod-MAPE}\fourthcropxmin\fourthcropymin\croplength{\fourthcropcolor}
\end{tikzpicture}
\\[-.25mm]
\TableMethodLabel{\firstmethoddisplaynamea}{\firstmethoddisplaynameb} & \TableMethodLabel{\secondmethoddisplaynamea}{\secondmethoddisplaynameb} & \TableMethodLabel{\thirdmethoddisplaynamea}{\thirdmethoddisplaynameb} & \TableMethodLabel{\secondmethoddisplaynamea}{\secondmethoddisplaynameb} & \TableMethodLabel{\fourthmethoddisplaynamea}{\fourthmethoddisplaynameb} & \TableMethodLabel{\fifthmethoddisplaynamea}{\fifthmethoddisplaynameb} & \TableMethodLabel{\sixthmethoddisplaynamea}{\sixthmethoddisplaynameb} & \TableMethodLabel{\fifthmethoddisplaynamea}{\fifthmethoddisplaynameb}
\end{tabularx}

%% file: figures/pool/mala_restore_new_layout.tex
%!TEX root = ../main.tex

% Rotated two-line method label (for right margin of table)
% #1 first line, #2 second line
\newcommand{\RotTableMethodLabel}[2]{%
    \smash{%
        \raisebox{0.5\height}{% optisch zentrieren; bei Bedarf feinjustieren
            \rotatebox[origin = c]{90}{%
                \tiny\begin{tabular}{@{}l@{}}
                    \hspace{.6cm}#1\\[.1ex]
                    \hspace{.6cm}#2
                \end{tabular}%
            }%
        }%
    }%
}

% #1 = x position
% #2 = y position
% #3 = anchor
% #3 = first line
% #4 = second line
\newcommand{\MethodLabelAt}[5]{%
  \node[anchor = #3] at (#1, #2) {%
    \shortstack[l]{%
      \contour{black}{\small\textcolor{white}{#4}}\\[-0.2ex]%
      \contour{black}{\small\textcolor{white}{#5}}%
    }%
  };%
}

% Two-line tiny label for tables (left-aligned)
\newcommand{\TableMethodLabel}[2]{%
  \begin{tabular}{@{}l@{}}%
    \tiny #1\\[-0.1ex]%
    \tiny #2%
  \end{tabular}%
}

% \pgfmathsetmacro\width{1024}
% \pgfmathsetmacro\height{768}
\pgfmathsetmacro\width{834}
\pgfmathsetmacro\height{469}
\pgfmathsetmacro\aspectratio{16 / 9}

\newcommand\scene{pool}
\newcommand\firstmethod{MALA\_Restore}
\newcommand\secondmethod{MALA\_Restore\_Vanilla\_Ours}
\newcommand\thirdmethod{MALA\_Restore\_Waste-Recycling}
\newcommand\firstmethoddisplaynamea{MALA Restore}
\newcommand\firstmethoddisplaynameb{\vphantom|}
\newcommand\secondmethoddisplaynamea{MALA Restore}
\newcommand\secondmethoddisplaynameb{Vanilla (Ours)}
\newcommand\thirdmethoddisplaynamea{MALA Restore}
\newcommand\thirdmethoddisplaynameb{Waste-Recycling}

\pgfmathsetmacro\croplength{.12}
\pgfmathsetmacro\firstcropxmin{.5}
\pgfmathsetmacro\firstcropymin{.2}
\pgfmathsetmacro\secondcropxmin{.85}
\pgfmathsetmacro\secondcropymin{.35}
\newcommand\firstcropcolor{red}
\newcommand\secondcropcolor{blue}

\newcommand{\SplitThreeImages}[3]{
    \begin{scope}
       \clip (0, 0) -- (.2 * \aspectratio, 0) -- (.3 * \aspectratio, 1) -- (0, 1) -- cycle;
        \path[fill overzoom image = results/#1.jpg] (0, 0) rectangle (\aspectratio, 1);
    \end{scope}
    \begin{scope}
       \clip (.2 * \aspectratio, 0) -- (0.6 * \aspectratio, 0) -- (0.7 * \aspectratio, 1) -- (0.3 * \aspectratio, 1) -- cycle;
        \path[fill overzoom image = results/#2.jpg] (0, 0) rectangle (\aspectratio, 1);
    \end{scope}
   \begin{scope}
       \clip (.6 * \aspectratio, 0) -- (\aspectratio, 0) -- (\aspectratio, 1) -- (0.7 * \aspectratio, 1) -- cycle;
       \path[fill overzoom image = results/#3.jpg] (0,0) rectangle (\aspectratio, 1);
   \end{scope}
   \begin{scope}
        \draw[black, thick] (0, 0) rectangle (\aspectratio, 1);
        \draw[draw=black,thick] (.2 * \aspectratio, 0) -- (.3 * \aspectratio, 1);
        \draw[draw=black,thick] (.6 * \aspectratio, 0) -- (.7 * \aspectratio, 1);
   \end{scope}
}

%   - #1: scene name (without extension; image is results/#1.jpg)
%   - #2: crop x (normalized, 0–1; 0=left)
%   - #3: crop y (normalized, 0–1; 0=bottom; note: image origin is lower left)
%   - #4: crop length (normalized)
%   - #5: color (for drawing the outline)
\newcommand{\CropImage}[5]{%
    \begingroup
        % Convert normalized crop to pixel values
        \pgfmathsetmacro\length{#4 * \width}
        \pgfmathsetmacro\xminval{#2 * \width}
        \pgfmathsetmacro\xmaxval{#2 * \width + \length}
        \pgfmathsetmacro\yminval{(1 - #3) * \height - \length}
        \pgfmathsetmacro\ymaxval{(1 - #3) * \height}

        \begin{scope}
            % Use fill with a rectangle, not a node
            \path[fill overzoom image = results/#1.jpg] (0, 0) rectangle (1, 1);
            % Now clip to viewport
            \clip (0, 0) rectangle (1, 1);
            \begin{scope}[xscale = 1, yscale = 1]
                \pgfmathsetmacro{\trimleft}{\xminval}
                \pgfmathsetmacro{\trimbottom}{\height - \ymaxval}  % Flip y-axis because trim starts from top-left
                \pgfmathsetmacro{\trimright}{\width - \xmaxval}
                \pgfmathsetmacro{\trimtop}{\yminval}
                
                \node at (0.5,0.5) {%
                    \includegraphics[
                        width = 1.78cm, height = 1.78cm,
                        trim = {\trimleft bp} {\trimbottom bp} {\trimright bp} {\trimtop bp},
                        clip
                    ]{results/#1.jpg}%
                };
            \end{scope}
        \end{scope}

        \begin{scope}
            \draw[#5, thick] (0, 0) rectangle (1, 1);
        \end{scope}
    \endgroup
}

\def\arraystretch{.5}
\setlength\tabcolsep{1pt}

\begin{tabularx}{\linewidth}{@{}cc@{}}
    \begin{minipage}[t]{.48\linewidth}
    \vspace{0pt}%
    \begin{tikzpicture}[scale = 5.5]
        \SplitThreeImages{\scene/\firstmethod}{\scene/\secondmethod}{\scene/\thirdmethod}
        \begin{scope}
            \draw[\firstcropcolor, ultra thick]
            (\aspectratio * \firstcropxmin, \firstcropymin)
            rectangle
            (\aspectratio * \firstcropxmin + \aspectratio * \croplength, \firstcropymin + \aspectratio * \croplength);
            
            \draw[\secondcropcolor, ultra thick]
            (\aspectratio * \secondcropxmin, \secondcropymin)
            rectangle
            (\aspectratio * \secondcropxmin + \aspectratio * \croplength, \secondcropymin + \aspectratio * \croplength);
        \end{scope}
        \begin{scope}
            \contourlength{.1em}
            \MethodLabelAt{.025}{.9}{west}{\firstmethoddisplaynamea}{\firstmethoddisplaynameb}
            \MethodLabelAt{.6}{.9}{west}{\secondmethoddisplaynamea}{\secondmethoddisplaynameb}
            \MethodLabelAt{1.26}{.9}{west}{\thirdmethoddisplaynamea}{\thirdmethoddisplaynameb}
        \end{scope}
    \end{tikzpicture}
    \end{minipage}
    &
    \hspace{1.25cm}
    \begin{minipage}[t]{.48\linewidth}
    \vspace{0pt}%
    \begin{tabular}{@{}ccccc@{}}
        % row 1
        \begin{tikzpicture}[scale = 1.78]
            \CropImage{\scene/\firstmethod}\firstcropxmin\firstcropymin\croplength{\firstcropcolor}
        \end{tikzpicture}
        &
        \begin{tikzpicture}[scale = 1.78]
            \CropImage{\scene/\firstmethod-MAPE}\firstcropxmin\firstcropymin\croplength{\firstcropcolor}
        \end{tikzpicture}
        &
        \begin{tikzpicture}[scale = 1.78]
            \CropImage{\scene/\firstmethod}\secondcropxmin\secondcropymin\croplength{\secondcropcolor}
        \end{tikzpicture}
        &
        \begin{tikzpicture}[scale = 1.78]
            \CropImage{\scene/\firstmethod-MAPE}\secondcropxmin\secondcropymin\croplength{\secondcropcolor}
        \end{tikzpicture}
        &
        \RotTableMethodLabel{\firstmethoddisplaynamea}{\firstmethoddisplaynameb}
        \\
        % row 2
        \begin{tikzpicture}[scale = 1.78]
            \CropImage{\scene/\secondmethod}\firstcropxmin\firstcropymin\croplength{\firstcropcolor}
        \end{tikzpicture}
        &
        \begin{tikzpicture}[scale = 1.78]
          \CropImage{\scene/\secondmethod-MAPE}\firstcropxmin\firstcropymin\croplength{\firstcropcolor}
        \end{tikzpicture}
        &
        \begin{tikzpicture}[scale = 1.78]
            \CropImage{\scene/\secondmethod}\secondcropxmin\secondcropymin\croplength{\secondcropcolor}
        \end{tikzpicture}
        &
        \begin{tikzpicture}[scale = 1.78]
            \CropImage{\scene/\secondmethod-MAPE}\secondcropxmin\secondcropymin\croplength{\secondcropcolor}
        \end{tikzpicture}
        &
        \RotTableMethodLabel{\secondmethoddisplaynamea}{\secondmethoddisplaynameb}
        \\
        % row 3
        \begin{tikzpicture}[scale = 1.78]
            \CropImage{\scene/\thirdmethod}\firstcropxmin\firstcropymin\croplength{\firstcropcolor}
        \end{tikzpicture}
        &
        \begin{tikzpicture}[scale = 1.78]
            \CropImage{\scene/\thirdmethod-MAPE}\firstcropxmin\firstcropymin\croplength{\firstcropcolor}
        \end{tikzpicture}
        &
        \begin{tikzpicture}[scale = 1.78]
            \CropImage{\scene/\thirdmethod}\secondcropxmin\secondcropymin\croplength{\secondcropcolor}
        \end{tikzpicture}
        &
        \begin{tikzpicture}[scale = 1.78]
            \CropImage{\scene/\thirdmethod-MAPE}\secondcropxmin\secondcropymin\croplength{\secondcropcolor}
        \end{tikzpicture}
        &
        \RotTableMethodLabel{\thirdmethoddisplaynamea}{\thirdmethoddisplaynameb}
    \end{tabular}
    \end{minipage}
\end{tabularx}

%% file: figures/torus/hmc_new_layout.tex
%!TEX root = ../main.tex

% Rotated two-line method label (for right margin of table)
% #1 first line, #2 second line
\newcommand{\RotTableMethodLabel}[2]{%
    \smash{%
        \raisebox{0.5\height}{% optisch zentrieren; bei Bedarf feinjustieren
            \rotatebox[origin = c]{90}{%
                \tiny\begin{tabular}{@{}l@{}}
                    \hspace{.6cm}#1\\[.1ex]
                    \hspace{.6cm}#2
                \end{tabular}%
            }%
        }%
    }%
}

% #1 = x position
% #2 = y position
% #3 = anchor
% #3 = first line
% #4 = second line
\newcommand{\MethodLabelAt}[5]{%
  \node[anchor = #3] at (#1, #2) {%
    \shortstack[l]{%
      \contour{black}{\small\textcolor{white}{#4}}\\[-0.2ex]%
      \contour{black}{\small\textcolor{white}{#5}}%
    }%
  };%
}

% Two-line tiny label for tables (left-aligned)
\newcommand{\TableMethodLabel}[2]{%
  \begin{tabular}{@{}l@{}}%
    \tiny #1\\[-0.1ex]%
    \tiny #2%
  \end{tabular}%
}

% \pgfmathsetmacro\width{1024}
% \pgfmathsetmacro\height{768}
\pgfmathsetmacro\width{685}
\pgfmathsetmacro\height{514}
\pgfmathsetmacro\aspectratio{16 / 9}

\newcommand\scene{torus}
\newcommand\firstmethod{HMC}
\newcommand\secondmethod{HMC\_Restore\_Vanilla\_Ours}
\newcommand\thirdmethod{HMC\_Waste-Recycling}
\newcommand\firstmethoddisplaynamea{HMC}
\newcommand\firstmethoddisplaynameb{\vphantom|}
\newcommand\secondmethoddisplaynamea{HMC}
\newcommand\secondmethoddisplaynameb{Vanilla (Ours)}
\newcommand\thirdmethoddisplaynamea{HMC}
\newcommand\thirdmethoddisplaynameb{Waste-Recycling}

\pgfmathsetmacro\croplength{.12}
\pgfmathsetmacro\firstcropxmin{.25}
\pgfmathsetmacro\firstcropymin{.3}
\pgfmathsetmacro\secondcropxmin{.075}
\pgfmathsetmacro\secondcropymin{.25}
\newcommand\firstcropcolor{red}
\newcommand\secondcropcolor{blue}

\newcommand{\SplitThreeImages}[3]{
    \begin{scope}
       \clip (0, 0) -- (.2 * \aspectratio, 0) -- (.3 * \aspectratio, 1) -- (0, 1) -- cycle;
        \path[fill overzoom image = results/#1.jpg] (0, 0) rectangle (\aspectratio, 1);
    \end{scope}
    \begin{scope}
       \clip (.2 * \aspectratio, 0) -- (0.6 * \aspectratio, 0) -- (0.7 * \aspectratio, 1) -- (0.3 * \aspectratio, 1) -- cycle;
        \path[fill overzoom image = results/#2.jpg] (0, 0) rectangle (\aspectratio, 1);
    \end{scope}
   \begin{scope}
       \clip (.6 * \aspectratio, 0) -- (\aspectratio, 0) -- (\aspectratio, 1) -- (0.7 * \aspectratio, 1) -- cycle;
       \path[fill overzoom image = results/#3.jpg] (0,0) rectangle (\aspectratio, 1);
   \end{scope}
   \begin{scope}
        \draw[black, thick] (0, 0) rectangle (\aspectratio, 1);
        \draw[draw=black,thick] (.2 * \aspectratio, 0) -- (.3 * \aspectratio, 1);
        \draw[draw=black,thick] (.6 * \aspectratio, 0) -- (.7 * \aspectratio, 1);
   \end{scope}
}

%   - #1: scene name (without extension; image is results/#1.jpg)
%   - #2: crop x (normalized, 0–1; 0=left)
%   - #3: crop y (normalized, 0–1; 0=bottom; note: image origin is lower left)
%   - #4: crop length (normalized)
%   - #5: color (for drawing the outline)
\newcommand{\CropImage}[5]{%
    \begingroup
        % Convert normalized crop to pixel values
        \pgfmathsetmacro\length{#4 * \width}
        \pgfmathsetmacro\xminval{#2 * \width}
        \pgfmathsetmacro\xmaxval{#2 * \width + \length}
        \pgfmathsetmacro\yminval{(1 - #3) * \height - \length}
        \pgfmathsetmacro\ymaxval{(1 - #3) * \height}

        \begin{scope}
            % Use fill with a rectangle, not a node
            \path[fill overzoom image = results/#1.jpg] (0, 0) rectangle (1, 1);
            % Now clip to viewport
            \clip (0, 0) rectangle (1, 1);
            \begin{scope}[xscale = 1, yscale = 1]
                \pgfmathsetmacro{\trimleft}{\xminval}
                \pgfmathsetmacro{\trimbottom}{\height - \ymaxval}  % Flip y-axis because trim starts from top-left
                \pgfmathsetmacro{\trimright}{\width - \xmaxval}
                \pgfmathsetmacro{\trimtop}{\yminval}
                
                \node at (0.5,0.5) {%
                    \includegraphics[
                        width = 1.78cm, height = 1.78cm,
                        trim = {\trimleft bp} {\trimbottom bp} {\trimright bp} {\trimtop bp},
                        clip
                    ]{results/#1.jpg}%
                };
            \end{scope}
        \end{scope}

        \begin{scope}
            \draw[#5, thick] (0, 0) rectangle (1, 1);
        \end{scope}
    \endgroup
}

\def\arraystretch{.5}
\setlength\tabcolsep{1pt}

\begin{tabularx}{\linewidth}{@{}cc@{}}
    \begin{minipage}[t]{.48\linewidth}
    \vspace{0pt}%
    \begin{tikzpicture}[scale = 5.5]
        \SplitThreeImages{\scene/\firstmethod}{\scene/\secondmethod}{\scene/\thirdmethod}
        \begin{scope}
            \draw[\firstcropcolor, ultra thick]
            (\aspectratio * \firstcropxmin, \firstcropymin - .05)
            rectangle
            (\aspectratio * \firstcropxmin + \aspectratio * \croplength, \firstcropymin - .05 + \aspectratio * \croplength);
            
            \draw[\secondcropcolor, ultra thick]
            (\aspectratio * \secondcropxmin, \secondcropymin - .05)
            rectangle
            (\aspectratio * \secondcropxmin + \aspectratio * \croplength, \secondcropymin - .05 + \aspectratio * \croplength);
        \end{scope}
        \begin{scope}
            \contourlength{.1em}
            \MethodLabelAt{.025}{.9}{west}{\firstmethoddisplaynamea}{\firstmethoddisplaynameb}
            \MethodLabelAt{.6}{.9}{west}{\secondmethoddisplaynamea}{\secondmethoddisplaynameb}
            \MethodLabelAt{1.26}{.9}{west}{\thirdmethoddisplaynamea}{\thirdmethoddisplaynameb}
        \end{scope}
    \end{tikzpicture}
    \end{minipage}
    &
    \hspace{1.25cm}
    \begin{minipage}[t]{.48\linewidth}
    \vspace{0pt}%
    \begin{tabular}{@{}ccccc@{}}
        % row 1
        \begin{tikzpicture}[scale = 1.78]
            \CropImage{\scene/\firstmethod}\firstcropxmin\firstcropymin\croplength{\firstcropcolor}
        \end{tikzpicture}
        &
        \begin{tikzpicture}[scale = 1.78]
            \CropImage{\scene/\firstmethod-MAPE}\firstcropxmin\firstcropymin\croplength{\firstcropcolor}
        \end{tikzpicture}
        &
        \begin{tikzpicture}[scale = 1.78]
            \CropImage{\scene/\firstmethod}\secondcropxmin\secondcropymin\croplength{\secondcropcolor}
        \end{tikzpicture}
        &
        \begin{tikzpicture}[scale = 1.78]
            \CropImage{\scene/\firstmethod-MAPE}\secondcropxmin\secondcropymin\croplength{\secondcropcolor}
        \end{tikzpicture}
        &
        \RotTableMethodLabel{\firstmethoddisplaynamea}{\firstmethoddisplaynameb}
        \\
        % row 2
        \begin{tikzpicture}[scale = 1.78]
            \CropImage{\scene/\secondmethod}\firstcropxmin\firstcropymin\croplength{\firstcropcolor}
        \end{tikzpicture}
        &
        \begin{tikzpicture}[scale = 1.78]
          \CropImage{\scene/\secondmethod-MAPE}\firstcropxmin\firstcropymin\croplength{\firstcropcolor}
        \end{tikzpicture}
        &
        \begin{tikzpicture}[scale = 1.78]
            \CropImage{\scene/\secondmethod}\secondcropxmin\secondcropymin\croplength{\secondcropcolor}
        \end{tikzpicture}
        &
        \begin{tikzpicture}[scale = 1.78]
            \CropImage{\scene/\secondmethod-MAPE}\secondcropxmin\secondcropymin\croplength{\secondcropcolor}
        \end{tikzpicture}
        &
        \RotTableMethodLabel{\secondmethoddisplaynamea}{\secondmethoddisplaynameb}
        \\
        % row 3
        \begin{tikzpicture}[scale = 1.78]
            \CropImage{\scene/\thirdmethod}\firstcropxmin\firstcropymin\croplength{\firstcropcolor}
        \end{tikzpicture}
        &
        \begin{tikzpicture}[scale = 1.78]
            \CropImage{\scene/\thirdmethod-MAPE}\firstcropxmin\firstcropymin\croplength{\firstcropcolor}
        \end{tikzpicture}
        &
        \begin{tikzpicture}[scale = 1.78]
            \CropImage{\scene/\thirdmethod}\secondcropxmin\secondcropymin\croplength{\secondcropcolor}
        \end{tikzpicture}
        &
        \begin{tikzpicture}[scale = 1.78]
            \CropImage{\scene/\thirdmethod-MAPE}\secondcropxmin\secondcropymin\croplength{\secondcropcolor}
        \end{tikzpicture}
        &
        \RotTableMethodLabel{\thirdmethoddisplaynamea}{\thirdmethoddisplaynameb}
    \end{tabular}
    \end{minipage}
\end{tabularx}

%% file: figures/bathroom2/mala_new_layout.tex
%!TEX root = ../main.tex

% Rotated two-line method label (for right margin of table)
% #1 first line, #2 second line
\newcommand{\RotTableMethodLabel}[2]{%
    \smash{%
        \raisebox{0.5\height}{% optisch zentrieren; bei Bedarf feinjustieren
            \rotatebox[origin = c]{90}{%
                \tiny\begin{tabular}{@{}l@{}}
                    \hspace{.6cm}#1\\[.1ex]
                    \hspace{.6cm}#2
                \end{tabular}%
            }%
        }%
    }%
}

% #1 = x position
% #2 = y position
% #3 = anchor
% #3 = first line
% #4 = second line
\newcommand{\MethodLabelAt}[5]{%
  \node[anchor = #3] at (#1, #2) {%
    \shortstack[l]{%
      \contour{black}{\small\textcolor{white}{#4}}\\[-0.2ex]%
      \contour{black}{\small\textcolor{white}{#5}}%
    }%
  };%
}

% Two-line tiny label for tables (left-aligned)
\newcommand{\TableMethodLabel}[2]{%
  \begin{tabular}{@{}l@{}}%
    \tiny #1\\[-0.1ex]%
    \tiny #2%
  \end{tabular}%
}

\pgfmathsetmacro\width{1024}
\pgfmathsetmacro\height{576}
\pgfmathsetmacro\aspectratio{16 / 9}

\newcommand\scene{bathroom2}
\newcommand\firstmethod{MALA}
\newcommand\secondmethod{MALA\_Vanilla\_Ours}
\newcommand\thirdmethod{MALA\_Waste-Recycling}
\newcommand\firstmethoddisplaynamea{MALA}
\newcommand\firstmethoddisplaynameb{\vphantom|}
\newcommand\secondmethoddisplaynamea{MALA}
\newcommand\secondmethoddisplaynameb{Vanilla (Ours)}
\newcommand\thirdmethoddisplaynamea{MALA}
\newcommand\thirdmethoddisplaynameb{Waste-Recycling}

\pgfmathsetmacro\croplength{.12}
\pgfmathsetmacro\firstcropxmin{.2}
\pgfmathsetmacro\firstcropymin{.2}
\pgfmathsetmacro\secondcropxmin{.35}
\pgfmathsetmacro\secondcropymin{.12}
\newcommand\firstcropcolor{red}
\newcommand\secondcropcolor{blue}

\newcommand{\SplitThreeImages}[3]{
    \begin{scope}
       \clip (0, 0) -- (.2 * \aspectratio, 0) -- (.3 * \aspectratio, 1) -- (0, 1) -- cycle;
        \path[fill overzoom image = results/#1.jpg] (0, 0) rectangle (\aspectratio, 1);
    \end{scope}
    \begin{scope}
       \clip (.2 * \aspectratio, 0) -- (0.6 * \aspectratio, 0) -- (0.7 * \aspectratio, 1) -- (0.3 * \aspectratio, 1) -- cycle;
        \path[fill overzoom image = results/#2.jpg] (0, 0) rectangle (\aspectratio, 1);
    \end{scope}
   \begin{scope}
       \clip (.6 * \aspectratio, 0) -- (\aspectratio, 0) -- (\aspectratio, 1) -- (0.7 * \aspectratio, 1) -- cycle;
       \path[fill overzoom image = results/#3.jpg] (0,0) rectangle (\aspectratio, 1);
   \end{scope}
   \begin{scope}
        \draw[black, thick] (0, 0) rectangle (\aspectratio, 1);
        \draw[draw=black,thick] (.2 * \aspectratio, 0) -- (.3 * \aspectratio, 1);
        \draw[draw=black,thick] (.6 * \aspectratio, 0) -- (.7 * \aspectratio, 1);
   \end{scope}
}

%   - #1: scene name (without extension; image is results/#1.jpg)
%   - #2: crop x (normalized, 0–1; 0=left)
%   - #3: crop y (normalized, 0–1; 0=bottom; note: image origin is lower left)
%   - #4: crop length (normalized)
%   - #5: color (for drawing the outline)
\newcommand{\CropImage}[5]{%
    \begingroup
        % Convert normalized crop to pixel values
        \pgfmathsetmacro\length{#4 * \width}
        \pgfmathsetmacro\xminval{#2 * \width}
        \pgfmathsetmacro\xmaxval{#2 * \width + \length}
        \pgfmathsetmacro\yminval{(1 - #3) * \height - \length}
        \pgfmathsetmacro\ymaxval{(1 - #3) * \height}

        \begin{scope}
            % Use fill with a rectangle, not a node
            \path[fill overzoom image = results/#1.jpg] (0, 0) rectangle (1, 1);
            % Now clip to viewport
            \clip (0, 0) rectangle (1, 1);
            \begin{scope}[xscale = 1, yscale = 1]
                \pgfmathsetmacro{\trimleft}{\xminval}
                \pgfmathsetmacro{\trimbottom}{\height - \ymaxval}  % Flip y-axis because trim starts from top-left
                \pgfmathsetmacro{\trimright}{\width - \xmaxval}
                \pgfmathsetmacro{\trimtop}{\yminval}
                
                \node at (0.5,0.5) {%
                    \includegraphics[
                        width = 1.78cm, height = 1.78cm,
                        trim = {\trimleft bp} {\trimbottom bp} {\trimright bp} {\trimtop bp},
                        clip
                    ]{results/#1.jpg}%
                };
            \end{scope}
        \end{scope}

        \begin{scope}
            \draw[#5, thick] (0, 0) rectangle (1, 1);
        \end{scope}
    \endgroup
}

\def\arraystretch{.5}
\setlength\tabcolsep{1pt}

\begin{tabularx}{\linewidth}{@{}cc@{}}
    \begin{minipage}[t]{.48\linewidth}
    \vspace{0pt}%
    \begin{tikzpicture}[scale = 5.5]
        \SplitThreeImages{\scene/\firstmethod}{\scene/\secondmethod}{\scene/\thirdmethod}
        \begin{scope}
            \draw[\firstcropcolor, ultra thick]
            (\aspectratio * \firstcropxmin, \firstcropymin)
            rectangle
            (\aspectratio * \firstcropxmin + \aspectratio * \croplength, \firstcropymin + \aspectratio * \croplength);
            
            \draw[\secondcropcolor, ultra thick]
            (\aspectratio * \secondcropxmin, \secondcropymin)
            rectangle
            (\aspectratio * \secondcropxmin + \aspectratio * \croplength, \secondcropymin + \aspectratio * \croplength);
        \end{scope}
        \begin{scope}
            \contourlength{.1em}
            \MethodLabelAt{.025}{.9}{west}{\firstmethoddisplaynamea}{\firstmethoddisplaynameb}
            \MethodLabelAt{.6}{.9}{west}{\secondmethoddisplaynamea}{\secondmethoddisplaynameb}
            \MethodLabelAt{1.26}{.9}{west}{\thirdmethoddisplaynamea}{\thirdmethoddisplaynameb}
        \end{scope}
    \end{tikzpicture}
    \end{minipage}
    &
    \hspace{1.25cm}
    \begin{minipage}[t]{.48\linewidth}
    \vspace{0pt}%
    \begin{tabular}{@{}ccccc@{}}
        % row 1
        \begin{tikzpicture}[scale = 1.78]
            \CropImage{\scene/\firstmethod}\firstcropxmin\firstcropymin\croplength{\firstcropcolor}
        \end{tikzpicture}
        &
        \begin{tikzpicture}[scale = 1.78]
            \CropImage{\scene/\firstmethod-MAPE}\firstcropxmin\firstcropymin\croplength{\firstcropcolor}
        \end{tikzpicture}
        &
        \begin{tikzpicture}[scale = 1.78]
            \CropImage{\scene/\firstmethod}\secondcropxmin\secondcropymin\croplength{\secondcropcolor}
        \end{tikzpicture}
        &
        \begin{tikzpicture}[scale = 1.78]
            \CropImage{\scene/\firstmethod-MAPE}\secondcropxmin\secondcropymin\croplength{\secondcropcolor}
        \end{tikzpicture}
        &
        \RotTableMethodLabel{\firstmethoddisplaynamea}{\firstmethoddisplaynameb}
        \\
        % row 2
        \begin{tikzpicture}[scale = 1.78]
            \CropImage{\scene/\secondmethod}\firstcropxmin\firstcropymin\croplength{\firstcropcolor}
        \end{tikzpicture}
        &
        \begin{tikzpicture}[scale = 1.78]
          \CropImage{\scene/\secondmethod-MAPE}\firstcropxmin\firstcropymin\croplength{\firstcropcolor}
        \end{tikzpicture}
        &
        \begin{tikzpicture}[scale = 1.78]
            \CropImage{\scene/\secondmethod}\secondcropxmin\secondcropymin\croplength{\secondcropcolor}
        \end{tikzpicture}
        &
        \begin{tikzpicture}[scale = 1.78]
            \CropImage{\scene/\secondmethod-MAPE}\secondcropxmin\secondcropymin\croplength{\secondcropcolor}
        \end{tikzpicture}
        &
        \RotTableMethodLabel{\secondmethoddisplaynamea}{\secondmethoddisplaynameb}
        \\
        % row 3
        \begin{tikzpicture}[scale = 1.78]
            \CropImage{\scene/\thirdmethod}\firstcropxmin\firstcropymin\croplength{\firstcropcolor}
        \end{tikzpicture}
        &
        \begin{tikzpicture}[scale = 1.78]
            \CropImage{\scene/\thirdmethod-MAPE}\firstcropxmin\firstcropymin\croplength{\firstcropcolor}
        \end{tikzpicture}
        &
        \begin{tikzpicture}[scale = 1.78]
            \CropImage{\scene/\thirdmethod}\secondcropxmin\secondcropymin\croplength{\secondcropcolor}
        \end{tikzpicture}
        &
        \begin{tikzpicture}[scale = 1.78]
            \CropImage{\scene/\thirdmethod-MAPE}\secondcropxmin\secondcropymin\croplength{\secondcropcolor}
        \end{tikzpicture}
        &
        \RotTableMethodLabel{\thirdmethoddisplaynamea}{\thirdmethoddisplaynameb}
    \end{tabular}
    \end{minipage}
\end{tabularx}

%% file: figures/glass_of_water/metropolis_restore_new_layout.tex
%!TEX root = ../main.tex

% Rotated two-line method label (for right margin of table)
% #1 first line, #2 second line
\newcommand{\RotTableMethodLabel}[2]{%
    \smash{%
        \raisebox{0.5\height}{% optisch zentrieren; bei Bedarf feinjustieren
            \rotatebox[origin = c]{90}{%
                \tiny\begin{tabular}{@{}l@{}}
                    \hspace{.6cm}#1\\[.1ex]
                    \hspace{.6cm}#2
                \end{tabular}%
            }%
        }%
    }%
}

% #1 = x position
% #2 = y position
% #3 = anchor
% #3 = first line
% #4 = second line
\newcommand{\MethodLabelAt}[5]{%
  \node[anchor = #3] at (#1, #2) {%
    \shortstack[l]{%
      \contour{black}{\small\textcolor{white}{#4}}\\[-0.2ex]%
      \contour{black}{\small\textcolor{white}{#5}}%
    }%
  };%
}

% Two-line tiny label for tables (left-aligned)
\newcommand{\TableMethodLabel}[2]{%
  \begin{tabular}{@{}l@{}}%
    \tiny #1\\[-0.1ex]%
    \tiny #2%
  \end{tabular}%
}

% \pgfmathsetmacro\width{1024}
% \pgfmathsetmacro\height{768}
\pgfmathsetmacro\width{1280}
\pgfmathsetmacro\height{720}
\pgfmathsetmacro\aspectratio{16 / 9}

\newcommand\scene{glass_of_water}
\newcommand\firstmethod{Metropolis\_Restore}
\newcommand\secondmethod{Metropolis\_Restore\_Vanilla\_Ours}
\newcommand\thirdmethod{Metropolis\_Restore\_Waste-Recycling}
\newcommand\firstmethoddisplaynamea{Metropolis Restore}
\newcommand\firstmethoddisplaynameb{\vphantom|}
\newcommand\secondmethoddisplaynamea{Metropolis Restore}
\newcommand\secondmethoddisplaynameb{Vanilla (Ours)}
\newcommand\thirdmethoddisplaynamea{Metropolis Restore}
\newcommand\thirdmethoddisplaynameb{Waste-Recycling}

\pgfmathsetmacro\croplength{.12}
\pgfmathsetmacro\firstcropxmin{.275}
\pgfmathsetmacro\firstcropymin{.025}
\pgfmathsetmacro\secondcropxmin{.75}
\pgfmathsetmacro\secondcropymin{.125}
\newcommand\firstcropcolor{red}
\newcommand\secondcropcolor{blue}

\newcommand{\SplitThreeImages}[3]{
    \begin{scope}
       \clip (0, 0) -- (.2 * \aspectratio, 0) -- (.3 * \aspectratio, 1) -- (0, 1) -- cycle;
        \path[fill overzoom image = results/#1.jpg] (0, 0) rectangle (\aspectratio, 1);
    \end{scope}
    \begin{scope}
       \clip (.2 * \aspectratio, 0) -- (0.6 * \aspectratio, 0) -- (0.7 * \aspectratio, 1) -- (0.3 * \aspectratio, 1) -- cycle;
        \path[fill overzoom image = results/#2.jpg] (0, 0) rectangle (\aspectratio, 1);
    \end{scope}
   \begin{scope}
       \clip (.6 * \aspectratio, 0) -- (\aspectratio, 0) -- (\aspectratio, 1) -- (0.7 * \aspectratio, 1) -- cycle;
       \path[fill overzoom image = results/#3.jpg] (0,0) rectangle (\aspectratio, 1);
   \end{scope}
   \begin{scope}
        \draw[black, thick] (0, 0) rectangle (\aspectratio, 1);
        \draw[draw=black,thick] (.2 * \aspectratio, 0) -- (.3 * \aspectratio, 1);
        \draw[draw=black,thick] (.6 * \aspectratio, 0) -- (.7 * \aspectratio, 1);
   \end{scope}
}

%   - #1: scene name (without extension; image is results/#1.jpg)
%   - #2: crop x (normalized, 0–1; 0=left)
%   - #3: crop y (normalized, 0–1; 0=bottom; note: image origin is lower left)
%   - #4: crop length (normalized)
%   - #5: color (for drawing the outline)
\newcommand{\CropImage}[5]{%
    \begingroup
        % Convert normalized crop to pixel values
        \pgfmathsetmacro\length{#4 * \width}
        \pgfmathsetmacro\xminval{#2 * \width}
        \pgfmathsetmacro\xmaxval{#2 * \width + \length}
        \pgfmathsetmacro\yminval{(1 - #3) * \height - \length}
        \pgfmathsetmacro\ymaxval{(1 - #3) * \height}

        \begin{scope}
            % Use fill with a rectangle, not a node
            \path[fill overzoom image = results/#1.jpg] (0, 0) rectangle (1, 1);
            % Now clip to viewport
            \clip (0, 0) rectangle (1, 1);
            \begin{scope}[xscale = 1, yscale = 1]
                \pgfmathsetmacro{\trimleft}{\xminval}
                \pgfmathsetmacro{\trimbottom}{\height - \ymaxval}  % Flip y-axis because trim starts from top-left
                \pgfmathsetmacro{\trimright}{\width - \xmaxval}
                \pgfmathsetmacro{\trimtop}{\yminval}
                
                \node at (0.5,0.5) {%
                    \includegraphics[
                        width = 1.78cm, height = 1.78cm,
                        trim = {\trimleft bp} {\trimbottom bp} {\trimright bp} {\trimtop bp},
                        clip
                    ]{results/#1.jpg}%
                };
            \end{scope}
        \end{scope}

        \begin{scope}
            \draw[#5, thick] (0, 0) rectangle (1, 1);
        \end{scope}
    \endgroup
}

\def\arraystretch{.5}
\setlength\tabcolsep{1pt}

\begin{tabularx}{\linewidth}{@{}cc@{}}
    \begin{minipage}[t]{.48\linewidth}
    \vspace{0pt}%
    \begin{tikzpicture}[scale = 5.5]
        \SplitThreeImages{\scene/\firstmethod}{\scene/\secondmethod}{\scene/\thirdmethod}
        \begin{scope}
            \draw[\firstcropcolor, ultra thick]
            (\aspectratio * \firstcropxmin, \firstcropymin)
            rectangle
            (\aspectratio * \firstcropxmin + \aspectratio * \croplength, \firstcropymin + \aspectratio * \croplength);
            
            \draw[\secondcropcolor, ultra thick]
            (\aspectratio * \secondcropxmin, \secondcropymin)
            rectangle
            (\aspectratio * \secondcropxmin + \aspectratio * \croplength, \secondcropymin + \aspectratio * \croplength);
        \end{scope}
        \begin{scope}
            \contourlength{.1em}
            \MethodLabelAt{.025}{.9}{west}{\firstmethoddisplaynamea}{\firstmethoddisplaynameb}
            \MethodLabelAt{.6}{.9}{west}{\secondmethoddisplaynamea}{\secondmethoddisplaynameb}
            \MethodLabelAt{1.26}{.9}{west}{\thirdmethoddisplaynamea}{\thirdmethoddisplaynameb}
        \end{scope}
    \end{tikzpicture}
    \end{minipage}
    &
    \hspace{1.25cm}
    \begin{minipage}[t]{.48\linewidth}
    \vspace{0pt}%
    \begin{tabular}{@{}ccccc@{}}
        % row 1
        \begin{tikzpicture}[scale = 1.78]
            \CropImage{\scene/\firstmethod}\firstcropxmin\firstcropymin\croplength{\firstcropcolor}
        \end{tikzpicture}
        &
        \begin{tikzpicture}[scale = 1.78]
            \CropImage{\scene/\firstmethod-MAPE}\firstcropxmin\firstcropymin\croplength{\firstcropcolor}
        \end{tikzpicture}
        &
        \begin{tikzpicture}[scale = 1.78]
            \CropImage{\scene/\firstmethod}\secondcropxmin\secondcropymin\croplength{\secondcropcolor}
        \end{tikzpicture}
        &
        \begin{tikzpicture}[scale = 1.78]
            \CropImage{\scene/\firstmethod-MAPE}\secondcropxmin\secondcropymin\croplength{\secondcropcolor}
        \end{tikzpicture}
        &
        \RotTableMethodLabel{\firstmethoddisplaynamea}{\firstmethoddisplaynameb}
        \\
        % row 2
        \begin{tikzpicture}[scale = 1.78]
            \CropImage{\scene/\secondmethod}\firstcropxmin\firstcropymin\croplength{\firstcropcolor}
        \end{tikzpicture}
        &
        \begin{tikzpicture}[scale = 1.78]
          \CropImage{\scene/\secondmethod-MAPE}\firstcropxmin\firstcropymin\croplength{\firstcropcolor}
        \end{tikzpicture}
        &
        \begin{tikzpicture}[scale = 1.78]
            \CropImage{\scene/\secondmethod}\secondcropxmin\secondcropymin\croplength{\secondcropcolor}
        \end{tikzpicture}
        &
        \begin{tikzpicture}[scale = 1.78]
            \CropImage{\scene/\secondmethod-MAPE}\secondcropxmin\secondcropymin\croplength{\secondcropcolor}
        \end{tikzpicture}
        &
        \RotTableMethodLabel{\secondmethoddisplaynamea}{\secondmethoddisplaynameb}
        \\
        % row 3
        \begin{tikzpicture}[scale = 1.78]
            \CropImage{\scene/\thirdmethod}\firstcropxmin\firstcropymin\croplength{\firstcropcolor}
        \end{tikzpicture}
        &
        \begin{tikzpicture}[scale = 1.78]
            \CropImage{\scene/\thirdmethod-MAPE}\firstcropxmin\firstcropymin\croplength{\firstcropcolor}
        \end{tikzpicture}
        &
        \begin{tikzpicture}[scale = 1.78]
            \CropImage{\scene/\thirdmethod}\secondcropxmin\secondcropymin\croplength{\secondcropcolor}
        \end{tikzpicture}
        &
        \begin{tikzpicture}[scale = 1.78]
            \CropImage{\scene/\thirdmethod-MAPE}\secondcropxmin\secondcropymin\croplength{\secondcropcolor}
        \end{tikzpicture}
        &
        \RotTableMethodLabel{\thirdmethoddisplaynamea}{\thirdmethoddisplaynameb}
    \end{tabular}
    \end{minipage}
\end{tabularx}

%% file: nomenclature.tex
\begin{table}
    \centering
    \small
    \captionsetup{skip = 4pt}
    \caption{
        Commonly used notation throughout the paper.
    }
    \label{tab:nomenclature}
    \vspace{-1.5mm}
    \setlength{\tabcolsep}{2.5pt}
    \fontsize{6.9pt}{7pt}
    \begin{tabularx}{\columnwidth}{l X}
        \toprule
        \small\textbf{Notation} & \small\textbf{Description} \\
        \midrule
        \hyperref[inline:target-distribution]{$\targetdistribution$}&Target distribution (p. \pageref{inline:target-distribution})\\
        \hyperref[inline:reference-measure]{$\referencemeasure$}&Reference measure (p. \pageref{inline:reference-measure})\\
        \hyperref[eq:target-distribution-has-density]{$\targetdensity$}&Density of $\targetdistribution$ with respect to $\referencemeasure$ (p. \pageref{eq:target-distribution-has-density})\\
        \hyperref[eq:targetdensity-assumption-1]{$\targetdensity_\referencemeasure$}&Normalization constant of $\targetdensity$ (p. \pageref{eq:targetdensity-assumption-1})\\
        \hyperref[eq:integral]{$\integral$}&Integral of $\integrand$ with respect to $\targetdistribution$ (p. \pageref{eq:integral})\\
        \hyperref[inline:proposal-kernel]{$\proposalkernel$}&Proposal kernel of the \gls{mh} algorithm (p. \pageref{inline:proposal-kernel})\\
        \hyperref[eq:proposal-kernel-has-density]{$\proposaldensity$}&Density of $\proposalkernel$ with respect to $\referencemeasure$ (p. \pageref{eq:proposal-kernel-has-density})\\
        \hyperref[inline:acceptance-function]{$\acceptancefunction$}&Acceptance function of the \gls{mh} algorithm (p. \pageref{inline:acceptance-function})\\
        \hyperref[eq:ergodic-theorem]{$\ergodicaverage$}&Ergodic average estimator (p. \pageref{eq:ergodic-theorem})\\
        \hyperref[eq:vanilla-estimation]{$\vanillaestimator$}&Vanilla Rao--Blackwellized estimator (p. \pageref{eq:vanilla-estimation})\\
        \hyperref[eq:waste-recycling]{$\wasterecyclingestimator$}&Waste-recycling estimator (p. \pageref{eq:waste-recycling})\\
        % \hyperref[inline:product]{$\otimes$}&Product of transition kernels (p. \pageref{inline:product})\\
        % \hyperref[inline:dirac-measure]{$\delta_\point$}&Dirac measure at $\point$ (p. \pageref{inline:dirac-measure})\\
        $\mathcal U_{[0,\;1)}$&Uniform distribution on $[0,1)$\\
    \end{tabularx}
\end{table}

%% file: main.bbl

\begin{thebibliography}{37}

%%% the Web address.

\ifx \showCODEN    \undefined \def \showCODEN     #1{\unskip}     \fi
\ifx \showISBNx    \undefined \def \showISBNx     #1{\unskip}     \fi
\ifx \showISBNxiii \undefined \def \showISBNxiii  #1{\unskip}     \fi
\ifx \showISSN     \undefined \def \showISSN      #1{\unskip}     \fi
\ifx \showLCCN     \undefined \def \showLCCN      #1{\unskip}     \fi
\ifx \shownote     \undefined \def \shownote      #1{#1}          \fi
\ifx \showarticletitle \undefined \def \showarticletitle #1{#1}   \fi
\ifx \showURL      \undefined \def \showURL       {\relax}        \fi
% invisible to TeX
\providecommand\bibfield[2]{#2}
\providecommand\bibinfo[2]{#2}
\providecommand\natexlab[1]{#1}
\providecommand\showeprint[2][]{arXiv:#2}

\bibitem[Bitterli(2016)]%
        {bitterli2016resources}
\bibfield{author}{\bibinfo{person}{Benedikt Bitterli}.} \bibinfo{year}{2016}\natexlab{}.
\newblock \bibinfo{title}{Rendering resources}.
\newblock
\newblock
\shownote{https://benedikt-bitterli.me/resources/}.


\bibitem[Blackwell(1947)]%
        {blackwell1947conditional}
\bibfield{author}{\bibinfo{person}{David Blackwell}.} \bibinfo{year}{1947}\natexlab{}.
\newblock \showarticletitle{Conditional Expectation and Unbiased Sequential Estimation}.
\newblock \bibinfo{journal}{\emph{The Annals of Mathematical Statistics}} \bibinfo{volume}{18}, \bibinfo{number}{1} (\bibinfo{year}{1947}), \bibinfo{pages}{105--110}.
\newblock


\bibitem[Ceperley et~al\mbox{.}(1977)]%
        {ceperley1977waste}
\bibfield{author}{\bibinfo{person}{D. Ceperley}, \bibinfo{person}{G.~V. Chester}, {and} \bibinfo{person}{M.~H. Kalos}.} \bibinfo{year}{1977}\natexlab{}.
\newblock \showarticletitle{Monte Carlo simulation of a many-fermion study}.
\newblock \bibinfo{journal}{\emph{Phys. Rev. B}}  \bibinfo{volume}{16} (\bibinfo{date}{Oct} \bibinfo{year}{1977}), \bibinfo{pages}{3081--3099}.
\newblock
Issue 7.
\href{https://doi.org/10.1103/PhysRevB.16.3081}{doi:\nolinkurl{10.1103/PhysRevB.16.3081}}


\bibitem[Delmas and Jourdain(2009)]%
        {delmas2009waste}
\bibfield{author}{\bibinfo{person}{Jean-François Delmas} {and} \bibinfo{person}{Benjamin Jourdain}.} \bibinfo{year}{2009}\natexlab{}.
\newblock \bibinfo{title}{Does waste-recycling really improve Metropolis-Hastings Monte Carlo algorithm?}
\newblock
\showeprint[arxiv]{math/0611949}~[math.PR]
\urldef\tempurl%
\url{https://arxiv.org/abs/math/0611949}
\showURL{%
\tempurl}


\bibitem[Douc et~al\mbox{.}(2018)]%
        {douc2018markov}
\bibfield{author}{\bibinfo{person}{Randal Douc}, \bibinfo{person}{Eric Moulines}, \bibinfo{person}{Pierre Priouret}, {and} \bibinfo{person}{Philippe Soulier}.} \bibinfo{year}{2018}\natexlab{}.
\newblock \bibinfo{booktitle}{\emph{Markov Chains}}.
\newblock \bibinfo{publisher}{Springer}.
\newblock


\bibitem[Douc and Robert(2011)]%
        {douc2011vanilla}
\bibfield{author}{\bibinfo{person}{Randal Douc} {and} \bibinfo{person}{Christian~P. Robert}.} \bibinfo{year}{2011}\natexlab{}.
\newblock \showarticletitle{A vanilla {Rao}–{Blackwellization} of {Metropolis}–{Hastings} algorithms}.
\newblock \bibinfo{journal}{\emph{The Annals of Statistics}} \bibinfo{volume}{39}, \bibinfo{number}{1} (\bibinfo{date}{feb} \bibinfo{year}{2011}).
\newblock


\bibitem[Duane et~al\mbox{.}(1987)]%
        {duane1987hmc}
\bibfield{author}{\bibinfo{person}{Simon Duane}, \bibinfo{person}{A.~D. Kennedy}, \bibinfo{person}{Brian~J. Pendleton}, {and} \bibinfo{person}{Duncan Roweth}.} \bibinfo{year}{1987}\natexlab{}.
\newblock \showarticletitle{Hybrid Monte Carlo}.
\newblock \bibinfo{journal}{\emph{Physics Letters B}} \bibinfo{volume}{195}, \bibinfo{number}{2} (\bibinfo{year}{1987}).
\newblock


\bibitem[Durmus and Eberle(2023)]%
        {eberle2023inexact}
\bibfield{author}{\bibinfo{person}{Alain~Oliviero Durmus} {and} \bibinfo{person}{Andreas Eberle}.} \bibinfo{year}{2023}\natexlab{}.
\newblock \showarticletitle{Asymptotic bias of inexact Markov Chain Monte Carlo methods in high dimension}.
\newblock  (\bibinfo{year}{2023}).
\newblock
\showeprint[arxiv]{2108.00682}


\bibitem[Ethier and Kurtz(2009)]%
        {ethier2009markov}
\bibfield{author}{\bibinfo{person}{Stewart~N. Ethier} {and} \bibinfo{person}{Thomas~G. Kurtz}.} \bibinfo{year}{2009}\natexlab{}.
\newblock \bibinfo{booktitle}{\emph{Markov Processes: Characterization and Convergence}}.
\newblock \bibinfo{publisher}{John Wiley \& Sons}.
\newblock


\bibitem[Gruson et~al\mbox{.}(2020)]%
        {gruson2020stratified}
\bibfield{author}{\bibinfo{person}{Adrien Gruson}, \bibinfo{person}{Rex West}, {and} \bibinfo{person}{Toshiya Hachisuka}.} \bibinfo{year}{2020}\natexlab{}.
\newblock \showarticletitle{Stratified {Markov} Chain {Monte} {Carlo} light transport}.
\newblock \bibinfo{journal}{\emph{Computer Graphics Forum}} \bibinfo{volume}{39}, \bibinfo{number}{2} (\bibinfo{year}{2020}).
\newblock


\bibitem[Hachisuka et~al\mbox{.}(2014)]%
        {hachisuka2014multiplexed}
\bibfield{author}{\bibinfo{person}{Toshiya Hachisuka}, \bibinfo{person}{Anton~S. Kaplanyan}, {and} \bibinfo{person}{Carsten Dachsbacher}.} \bibinfo{year}{2014}\natexlab{}.
\newblock \showarticletitle{Multiplexed {Metropolis} light transport}.
\newblock \bibinfo{journal}{\emph{ACM Transactions on Graphics}} \bibinfo{volume}{33}, \bibinfo{number}{4} (\bibinfo{year}{2014}).
\newblock


\bibitem[Hastings(1970)]%
        {hastings1970monte}
\bibfield{author}{\bibinfo{person}{W.~K. Hastings}.} \bibinfo{year}{1970}\natexlab{}.
\newblock \showarticletitle{Monte {Carlo} sampling methods using {Markov} chains and their applications}.
\newblock \bibinfo{journal}{\emph{Biometrika}} \bibinfo{volume}{57}, \bibinfo{number}{1} (\bibinfo{date}{4} \bibinfo{year}{1970}).
\newblock


\bibitem[Holl et~al\mbox{.}(2025)]%
        {holl2025jrlt}
\bibfield{author}{\bibinfo{person}{Sascha Holl}, \bibinfo{person}{Gurprit Singh}, {and} \bibinfo{person}{Hans-Peter Seidel}.} \bibinfo{year}{2025}\natexlab{}.
\newblock \showarticletitle{Jump {Restore} {Light} {Transport}}.
\newblock  (\bibinfo{year}{2025}).
\newblock


\bibitem[Kajiya(1986)]%
        {kajiya1986rendering}
\bibfield{author}{\bibinfo{person}{James~T. Kajiya}.} \bibinfo{year}{1986}\natexlab{}.
\newblock \showarticletitle{The rendering equation}.
\newblock \bibinfo{journal}{\emph{SIGGRAPH Comput. Graph.}} \bibinfo{volume}{20}, \bibinfo{number}{4} (\bibinfo{date}{aug} \bibinfo{year}{1986}), \bibinfo{pages}{143–150}.
\newblock
\showISSN{0097-8930}
\href{https://doi.org/10.1145/15886.15902}{doi:\nolinkurl{10.1145/15886.15902}}


\bibitem[Kallenberg(2021)]%
        {kallenberg2021probability}
\bibfield{author}{\bibinfo{person}{Olav Kallenberg}.} \bibinfo{year}{2021}\natexlab{}.
\newblock \bibinfo{booktitle}{\emph{Foundations of Modern Probability} (\bibinfo{edition}{3} ed.)}.
\newblock \bibinfo{publisher}{Springer Nature Switzerland AG 2021}.
\newblock


\bibitem[Kelemen et~al\mbox{.}(2002)]%
        {kelemen2002simple}
\bibfield{author}{\bibinfo{person}{Csaba Kelemen}, \bibinfo{person}{László Szirmay-Kalos}, \bibinfo{person}{György Antal}, {and} \bibinfo{person}{Ferenc Csonka}.} \bibinfo{year}{2002}\natexlab{}.
\newblock \showarticletitle{A simple and robust mutation strategy for the {Metropolis} light transport algorithm}.
\newblock \bibinfo{journal}{\emph{Computer Graphics Forum}} \bibinfo{volume}{21}, \bibinfo{number}{3} (\bibinfo{year}{2002}).
\newblock


\bibitem[Klenke(2020)]%
        {klenke2020probability}
\bibfield{author}{\bibinfo{person}{Achim Klenke}.} \bibinfo{year}{2020}\natexlab{}.
\newblock \bibinfo{booktitle}{\emph{Probability Theory: A Comprehensive Course} (\bibinfo{edition}{3} ed.)}.
\newblock \bibinfo{publisher}{Springer}.
\newblock


\bibitem[Lafortune and Willems(1996)]%
        {lafortune1996rendering}
\bibfield{author}{\bibinfo{person}{Eric~P. Lafortune} {and} \bibinfo{person}{Yves~D. Willems}.} \bibinfo{year}{1996}\natexlab{}.
\newblock \showarticletitle{Rendering Participating Media with Bidirectional Path Tracing}. In \bibinfo{booktitle}{\emph{Rendering Techniques '96}}, \bibfield{editor}{\bibinfo{person}{Xavier Pueyo} {and} \bibinfo{person}{Peter Schr{\"o}der}} (Eds.). \bibinfo{publisher}{Springer Vienna}, \bibinfo{address}{Vienna}, \bibinfo{pages}{91--100}.
\newblock


\bibitem[Li et~al\mbox{.}(2015)]%
        {li2015anisotropic}
\bibfield{author}{\bibinfo{person}{Tzu-Mao Li}, \bibinfo{person}{Jaakko Lehtinen}, \bibinfo{person}{Ravi Ramamoorthi}, \bibinfo{person}{Wenzel Jakob}, {and} \bibinfo{person}{Fr\'{e}do Durand}.} \bibinfo{year}{2015}\natexlab{}.
\newblock \showarticletitle{Anisotropic gaussian mutations for {Metropolis} light transport through {Hessian}-{Hamiltonian} dynamics}.
\newblock \bibinfo{journal}{\emph{ACM Transactions on Graphics}} \bibinfo{volume}{34}, \bibinfo{number}{6} (\bibinfo{year}{2015}).
\newblock


\bibitem[Luan et~al\mbox{.}(2020a)]%
        {luan2020langevin}
\bibfield{author}{\bibinfo{person}{Fujun Luan}, \bibinfo{person}{Shuang Zhao}, \bibinfo{person}{Kavita Bala}, {and} \bibinfo{person}{Ioannis Gkioulekas}.} \bibinfo{year}{2020}\natexlab{a}.
\newblock \showarticletitle{Langevin {Monte} {Carlo} rendering with gradient-based adaptation}.
\newblock \bibinfo{journal}{\emph{ACM Transactions on Graphics}} \bibinfo{volume}{39}, \bibinfo{number}{4} (\bibinfo{year}{2020}).
\newblock


\bibitem[Luan et~al\mbox{.}(2020b)]%
        {lmc}
\bibfield{author}{\bibinfo{person}{Fujun Luan}, \bibinfo{person}{Shuang Zhao}, \bibinfo{person}{Kavita Bala}, {and} \bibinfo{person}{Ioannis Gkioulekas}.} \bibinfo{year}{2020}\natexlab{b}.
\newblock \bibinfo{booktitle}{\emph{lmc}}.
\newblock
\urldef\tempurl%
\url{https://github.com/luanfujun/Langevin-MCMC}
\showURL{%
\tempurl}


\bibitem[McKimm et~al\mbox{.}(2024)]%
        {mckimm2024adaptive}
\bibfield{author}{\bibinfo{person}{Hector McKimm}, \bibinfo{person}{Andi~Q. Wang}, \bibinfo{person}{Murray Pollock}, \bibinfo{person}{Christian~P. Robert}, {and} \bibinfo{person}{Gareth~O. Roberts}.} \bibinfo{year}{2024}\natexlab{}.
\newblock \bibinfo{title}{Sampling using Adaptive Regenerative Processes}.
\newblock
\showeprint[arxiv]{2210.09901}
\urldef\tempurl%
\url{https://arxiv.org/abs/2210.09901}
\showURL{%
\tempurl}


\bibitem[Metropolis et~al\mbox{.}(1953)]%
        {metropolis1953equation}
\bibfield{author}{\bibinfo{person}{Nicholas Metropolis}, \bibinfo{person}{Arianna~W. Rosenbluth}, \bibinfo{person}{Marshall~N. Rosenbluth}, \bibinfo{person}{Augusta~H. Teller}, {and} \bibinfo{person}{Edward Teller}.} \bibinfo{year}{1953}\natexlab{}.
\newblock \showarticletitle{Equation of state calculations by fast computing machines}.
\newblock \bibinfo{journal}{\emph{The Journal of Chemical Physics}} \bibinfo{volume}{21}, \bibinfo{number}{6} (\bibinfo{date}{6} \bibinfo{year}{1953}).
\newblock


\bibitem[Meyn and Tweedie(1993)]%
        {meyn1993markov}
\bibfield{author}{\bibinfo{person}{Sean~P. Meyn} {and} \bibinfo{person}{Richard~L. Tweedie}.} \bibinfo{year}{1993}\natexlab{}.
\newblock \bibinfo{booktitle}{\emph{Markov Chains and Stochastic Stability}}.
\newblock \bibinfo{publisher}{Springer}.
\newblock


\bibitem[Pharr(2025)]%
        {pharr2025scenes}
\bibfield{author}{\bibinfo{person}{Matt Pharr}.} \bibinfo{year}{2025}\natexlab{}.
\newblock \bibinfo{title}{{pbrt-v4-scenes}: Example scenes for pbrt-v4}.
\newblock \bibinfo{howpublished}{\url{https://github.com/mmp/pbrt-v4-scenes}}.
\newblock
\newblock
\shownote{GitHub repository, accessed December 2025}.


\bibitem[Pharr et~al\mbox{.}(2021)]%
        {pharr2023pbrt}
\bibfield{author}{\bibinfo{person}{Matt Pharr}, \bibinfo{person}{Wenzel Jakob}, {and} \bibinfo{person}{Greg Humphreys}.} \bibinfo{year}{2021}\natexlab{}.
\newblock \bibinfo{booktitle}{\emph{Physically Based Rendering, fourth edition}}.
\newblock \bibinfo{publisher}{The MIT Press}.
\newblock
\urldef\tempurl%
\url{https://pbr-book.org/}
\showURL{%
\tempurl}


\bibitem[Pharr et~al\mbox{.}(2023)]%
        {pbrt}
\bibfield{author}{\bibinfo{person}{Matt Pharr}, \bibinfo{person}{Wenzel Jakob}, {and} \bibinfo{person}{Greg Humphreys}.} \bibinfo{year}{2023}\natexlab{}.
\newblock \bibinfo{booktitle}{\emph{pbrt}}.
\newblock
\urldef\tempurl%
\url{https://github.com/mmp/pbrt-v4}
\showURL{%
\tempurl}


\bibitem[Rao(1945)]%
        {rao1945information}
\bibfield{author}{\bibinfo{person}{C.~Radhakrishna Rao}.} \bibinfo{year}{1945}\natexlab{}.
\newblock \showarticletitle{Information and the Accuracy Attainable in the Estimation of Statistical Parameters}.
\newblock \bibinfo{journal}{\emph{Bulletin of the Calcutta Mathematical Society}}  \bibinfo{volume}{37} (\bibinfo{year}{1945}), \bibinfo{pages}{81--91}.
\newblock


\bibitem[Rioux-Lavoie et~al\mbox{.}(2020)]%
        {rioux2020delayed}
\bibfield{author}{\bibinfo{person}{Damien Rioux-Lavoie}, \bibinfo{person}{Joey Litalien}, \bibinfo{person}{Adrien Gruson}, \bibinfo{person}{Toshiya Hachisuka}, {and} \bibinfo{person}{Derek Nowrouzezahrai}.} \bibinfo{year}{2020}\natexlab{}.
\newblock \showarticletitle{Delayed rejection {Metropolis} light transport}.
\newblock \bibinfo{journal}{\emph{ACM Transactions on Graphics}} \bibinfo{volume}{39}, \bibinfo{number}{3} (\bibinfo{date}{may} \bibinfo{year}{2020}).
\newblock
\href{https://doi.org/10.1145/3388538}{doi:\nolinkurl{10.1145/3388538}}


\bibitem[Robert and Roberts(2021)]%
        {robert2021rao}
\bibfield{author}{\bibinfo{person}{Christian~P. Robert} {and} \bibinfo{person}{Gareth~O. Roberts}.} \bibinfo{year}{2021}\natexlab{}.
\newblock \bibinfo{title}{Rao-{Blackwellization} in the {MCMC} era}.
\newblock
\showeprint[arxiv]{2101.01011}
\urldef\tempurl%
\url{https://arxiv.org/abs/2101.01011}
\showURL{%
\tempurl}


\bibitem[Roberts and Tweedie(1996)]%
        {roberts1996mala}
\bibfield{author}{\bibinfo{person}{Gareth~O. Roberts} {and} \bibinfo{person}{Richard~L. Tweedie}.} \bibinfo{year}{1996}\natexlab{}.
\newblock \showarticletitle{Exponential convergence of Langevin distributions and their discrete approximations}.
\newblock \bibinfo{journal}{\emph{Bernoulli}} \bibinfo{volume}{2}, \bibinfo{number}{4} (\bibinfo{year}{1996}).
\newblock


\bibitem[Rudolf and Sprungk(2017)]%
        {rudolf2017importance}
\bibfield{author}{\bibinfo{person}{Daniel Rudolf} {and} \bibinfo{person}{Björn Sprungk}.} \bibinfo{year}{2017}\natexlab{}.
\newblock \showarticletitle{On a Metropolis-Hastings importance sampling estimator}.
\newblock \bibinfo{journal}{\emph{PAMM}}  \bibinfo{volume}{17} (\bibinfo{year}{2017}).
\newblock


\bibitem[Tierney(1998)]%
        {tierney1998note}
\bibfield{author}{\bibinfo{person}{Luke Tierney}.} \bibinfo{year}{1998}\natexlab{}.
\newblock \showarticletitle{A note on Metropolis-Hastings kernels for general state spaces}.
\newblock \bibinfo{journal}{\emph{The Annals of Applied Probability}}  \bibinfo{volume}{8} (\bibinfo{year}{1998}).
\newblock


\bibitem[Veach(1997)]%
        {veach1997thesis}
\bibfield{author}{\bibinfo{person}{Eric Veach}.} \bibinfo{year}{1997}\natexlab{}.
\newblock \emph{\bibinfo{title}{Robust Monte Carlo Methods for Light Transport Simulation}}.
\newblock \bibinfo{thesistype}{Ph.\,D. Dissertation}. \bibinfo{school}{Stanford University}.
\newblock


\bibitem[Veach and Guibas(1995)]%
        {veach1994bidirectional}
\bibfield{author}{\bibinfo{person}{Eric Veach} {and} \bibinfo{person}{Leonidas Guibas}.} \bibinfo{year}{1995}\natexlab{}.
\newblock \showarticletitle{Bidirectional Estimators for Light Transport}. In \bibinfo{booktitle}{\emph{Photorealistic Rendering Techniques}}, \bibfield{editor}{\bibinfo{person}{Georgios Sakas}, \bibinfo{person}{Stefan M{\"u}ller}, {and} \bibinfo{person}{Peter Shirley}} (Eds.). \bibinfo{publisher}{Springer Berlin Heidelberg}, \bibinfo{address}{Berlin, Heidelberg}, \bibinfo{pages}{145--167}.
\newblock


\bibitem[Wang et~al\mbox{.}(2021)]%
        {wang2021regeneration}
\bibfield{author}{\bibinfo{person}{Andi~Q. Wang}, \bibinfo{person}{Murray Pollock}, \bibinfo{person}{Gareth~O. Roberts}, {and} \bibinfo{person}{David Steinsaltz}.} \bibinfo{year}{2021}\natexlab{}.
\newblock \showarticletitle{Regeneration-enriched {Markov} processes with application to {Monte} {Carlo}}.
\newblock \bibinfo{journal}{\emph{The Annals of Applied Probability}} \bibinfo{volume}{31}, \bibinfo{number}{2} (\bibinfo{year}{2021}).
\newblock


\bibitem[Çinlar(2011)]%
        {cinlar2011probability}
\bibfield{author}{\bibinfo{person}{Erhan Çinlar}.} \bibinfo{year}{2011}\natexlab{}.
\newblock \bibinfo{booktitle}{\emph{Probability and Stochastics}}.
\newblock \bibinfo{publisher}{Springer Science+Business Media}.
\newblock


\end{thebibliography}
